\documentclass[fleqn]{elsart}
\usepackage{graphicx,color}
\usepackage{amsmath,amssymb}
\usepackage{slashbox}
\newcommand{\cal}{\mathcal}
\newcommand{\be}{\begin{equation}}
\newcommand{\ee}{\end{equation}}
\newcommand{\bea}{\begin{eqnarray}}
\newcommand{\eea}{\end{eqnarray}}
\newcommand{\Slash}[1]{{\ooalign{\hfil/\hfil\crcr$#1$}}}
\newcommand{\tr}{{\rm tr}}

\newcommand{\Nc}{N_{\rm c}}
\newcommand{\Nf}{N_{\rm f}}

\newcommand{\lqcd}{\Lambda_{\rm QCD}}
\newcommand{\lf}{\Lambda_{\rm f}}
\newcommand{\lc}{\Lambda_{\rm c}}
\newcommand{\vp}{\vec{p}}
\newcommand{\vq}{\vec{q}}
\newcommand{\vk}{\vec{k}}
\newcommand{\vl}{\vec{l}}

\newcommand{\vx}{\vec{x}}

\newcommand{\va}{\vec{a}}

\newcommand{\la}{\langle}
\newcommand{\ra}{\rangle}

\newcommand{\para}{\parallel}
\newcommand{\pF}{p_{\rm F}}
\newcommand{\ptot}{\vp_{\rm total}}

\newcommand{\calE}{\mathcal{E}}
\newcommand{\Np}{N_{\rm p}}
\newcommand{\rmd}{\mathrm{d}}
\newcommand{\rmi}{\mathrm{i}}
\newcommand{\rme}{\mathrm{e}}

\begin{document}
\vspace*{-5mm}
\begin{flushright}
{\scriptsize RBRC-911, RIKEN-MP-24}
\end{flushright}
\vspace{5mm}
\begin{frontmatter}
\title{Interweaving Chiral Spirals}
\author[bnl]{Toru Kojo},
\author[riken]{Yoshimasa Hidaka},
\author[keio]{Kenji Fukushima},
\author[bnl,bnl1]{Larry D. McLerran},
\author[bnl1]{Robert D. Pisarski}
\address[bnl]{RIKEN BNL Research Center,
 Brookhaven National Laboratory, Upton, NY-11973, USA}
\address[riken]{Mathematical Physics Laboratory,
 RIKEN Nishina Center, Saitama 351-0198, Japan}
\address[keio]{Department of Physics,
 Keio University, Kanagawa 223-8522, Japan}
\address[bnl1]{Department of Physics,
 Brookhaven National Laboratory, Upton, NY-11973, USA}
%
\begin{abstract}
We elaborate how to construct interweaving chiral spirals
in (2+1) dimensions,
defined as a superposition of chiral spirals oriented in different directions.
We divide a two-dimensional Fermi sea into distinct wedges,
characterized by the opening angle $2\Theta$ and depth
$Q\simeq \pF$, where $\pF$ is the Fermi momentum.  In each wedge,
the energy is lowered by forming a single chiral spiral.
The optimal values for
$\Theta$ and $Q$ are chosen by balancing this gain in energy
versus the cost of deforming the Fermi surface (which dominates at
large $\Theta$) and patch-patch interactions (dominant at small
$\Theta$).  Using a non-local
four-Fermi interaction model, we estimate the gain and cost in energy
by expanding in terms of $1/\Nc$ (where $\Nc$ is the number
of colors), $\lqcd/Q$, and $\Theta$.
Due to a form factor in our non-local model,
at small $1/\Nc$ the mass gap (chiral condensate) is large,
and the interaction among quarks and the condensate 
\textit{local} in momentum space.
Consequently, interactions between different patches are localized
near their boundaries, and it is simple to embed many chiral spirals.  We
identify the dominant and subdominant terms at high density and
catagorize formulate an expansion in terms of
$\lqcd/Q$ or $\Theta$.  The kinetic term in the
transverse directions is subdominant, so that 
techniques from (1+1)-dimensional systems can be utilized.  
To leading order in $1/\Nc$ and $\lqcd/Q$,
the total gain in energy is
$\sim \pF \lqcd^2$ with $\Theta \sim (\lqcd/\pF)^{3/5}$.  Since
$\Theta$ decreases with increasing $\pF$, there should be phase
transitions associated with the change in the wedge number.
We also argue the effects of subdominant terms at lower density where
the large-$\Nc$ approximation is more reliable.
\end{abstract}
\end{frontmatter}

\section{Introduction and Our Central Results}
In recent works \cite{McLerran:2007qj,Hidaka:2008yy,Glozman:2007tv,%
Fukushima:2008wg,Andronic:2009gj,Kojo:2011fh},
it has been argued that there is a new state of QCD, Quarkyonic matter, at high
baryon density and low to intermediate temperatures.

This novel state exists at 
densities large compared to the QCD scale, so that the Fermi sea
is best thought of in terms of quark degrees of freedom; 
it is, nevertheless, confining. 
It may be thought of
as a Fermi sea of approximately free quarks, but with thermal and
Fermi surface excitations made of color-confined mesons and baryons.
The name ``Quarkyonic'' expresses this dualism.

While the arguments for the existence of Quarkyonic matter are rigorous only
in the limit of large number of colors, for
three colors this may not be such a bad
approximation, at least for some range of density.  The inter-quark
potential inferred from the charmonium spectrum is linear out to
distances of $\sim$ fm, indicating that the production of
quark anti-quark pairs
is not very efficient in tempering its growth.  One way of
understanding is the large-$\Nc$ limit where quark pairs are
suppressed by $1/\Nc$ \cite{'tHooft:1973jz}.  
Similarly, in numerical studies of lattice QCD,
the (pseudo-)critical temperature of the 
phase transition is a slowly varying function of baryon density, 
certainly for small density \cite{Kaczmarek:2011zz}. 

At high baryon density, one might expect that
chiral symmetry is restored while quark
confinement survives.  In fact for a spatially homogeneous chiral
condensate, several computations have confirmed this expectation
\cite{Glozman:2007tv,Fukushima:2008wg}.  This conclusion was
challenged by later analysis \cite{Schaefer:2007pw} and
by simple phenomenological arguments which suggest that chiral
symmetry is broken in a confining phase of QCD \cite{Casher:1979vw}.

\begin{figure}[tb]
\vspace{0.0cm}
\begin{center}
\scalebox{0.6}[0.6] {
\hspace{-0.6cm}
  \includegraphics[scale=.30]{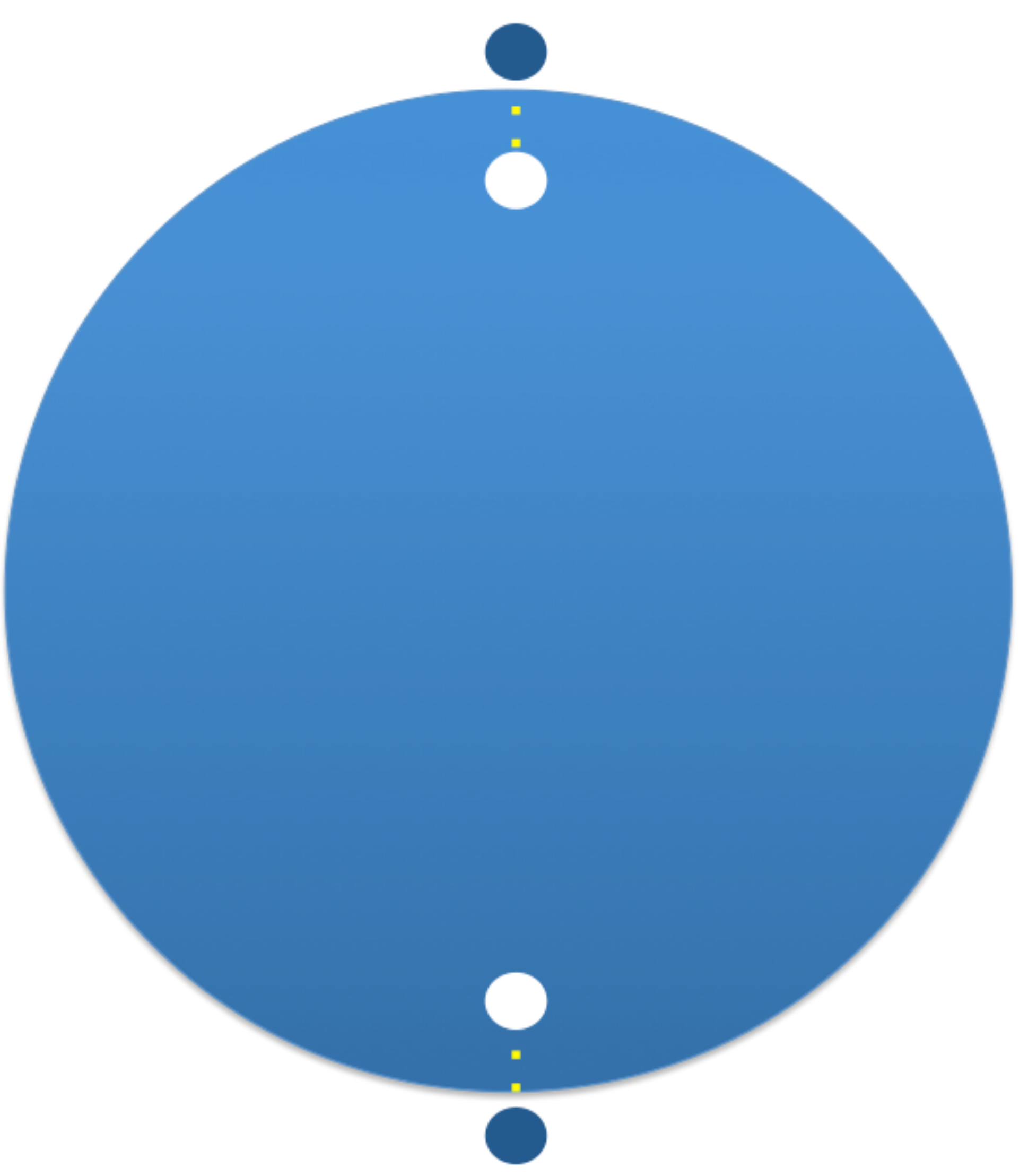} }
\scalebox{0.6}[0.6] {
\hspace{4.5cm}
  \includegraphics[scale=.30]{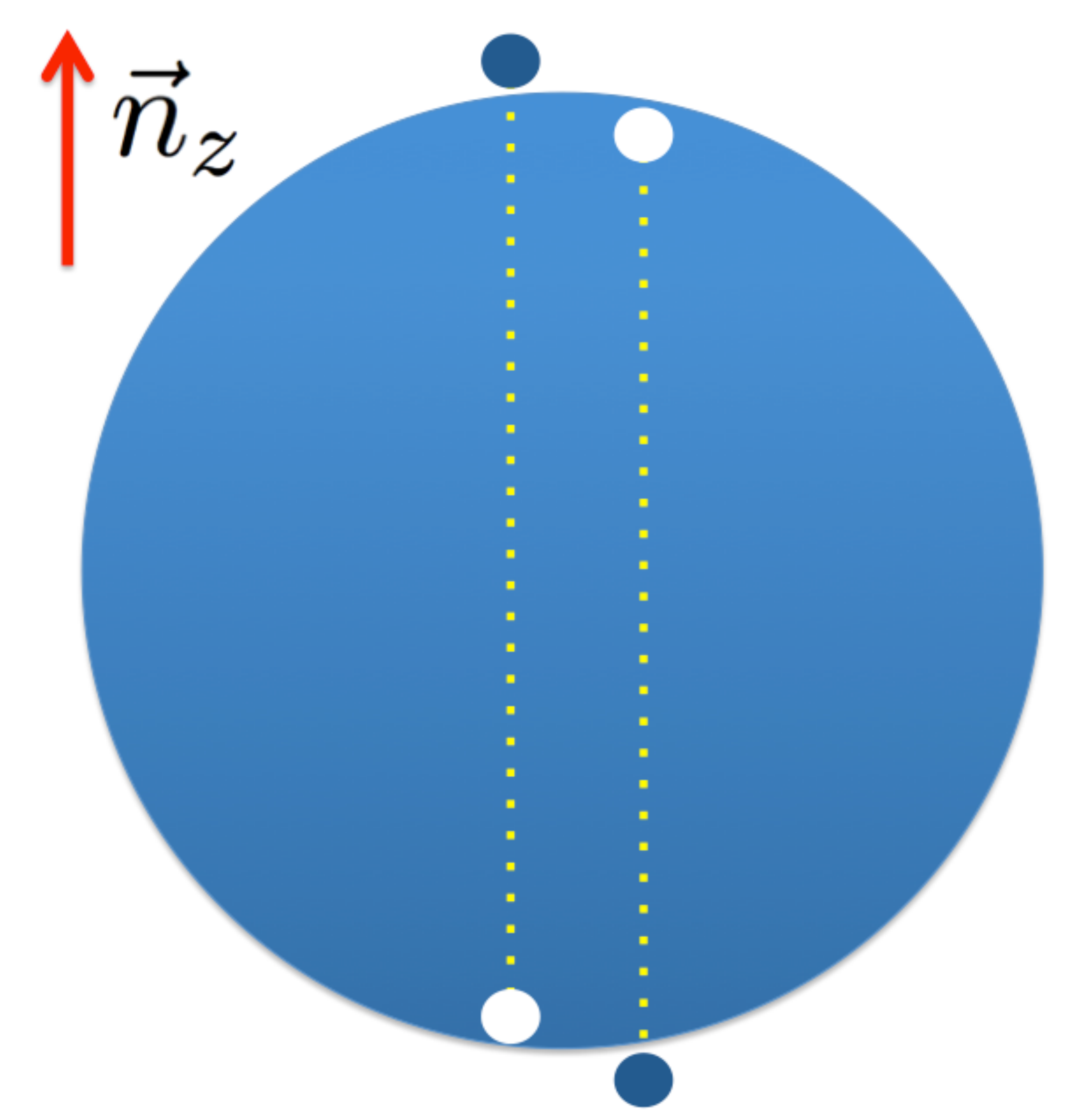} }
\end{center}
\begin{center}
\scalebox{0.6}[0.6] {
\hspace{-0.6cm}
  \includegraphics[scale=.30]{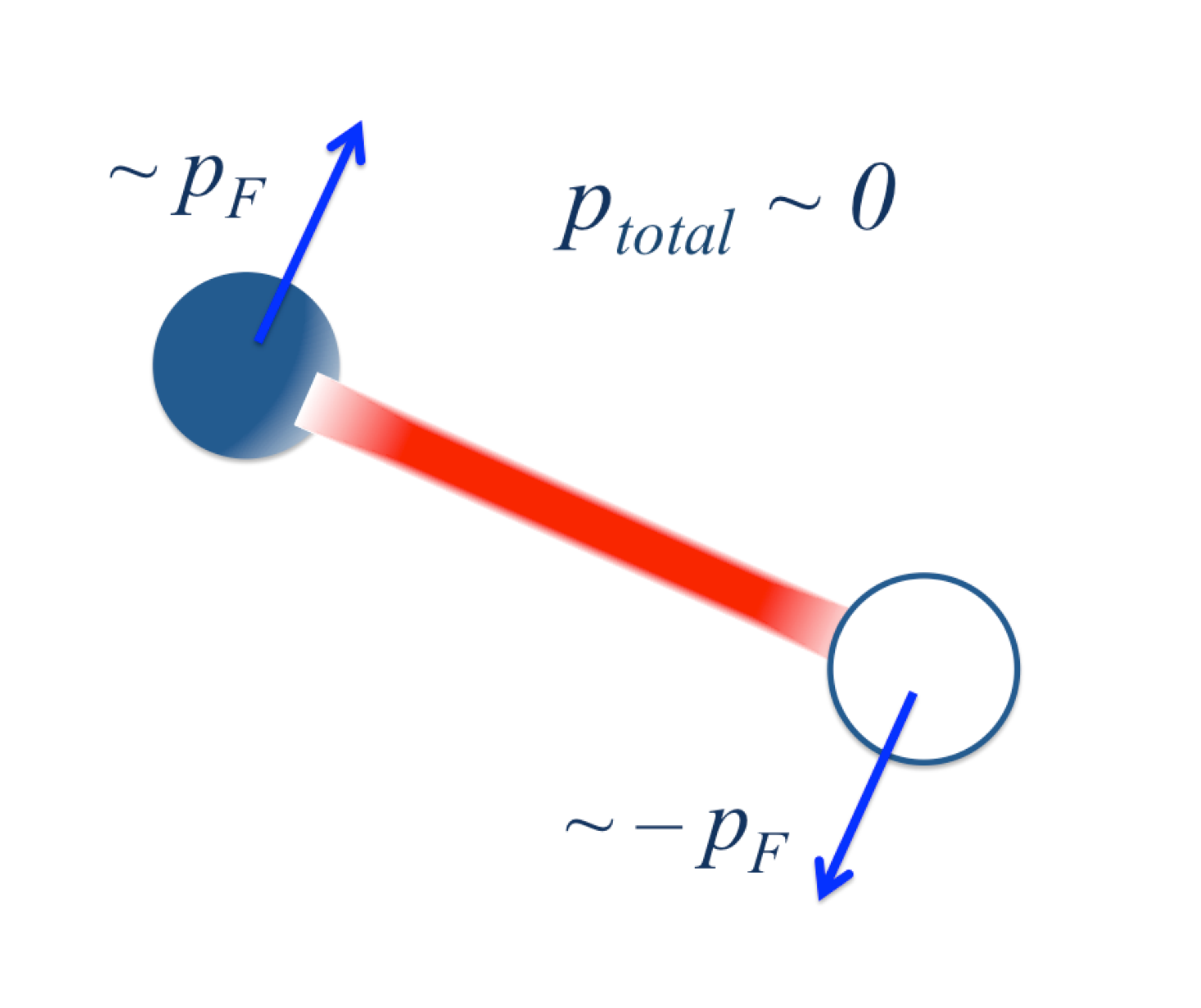} }
\scalebox{0.6}[0.6] {
\hspace{1.5cm}
  \includegraphics[scale=.30]{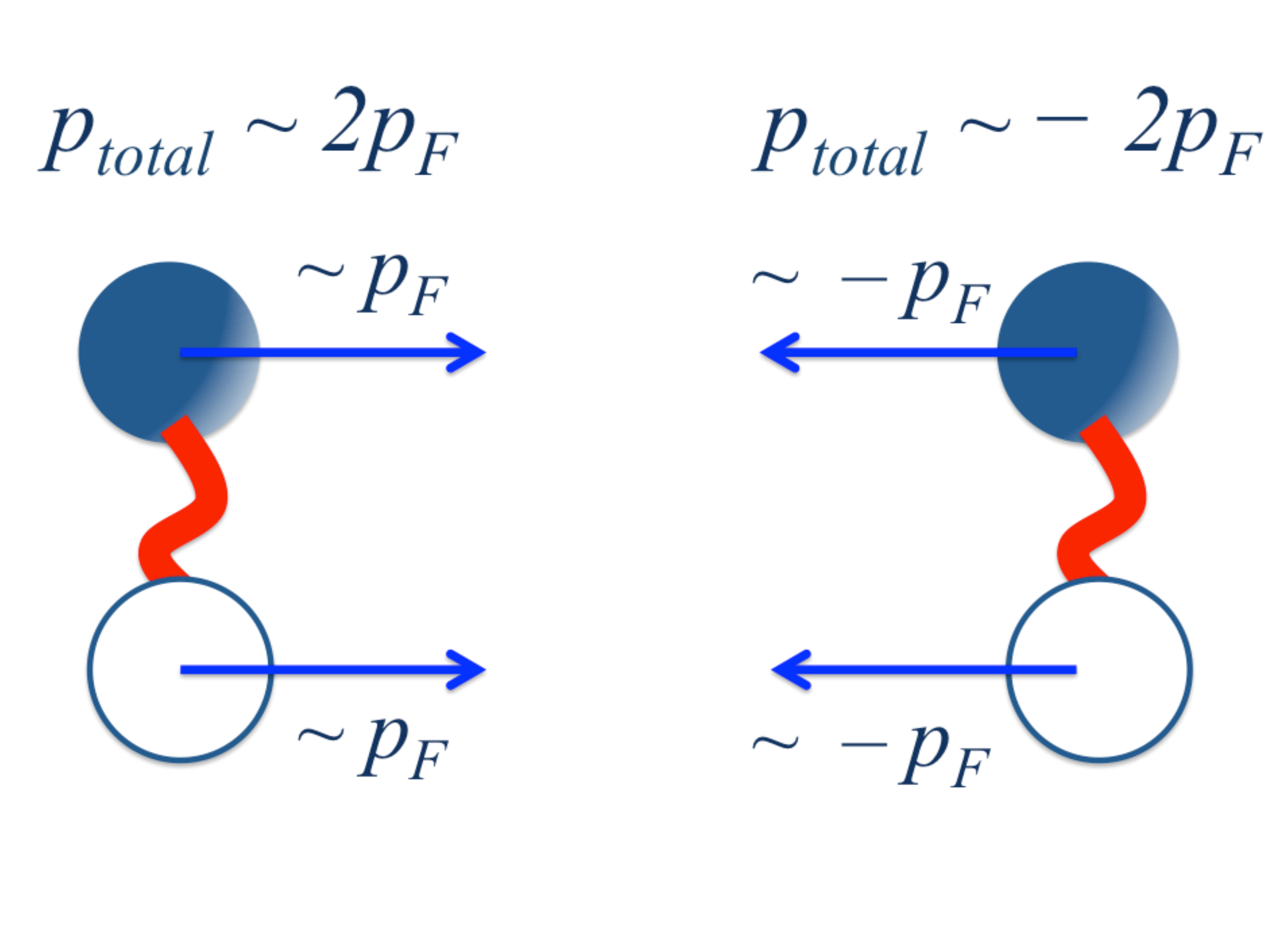} }
\end{center}
\vspace{0.0cm}
\caption{
Particle-hole pairing with a confining interaction.
(Left):
A homogeneous chiral condensate, with total momentum
$\ptot = \vec{0}$.  The relative momentum between a particle and a
hole is large.
(Right): 
An inhomogeneous chiral condensate with
$\ptot \simeq \pm 2 \pF \vec{n}_z$.  The relative momentum is small.
A superposition of pairs with momenta
$\ptot \simeq  2 \pF \vec{n}_z$ and $\ptot \simeq  -2 \pF \vec{n}_z$
creates chiral spirals.
}
\label{fig:ph}
\end{figure}

For a spatially homogeneous phase, the restoration of chiral symmetry
is understood as
follows.  In a homogenous phase, the scalar mesons that condense to
form the chiral condensate have zero net momentum.  Usually a chiral
condensate, composed of quarks and anti-quarks, is not energetically
favored, since popping an anti-quark up from the Dirac sea, to above
the Fermi sea, costs 
$\mu_{\rm q} \simeq \pF$, where $\mu_{\rm q}$ is the quark
chemical potential, and $\pF$ the quark
Fermi momentum.  Another way of forming
a homogeneous chiral condensate is to pair up quarks with
quark-holes near the Fermi surface;  see the left panel in
Fig.~\ref{fig:ph}.  In the presence of a Fermi sea, 
to make a scalar with zero net momentum one pairs
a quark with momentum $\vec{p}_F$ with a quark hole with 
momentum $- \vec{p}_F$.
The relative momentum of the quark and the quark-hole is large,
so that in a confining theory, the string tension of the bound 
requires that the excitation energy of such a bound state is of order 
$2\mu_{\rm q}$ relative to that of the scalar meson 
in vacuum\footnote{Presumably this is a sufficient condition not to have
homogeneous particle-hole condensation.  Actually, even without
confinement, it is likely that the condensation of
chiral density waves is favored.  
See discussions below Eq.~(\ref{eq4fermi+-1}).}.
Since it is unlikely that such highly excited scalar mesons condense,
then chiral symmetry restoration occurs.

Another possibility is that charge density waves form through
the condensation of particles and holes
\cite{Deryagin:1992rw,Shuster:1999tn,Park:1999bz,Sadzikowski:2000ap,%
Nickel:2009ke}, 
similar to $p$-wave pion condensation in
nuclear matter \cite{Migdal:1978az}.
Early studies using perturbative gluon propagators
\cite{Shuster:1999tn,Park:1999bz} argued that the charge density waves
are only realized if the number of colors is very large.  
These arguments,
however, do not take confinement into account.  More
precisely, the attractive force in the infrared (IR) sector
is not strong enough to overtake screening.  In a recent
paper \cite{Kojo:2009ha}, several of us have argued that in Quarkyonic
matter, translationally non-invariant chiral condensates form as chiral
spirals.  The argument for a translationally non-invariant condensate
follows again from a particle-hole pair near the Fermi surface;  see
the right panel in Fig.~\ref{fig:ph}.  A difference from the
homogeneous condensate is that quarks and quark-holes co-move in the
same spatial direction, and thereby exchange only small momenta of the
order of $\lqcd$, where $\lqcd$ denotes the typical scale of QCD. 
In contrast to the homogeneous case,
the bound sate which forms does not 
cost much energy, and condensation is possible.
One finds that the optimal mode of condensation is
a linear combination of the chiral condensate, 
$\la \overline \psi \psi \ra$, and an excitation
which has spin-one, is an isosinglet, and has odd-parity,
$\la \overline \psi \sigma^{0z} \psi \ra$.  Here $z$ is
the direction of motion of the wave \cite{Shuster:1999tn,Kojo:2009ha},
and we call the ``longitudinal'' direction; those directions 
orthogonal to $z$ are the ``transverse'' directions.
The chiral spiral is characterized by a spatial oscillation between
these two modes.  This combination can be naturally interpreted as a
superposition of particle-hole condensates with momenta
$\sim \pm 2\pF \vec{n}_z$.  At high density, like heavy-quark
symmetry, there emerges an approximate symmetry of
$SU(2\Nf)_+ \times SU(2\Nf)_-$ \cite{Shuster:1999tn}, where
$\pm$ expresses (1+1)-dimensional chirality that characterizes the
moving directions along the $z$ axis.  After the formation
of chiral spirals, there are $(2\Nf)^2$ Nambu-Goldstone (NG) modes:
$(2\Nf)^2-1$ isospin-spin excitation modes and one phonon mode,
associated with spontaneous breaking of spin-chiral and translational
symmetry, respectively\footnote{The formation of a single chiral spiral breaks
  rotational symmetry in addition to translational symmetry.
  Since the translation and rotation are not independent, only one
  phonon mode appears as an NG mode \cite{Casalbuoni:2001gt}.}.
These results were derived by the dimensional reduction from the
(3+1)-dimensional self-consistent equations to those in the
(1+1)-dimensional 't~Hooft model for degrees of freedom near the Fermi
surface.

\begin{figure}[tb]
\vspace{0.0cm}
\begin{center}
\scalebox{1.0}[1.0] {
\hspace{0.2cm}
  \includegraphics[scale=.35]{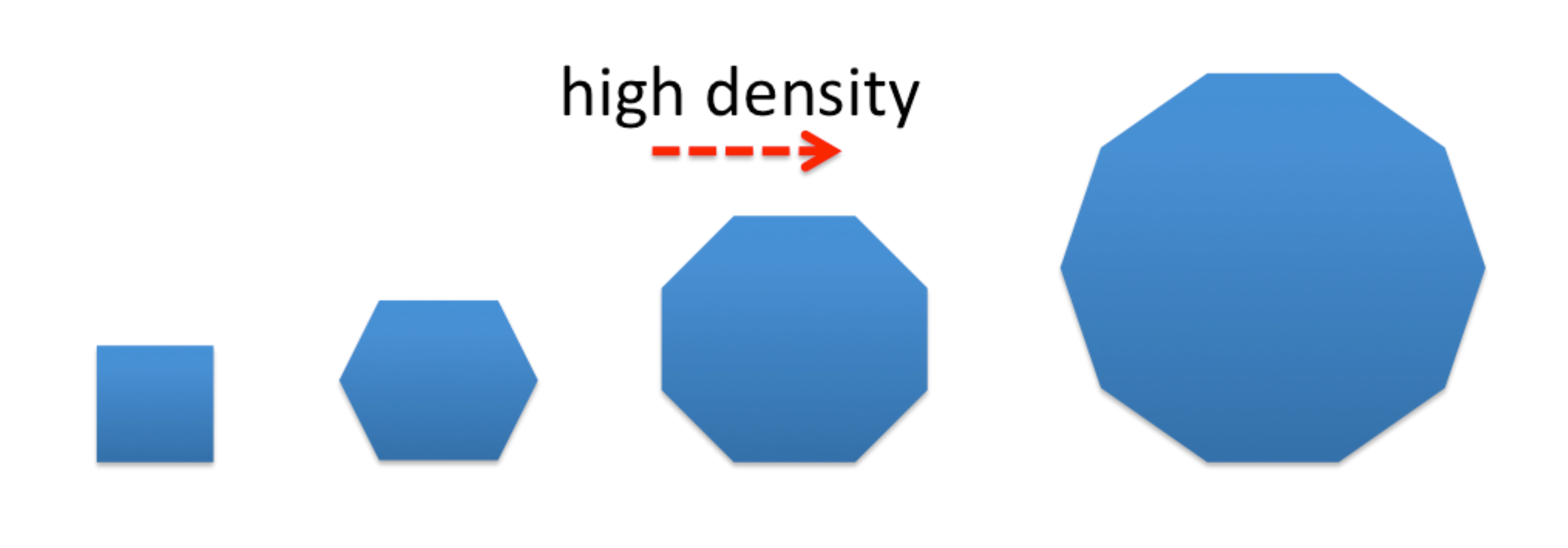} }
\end{center}
\caption{The two-dimensional slice of the Fermi sea, suggested in
  Ref.~\cite{Kojo:2010fe}.  Rotational symmetry is spontaneously
  broken from continuous to discrete one.  The number of patches
  accompanying chiral spirals in different directions increases with
  increasing density.
}
\label{fig:multi}
\vspace{0.2cm}
\end{figure}

The stability analysis of Ref.~\cite{Kojo:2009ha}
showed that Quarkyonic matter in the absence of a chiral condensate
was unstable with respect to the formation of a (1+1)-dimensional
chiral spiral.  Furthermore, it was suggested that many chiral spirals
of different spatial orientations interweave to form a more
complicated condensate \cite{Kojo:2010fe}.  This corresponds to a
transition from a spherical Fermi surface into patches, inducing
breaking of continuous rotational symmetry down to discrete one (see
Fig.~\ref{fig:multi}).  The number of patches increases as the density
grows up and such phase transitions continue to occur until the
screening effect on gluons strongly reduces the IR attraction between
a pair of 
a quark and a quark-hole.
Such reduction happens
around the density scale, $\mu_q \sim \Nc^{1/2} \lqcd$
\cite{McLerran:2007qj}.

\begin{figure}[tb]
\vspace{0.0cm}
\begin{center}
\scalebox{1.0}[1.0] {
\hspace{0.0cm}
  \includegraphics[scale=.26]{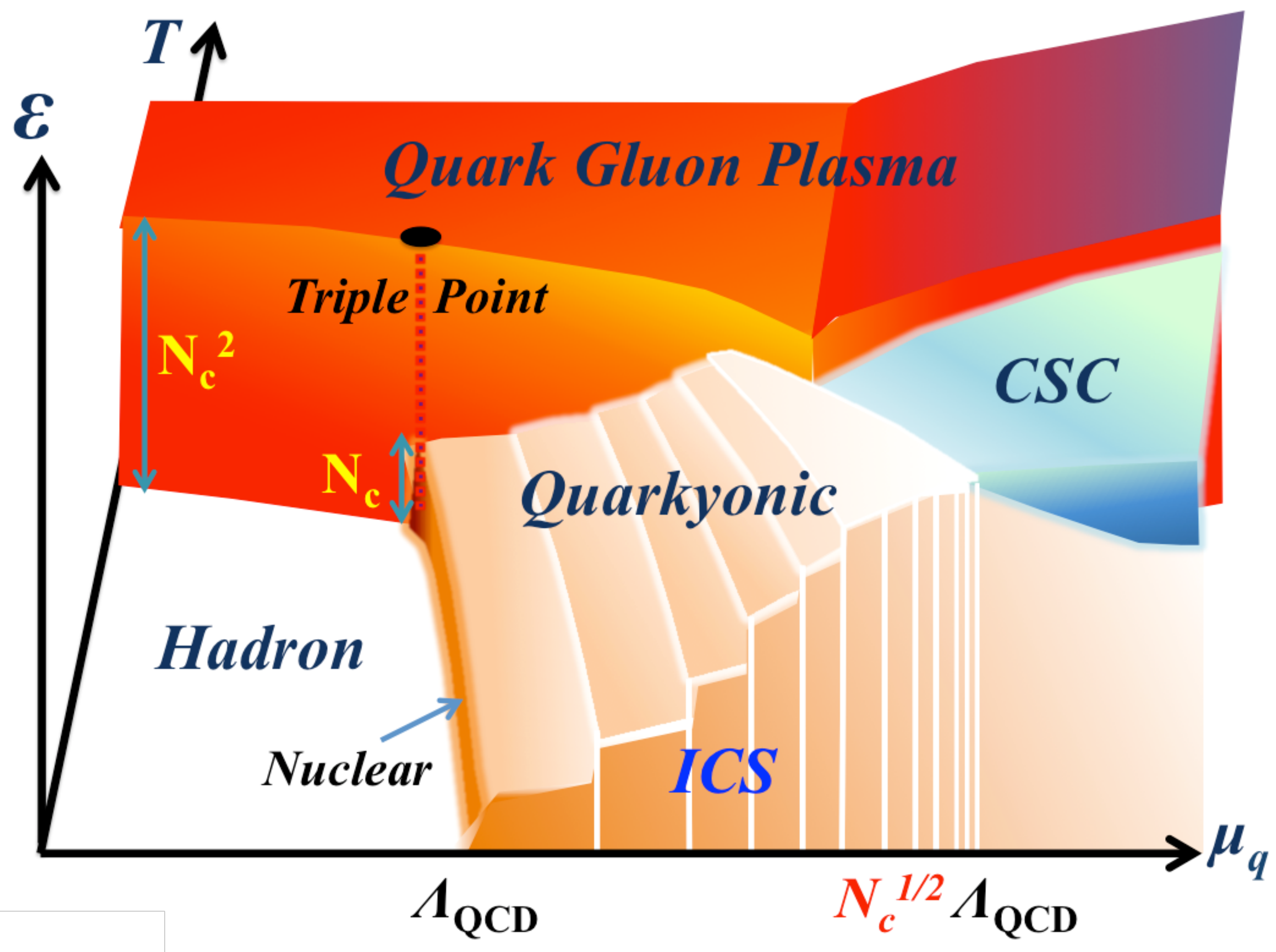} }
\end{center}
\caption{The three dimensional plot of the phase diagram in 
$\mu_q$-$T$-${\cal E}$ space
(${\cal E}$ is the energy density of the system).
Chiral symmetry in Quarkyonic matter
is broken by the formation of the 
interweaving chiral spirals (ICS).
The stair-like growth of the energy density
along $\mu_q$ axis expresses
the discontinuous change in the shape of the
Fermi sea (Fig.\ref{fig:multi}).
}
\label{fig:Phase}
\vspace{0.2cm}
\end{figure}

These behavior are 
sketched in a the phase diagram in $\mu_q$-$T$-${\cal E}$ space,
as shown in Fig.~\ref{fig:Phase}
(${\cal E}$ is the energy density of the 
system)\footnote{Here we consider chiral limit for the light flavors
and ignore the electroweak interactions.}.
Quarkyonic matter starts to appear just above 
$\mu_q \sim \lqcd$, where a transition from nuclear to 
quark matter quickly occurs,
and continues to exist up to $\mu_q \sim \Nc^{1/2} \lqcd$.
In the Quarkyonic region,
the stair-like growth of the energy density
along $\mu_q$ axis reflects
the discontinuous changes in the shape of the Fermi sea
(Fig.~\ref{fig:multi}).
At larger $\mu_q$,
the screening of the IR attraction
reduces the size of the chiral spiral condensate,
and accordingly,
the interval of stair in $\mu_q$ axis
and jumps in ${\cal E}$
become smaller.
The shape of the Fermi sea smoothly approaches spherical one. 

While Refs.~\cite{Kojo:2009ha,Kojo:2010fe} have argued for chiral
spirals using the confining interactions, several aspects have not
been explicitly shown due to technical difficulties and/or conceptual
uncertainties in treating the deep IR structure of confining forces.
On the other hand, while we postulate that the confining force could
give a sufficient condition to drive chiral symmetry breaking, it is
certainly not a necessary condition.  The aforementioned phenomena may
appear in a wider class of models that encompass chiral symmetry
breaking even without confinement.  Indeed, what is relevant for
interweaving chiral spirals are Fermi surface effects and the IR
enhancement of the interaction, but not precise knowledge about the
deep IR region.  Taking this viewpoint, we will characterize chiral
symmetry breaking at high density by a simple, tractable model in
which we can explore analytic insights.

We will use an effective field theory to describe QCD at large $\Nc$.
This model will be apparently similar to the Nambu--Jona-Lasinio (NJL)
model \cite{Nambu:1961tp,Vogl:1991qt} in which the dominant
interacting process in the large-$\Nc$ limit is the scattering of
particles and the condensate.  The crucial difference of our model
from the usual NJL model is that the interaction vertex has a form
factor that mimics the IR enhancement of the non-perturbative gluon
propagator.  We denote a form-factor scale of the model as $\lf$
($\sim \lqcd$), beyond which interactions are negligible compared to
the IR interaction.

\begin{figure}[tb]
\vspace{0.0cm}
\begin{center}
\scalebox{0.6}[0.6] {
  \includegraphics[scale=.40]{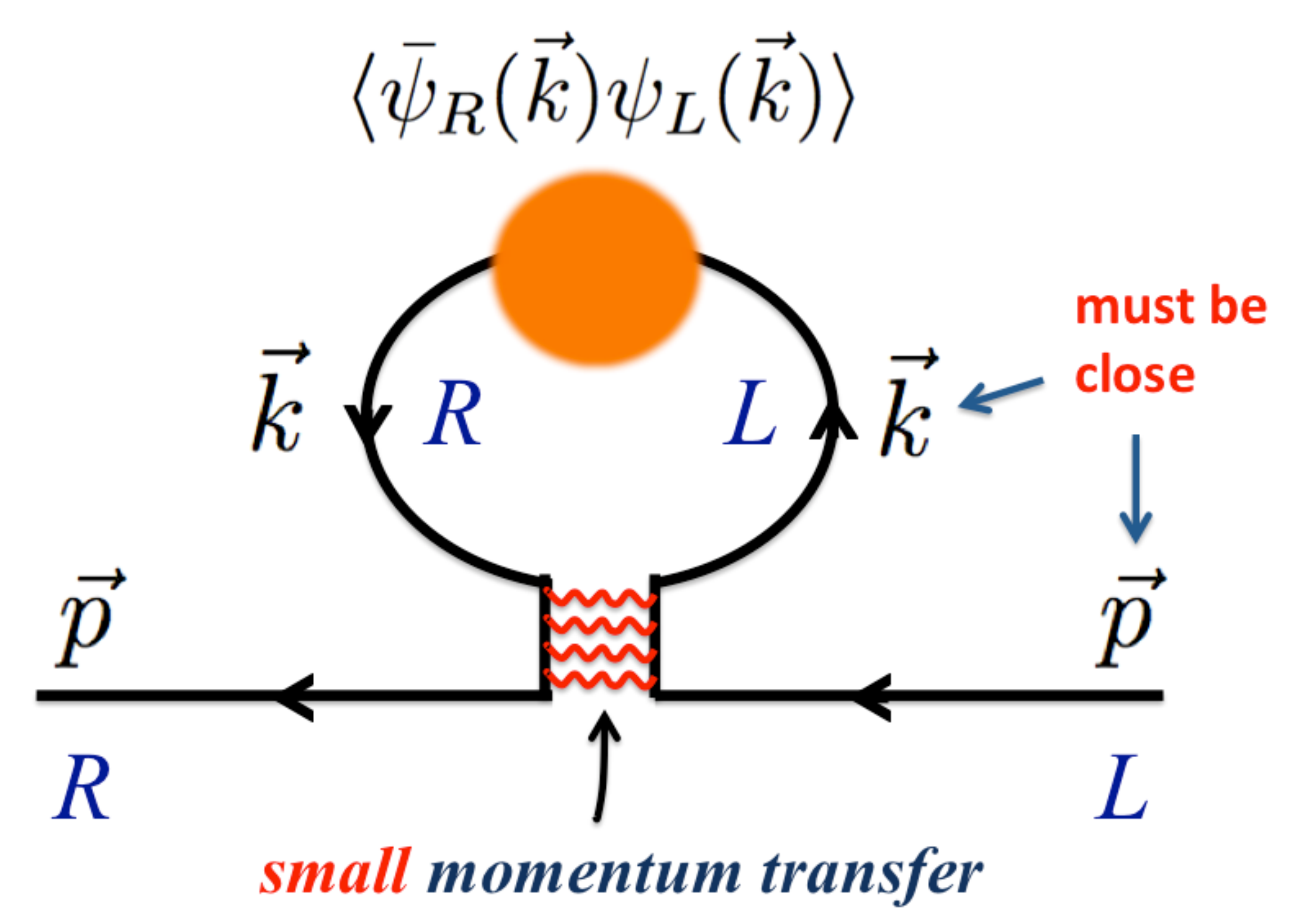} }
\scalebox{0.6}[0.6] {
  \includegraphics[scale=.40]{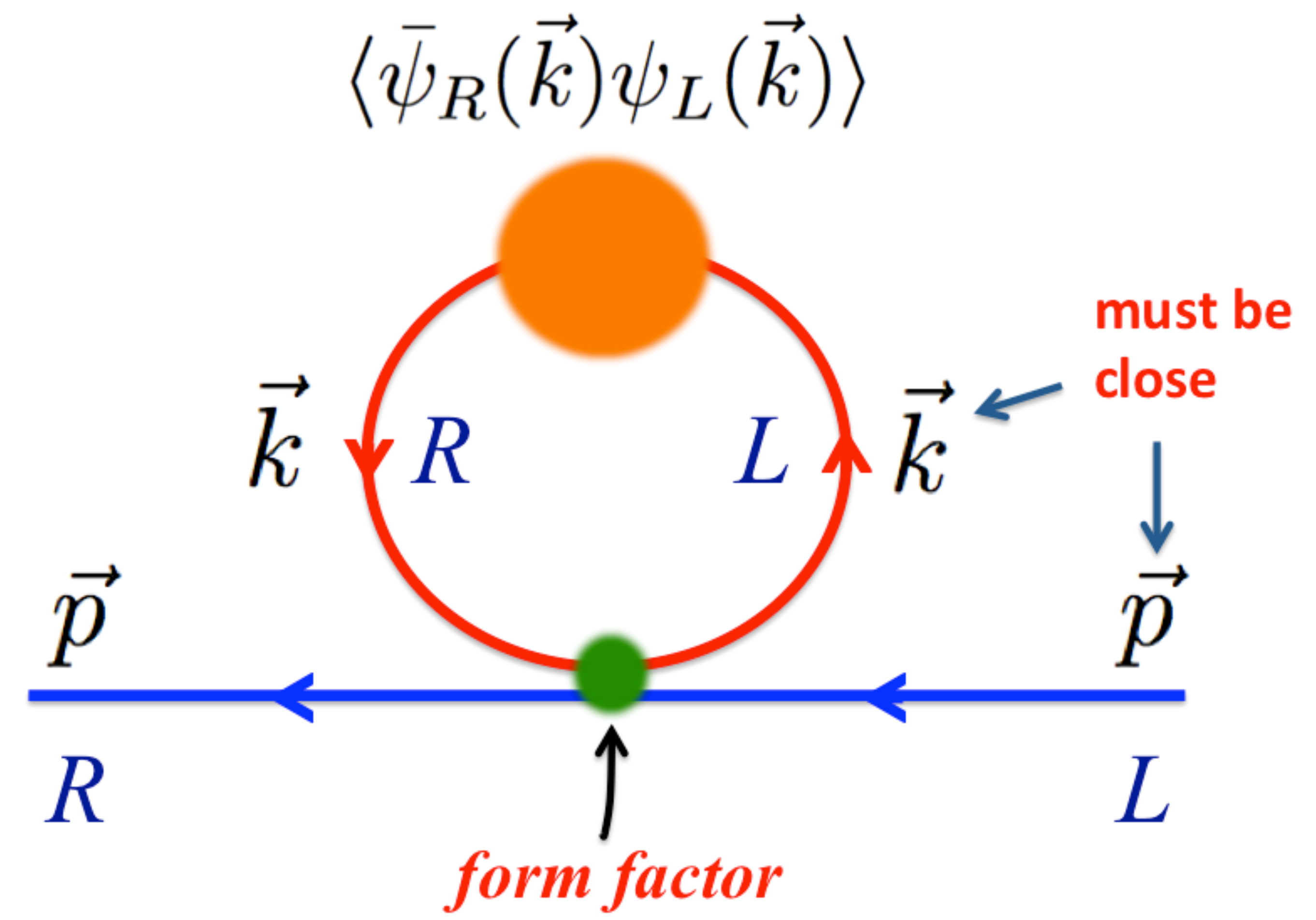} }
\end{center}
\vspace{0.0cm}
\caption{
The leading 
self-energy diagram at zero density.  At large $\Nc$, we have only to
keep the rainbow ladder.\ \
(Left) The diagram in terms of QCD dynamics.  Integrating the temporal
component out, we can interpret the loop with momentum $\vk$ as the
condensate, $\la \bar{\psi}(\vk) \psi(\vk) \ra$, which is made of
particle-antiparticle with momentum $\vk$.\ \
(Right) The corresponding diagram in our model.  The soft-gluon
exchange part in QCD is replaced with the form-factor function whose
strength damps as $\vk$ and $\vp$ go far apart.  The diagrams for the
$1/\Nc$ corrections will be given in Fig.~\ref{fig:1nc} in
Sec.~\ref{secdiscussion}.
}
\label{fig:SDeq}
\end{figure}

\begin{figure}[tb]
\vspace{0.0cm}
\begin{center}
\scalebox{0.6}[0.6] {
  \includegraphics[scale=.30]{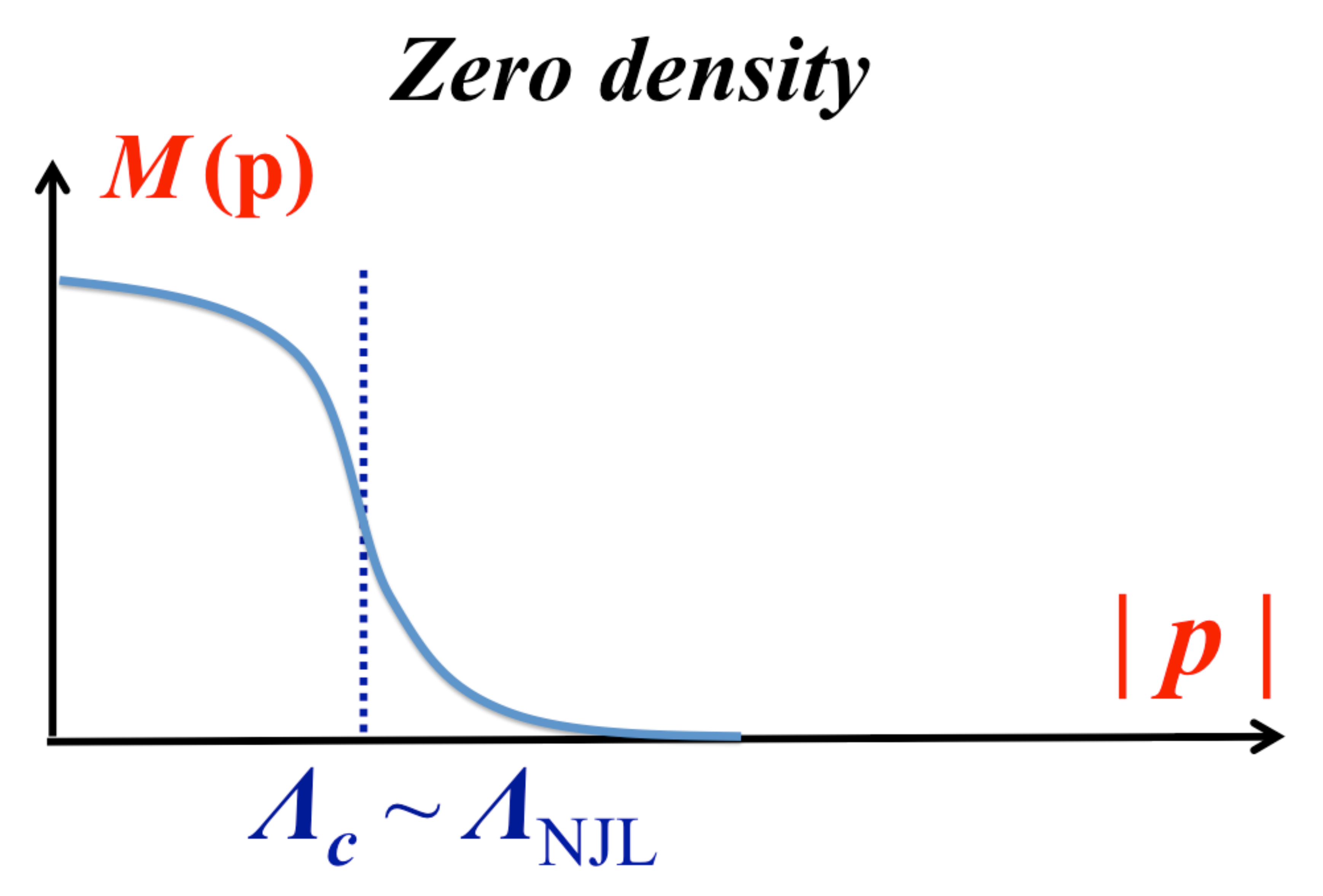} }
\scalebox{0.6}[0.6] {
  \includegraphics[scale=.30]{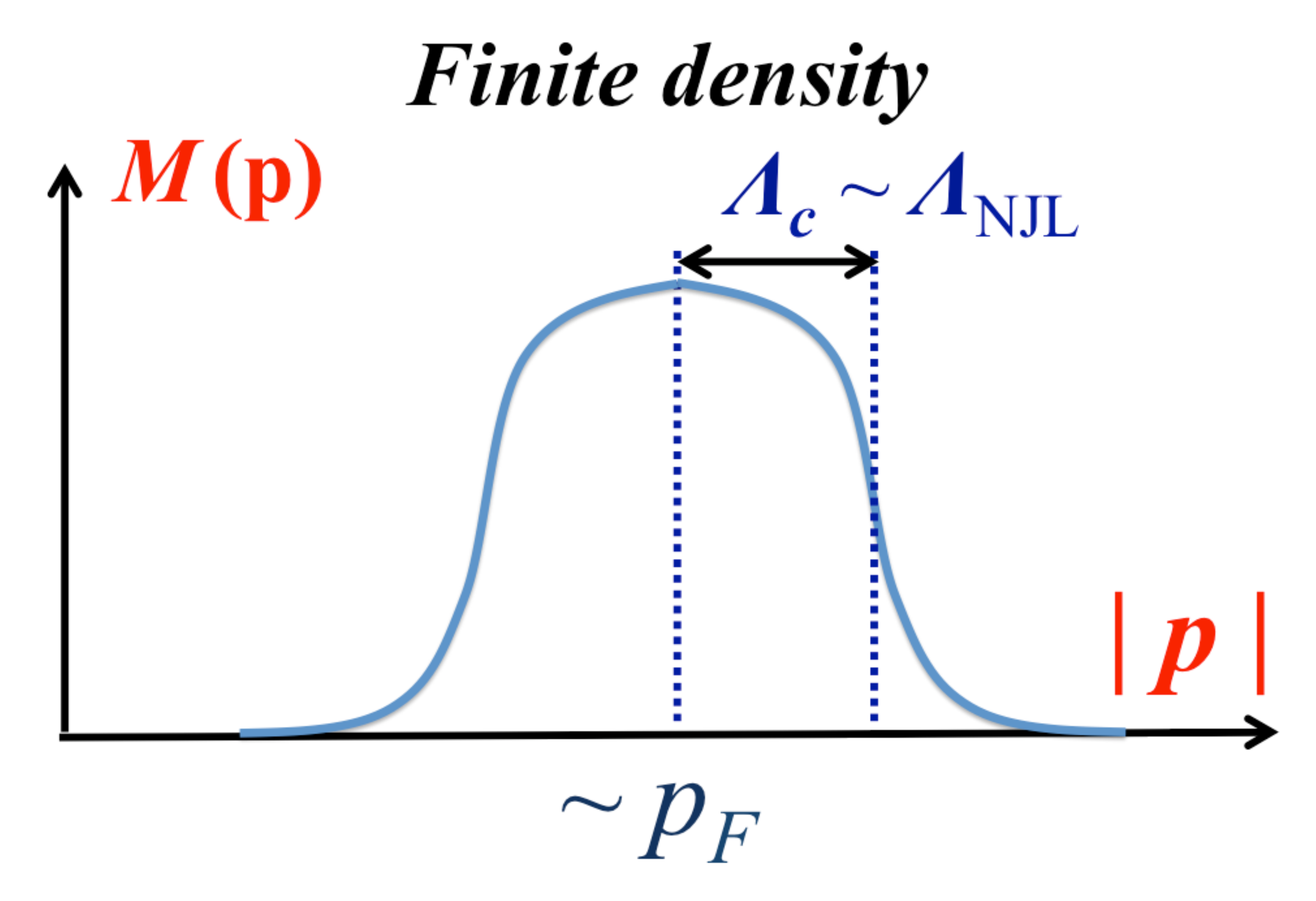} }
\end{center}
\vspace{0.0cm}
\caption{
Schematic figures for the mass gap function.\ \
(Left) The zero density case.\ \
(Right) The finite-density case with chiral symmetry breaking near the
Fermi surface.
}
\label{fig:density}
\end{figure}

Form-factor effects lead to the following consequences at large
$\Nc$.  The interaction between a quark and a condensate becomes
strongly momentum dependent; see Fig.~\ref{fig:SDeq}.  They decouple
one another\footnote{
This sort of picture has been discussed for the high-lying mesons and
baryons \cite{Glozman:1999tk}.  See also Ref.~\cite{Shifman:2007xn}
for some caveats on this picture.}
if the relative momentum between the quark and the condensate is much
larger than $\lf$, reflecting composite nature of the quark
condensate.  As a consequence, the quark mass gap damps when the quark
momentum is far away from the domain of condensation.  In vacuum, the
damping scale of the mass function $\lc$ may play a similar role to
the ultraviolet (UV) cutoff, $\Lambda_{{\rm NJL}}$, in the usual NJL
model (see the left panel in Fig.~\ref{fig:density}).  In this sense,
the form factor can naturally remove the UV cutoff artifact of the
NJL-type model.
%

All of these aspects are crucial when we consider dense quark matter.
Since condensation phenomena should happen near the Fermi surface, the
effective UV cutoff should appear in the distance from the Fermi
surface, not that from vacuum (see the right panel in
Fig.~\ref{fig:density}).  If we fixed the effective UV cutoff to be
the same as the vacuum value by hand, arguments on chiral symmetry
breaking would not make sense for
$\mu_{\rm q} \ge \Lambda_{{\rm NJL}}$.  Indeed, because of Pauli
blocking, phase space for quarks contributing to the condensate would
disappear, leading to chiral symmetry restoration as a cutoff artifact.
Therefore it would be desirable
to derive an effective UV cutoff dynamically for each density.  We
claim that the introduction of the form factor gives a natural
extension of the treatment of the zero-density NJL model to that at
finite density. 
Also we make a remark that the so-called Debye
cutoff frequency is introduced in this way from the Fermi surface in
the standard BCS theory.

With this modified NJL model at hand, the purpose of this paper is to
give detailed and analytic insights on the interweaving of chiral
spirals and on the associated breakdown of the rotational invariance.
A key observation in our model treatment with a form factor will be that at
large $\Nc$, the particle-condensate interaction
(i.e.\ the interaction of particles scattering off the
  condensate)
happens {\it locally in momentum space}\footnote{
The $1/\Nc$ corrections (as shown in Fig.~\ref{fig:1nc}) will violate
this locality; see discussions in Sec.~\ref{secdiscussion}.},
which will allow the system to simultaneously embed many chiral
spirals at sufficiently high density.

For the sake of simplicity, we will work in (2+1) dimensions, where the 
original Fermi surface is
a circle and takes a simple geometric structure even after formation
of many chiral spirals.  An extension of this study to higher
dimensions might be technically difficult but conceptually
straightforward.
  
Speaking precisely, in three space-time dimensions
there is no true chiral symmetry, since there is no
$\gamma_5$ matrix for two component spinors.  
Using four component spinors, there is flavor symmetry
breaking.  This technicality 
does not change any of our main considerations.
(Further discussions is given in Sec.\ref{Spinors}.)

\begin{figure}[tb]
\vspace{0.2cm}
\begin{center}
\scalebox{1.0}[1.0] {
  \includegraphics[scale=.80]{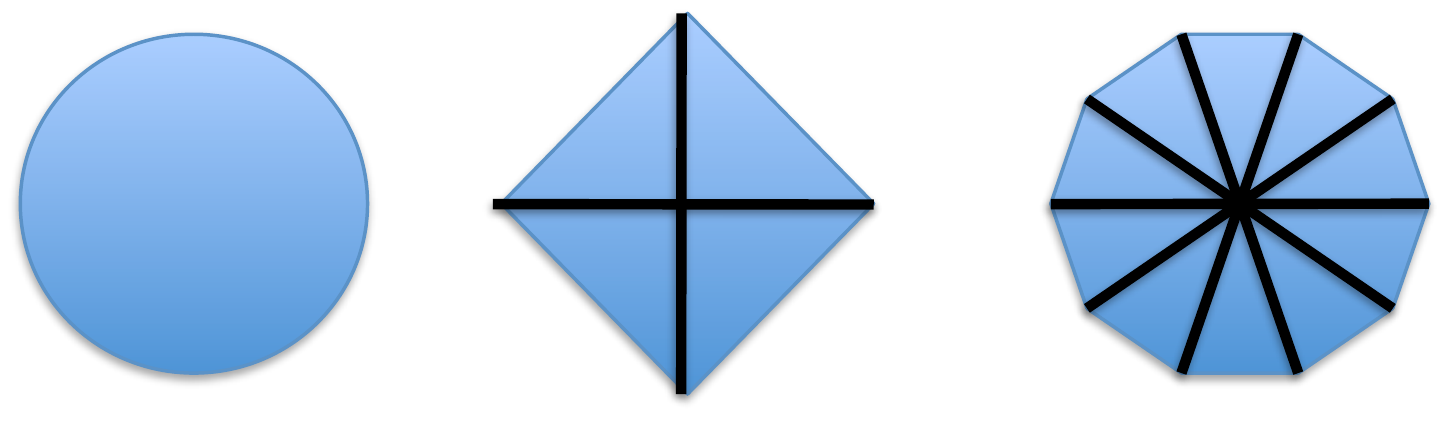} }
\end{center}
\vspace{0.2cm}
\caption{The leftmost is the Fermi circle in spatial two dimensions.
  The middle figure is a square approximation to the circle.  The
  wedges are shown corresponding to discrete sub-area of the circle.
  The rightmost figure is a higher-order polygon approximation to the
  circle.
}
\label{fermisphere}
\vspace{0.2cm}
\end{figure}

\begin{figure}[tb]
\vspace{0.2cm}
\begin{center}
\scalebox{1.0}[1.0] {
  \includegraphics[scale=.25]{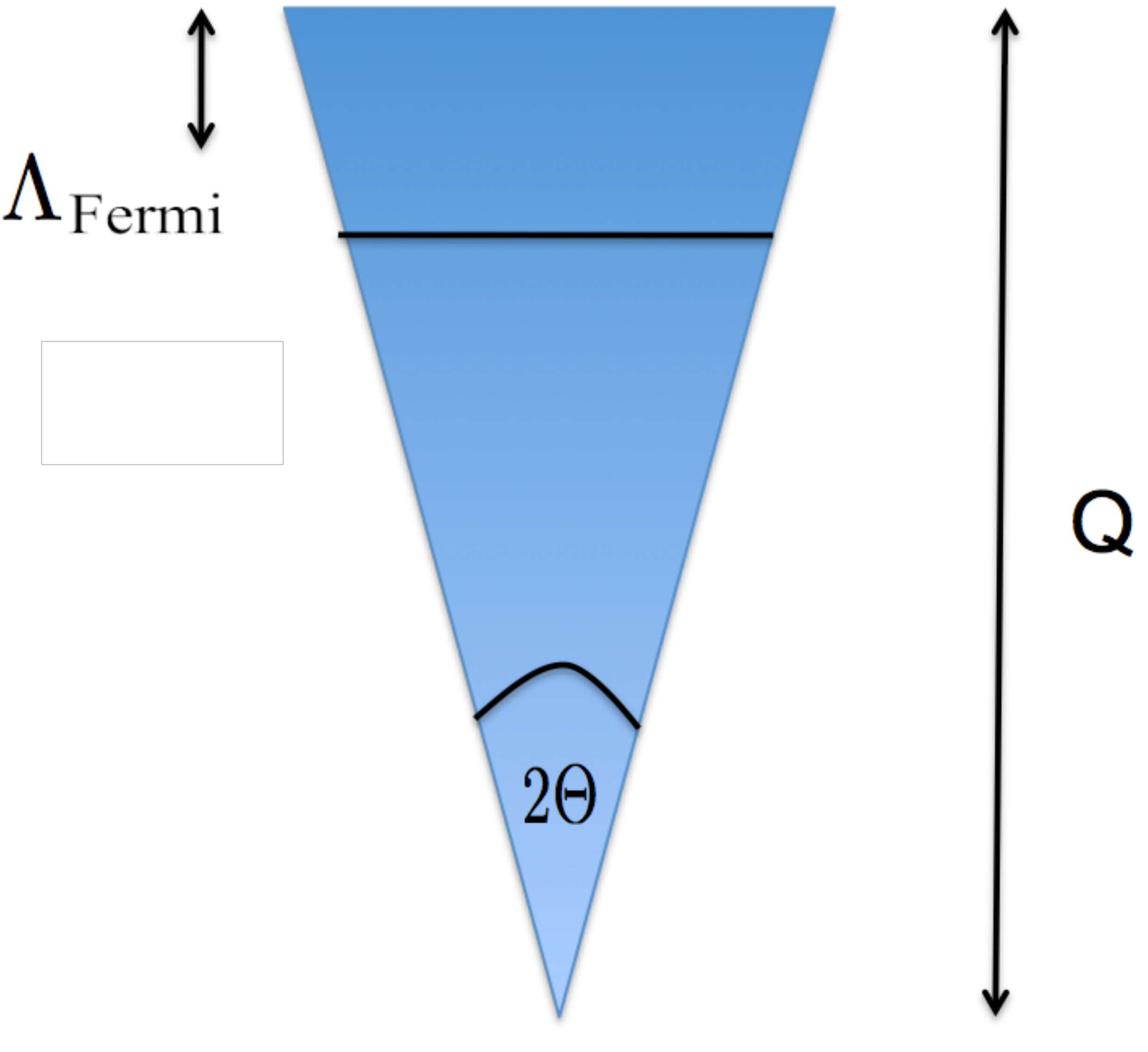} }
\end{center}
\vspace{0.2cm}
\caption{The basic wedge from which the polygon is constructed.  The
  wedge has an opening angle $2\Theta$, a depth $Q$, and a surface
  thickness $\Lambda_{{\rm Fermi}}$.
}
\label{wedge}
\vspace{0.2cm}
\end{figure}

We approximate the patched Fermi surface by a polygon of degree $\Np$.
In Fig.~\ref{fermisphere} we show the two-dimensional Fermi surface
and some polygon approximations to it.  We will look for an energy
minimum at a non-zero value of $\Np$, and we will find that there
exists such a minimum that depends upon density.  We can think of
each sub-sector that constitutes the polygon shape
as a wedge.  The wedge is characterized by an opening angle $2\Theta$
and a depth $Q$ as shown in Fig.~\ref{wedge}.  The depth $Q$ will be
of the order of the Fermi momentum $\pF$, and the opening angle of the
wedge is constrained\footnote{
Here a factor $2$ is included since we will take one patch as a set of
one wedge and the other wedge in the opposite side of the Fermi sea.
See Sec.~\ref{secdecomp} for details.}
by $2\times 2\Np \Theta = 2\pi$.
The surface thickness $\Lambda_{\rm Fermi}$ ($\sim \lf$) characterizes
the momentum scale for which non-perturbative Fermi surface effects
are important.

The use of the wedge shape is motivated by the following reasons:
In order to maximize the energy gain from condensation effects, each
patch should contain only one chiral spiral by aligning total momenta
of a bunch of particle-hole pairs.  If we had a misalignment,
interplay
among chiral spirals would
reduce the size of the gap, as exemplified in Sec.~\ref{1+1Dexample}.
So we have to look for the most effective shape to achieve the
alignment of total momenta.  An obvious candidate is the flat Fermi
surface with which particles and holes participating in the condensate
can stay close to the Fermi surface, saving the virtual excitation
energies.  Other shapes require some of particles and holes with
larger excitation energies, so are not effective to create a bigger
condensate.

What is the principle to determine the number of patches?  It is
essentially determined by the balance between the kinetic-energy cost
and the condensation-energy gain.  Without condensation effects, the
Fermi surface would simply take a circle shape that minimizes the
kinetic energy.  Once condensation effects are turned on, the
associated energy gain can overtake the kinetic-energy cost, changing
the shape of the Fermi sea from a circle to a polygon with discrete
number of wedges.  The condensation effects appear in two places.
One is the energy reduction of single-particle contributions inside of
one wedge, as is the case in the single chiral spiral problem.  The
other is 
the inter-patch interactions
among different chiral spirals.  The
latter will provide an energy cost, that is, chiral spirals destroy
one another if their wavevectors are different (as explained in
Sec.~\ref{sec:interference}, and also some results from the existing
literatures will be addressed in Sec.~\ref{comparison}).  This means
that we should divide the Fermi sea into not too many wedges to get
the largest energy gain.
Therefore the size of one wedge tends to be as large as possible until the
kinetic-energy cost becomes too big.

Let us outline how to optimize the shape of the Fermi sea.  In the
following, we will consider the canonical ensemble, in which the
particle number is fixed.  Then the Fermi volume
must be conserved, so it follows that
\begin{equation}
 \pF^2 \Theta = Q^2  \tan \Theta\,.
\end{equation}
Here the LHS is a fermion number for the plain circle shape of the
Fermi sea, while the RHS is for the deformed Fermi sea.
For small $\Theta$ we can approximate the RHS by the Taylor
expansion in terms of $\Theta$, with which we can solve $Q(\Theta)$ as
%
%
%
\begin{equation}
Q(\Theta) = \pF \big( 
1 - \frac{\Theta^2}{6} - \frac{\Theta^4}{40} +
\cdots \big)\,.
\end{equation}
Thus, the size of one patch in the transverse direction is
\begin{equation}
\text{(one patch size)}
= Q(\Theta) \tan\Theta
= \pF \Theta \: 
\big ( 1 + \frac{1}{6} \Theta^2+ \frac{19}{360}\Theta^4 + \cdots \big)\,.
\end{equation}
Now that $Q$ is solved as a function of $\Theta$, the multiple chiral
spiral states are characterized by $\pF$, $\Theta$, and
the single-particle mass gap, $M$.  We will perform the energy
minimization by taking $\Theta$ and $M$ as variational parameters.

Provided that the energy minimum exists for
$\Theta \ll 1$, we can expand the energy density by powers of
$\Theta$.  It will turn out that the expression takes the following
form,
\begin{align}
\calE(M,\Theta) 
&= \Np \cdot \calE^{\text{1-patch}}(M,\Theta) \notag \\
&= \frac{\calE_{-1}(M)}{\Theta} + \calE_0(M) + \calE_2(M) \Theta^2
 + \calE_4 (M) \Theta^4 + \cdots \,.
\end{align}
The $1/\Theta$ term can appear simply because $\Np = \pi/2\Theta$, 
but it will disappear for vanishing $M$, that is,
$\calE_{-1}(M\to0)=0$.  
Expanding $\calE_n(M)$ by powers of $M/\pF$, we expect
\begin{equation}
\calE_n(M) 
= c_n^{(0)} \pF^3 + c_n^{(1)} \pF^2 M + c_n^{(2)} \pF M^2 + \cdots \,,
\end{equation}
where we did not write possible non-analytic terms explicitly.  (In
what follows, solving the gap equation, in fact, we can arrive at the
above form of the expression.)  At sufficiently high density, one has
only to keep
the leading term in the $M/\pF$-expansion
for each $\calE_n(M)$.

Let us first discuss terms insensitive to details of condensation
effects, computing at $M=0$.
The first non-vanishing contribution of $O(\pF^3)$ with
non-trivial $\Theta$ dependence should arise from the
kinetic-energy cost for the deformation of the Fermi surface.  The
energy at $M=0$ is an increasing function of $\Theta$,
\begin{align}
\calE(M=0, \Theta) 
&= \Np \cdot 2\Nc \cdot 4
\int_0^{ Q }  \frac{ \rmd p_{\para} }{2\pi}
\int_0^{ p_\para \tan \Theta} \frac{\rmd p_\perp}{2\pi}\:
\sqrt{ p_\para^2 +  p_\perp^2 } 
\notag\\
&= \Nc \cdot
\frac{\pF^3}{3\pi} \:
\bigg( 1 + \frac{1}{30} \Theta^4 + O(\Theta^6) \bigg)\,,
\end{align}
where the first term gives the trivial contribution which should be
subtracted out\footnote{  
In the first line of the equation,
the factor $2\Nc$ is for degeneracy factors of colors and spins
for four component spinors.
The second factor $4$ arises 
because one patch is made of two opposite wedges
and the $p_\perp$ integral with 
the opening angle is $2\Theta$.}.
It is extremely important to notice that the
non-trivial deformation energy does not appear until $O(\Theta^4)$.
The volume conservation of the Fermi sea cancels the $\Theta^2$-term
out from the average kinetic energy.

Terms beyond $O(\Theta^4)$ are much smaller than $\pF^3\Theta^4$ and
irrelevant in our minimization procedure.  Thus, the energy minimum
will be found by balancing the $\pF^3 \Theta^4$ term with
condensation effects which yield terms with smaller powers of
$\Theta$ than the $\pF^3 \Theta^4$ term.  In the following we
concentrate on the estimation of such condensation terms.

The condensation effects depend on interaction properties of models.
The point of our model is that at large $\Nc$ the single-particle
dispersion of a fermion with momentum $\vp$ is determined only by
condensates within the domain of the size $\sim \lf^2$ around $\vp$.
It means that if we consider the gap for single particles farther from
patch boundaries than $\lf$, it will be affected only by the single
chiral spiral, not by adjacent chiral spirals.

The above argument suggests that we have to
evaluate the mass gap differently depending on the different domains
of $\Theta$.
then there are intersection points of more than one patch that have
interactions among them.
To solve the gap equation, hence, we have to take into account the
influence of several chiral spirals simultaneously.  This is a rather
technically complicated problem.  Fortunately, the energy minimum in
our problem will be outside of this $\Theta$ domain.

We can self-consistently show that the transverse size of one patch is
much larger than $\lf$, i.e.
\begin{equation}
\pF \Theta \gg \lf\,. 
\label{eq:pfgg}
\end{equation}
Once this condition is satisfied, the single-particle gap in one patch
can be determined independently from $\Theta$, except in the region
close to the 
patch boundaries.  We denote such a solution as $M= M_0 \sim \lf$.
Then the energy gain from condensation effects should be
\begin{equation}
\hspace{-0.5cm}
\text{(energy gain)} 
\sim  \Np \cdot \Nc\,(\lf\, \pF \tan\Theta)\, M_0
\sim \Nc M_0\, \lf\, \pF \,\bigg( 1 + \frac{\Theta^2}{3} + \cdots
 \bigg)\,,
\label{con1}
\end{equation}
where $\lf\, \pF \tan\Theta$ is the one-patch phase space where
the condensation occurs.  
One important observation here is that, while
the gap is insensitive to $\Theta$, the phase space has $\Theta$
dependence, so that the leading term is $\Theta$ independent after
multiplying a patch number $\Np$.

Let us see contributions at the intersection point of two
adjacent patches.  The point is that a particle from one patch and
condensates from other patches interact within a limited domain of
$\sim \lf^2$ near the intersection points.  Its phase space is
independent of $\Theta$.  Therefore the contribution from the
intersection points is
\begin{equation}
\text{(energy cost)}
\sim \Np \cdot \Nc
 \lf^2 \, f(M_{\rm B})
\sim \frac{\Nc}{\Theta} \cdot \lf^2 \, f(M_{\rm B})\,,
\end{equation}
where $f(M_{\rm B})$ is some function of the order
$M_{\rm B} \sim \lf$ with $M_{\rm B}$ be the mass gap near the
boundary, and vanishes as $M_{\rm B}\rightarrow 0$.  The contribution
must be an energy cost.  The reason is that in Eq.~(\ref{con1}) we
overestimated the energy gain which should be reduced around the patch
boundaries.  The misalignment of chiral spiral wavevectors tends to
destroy the different chiral spirals one another, and reduces the size
of the gap at the intersection points.  Diagrammatically, this
contribution will appear as interactions among chiral spiral mean
fields in different domains.  The presence of this term becomes more
important for smaller $\Theta$.

\begin{figure}[tb]
\vspace{0.0cm}
\begin{center}
\scalebox{1.0}[1.0] {
  \includegraphics[scale=.25]{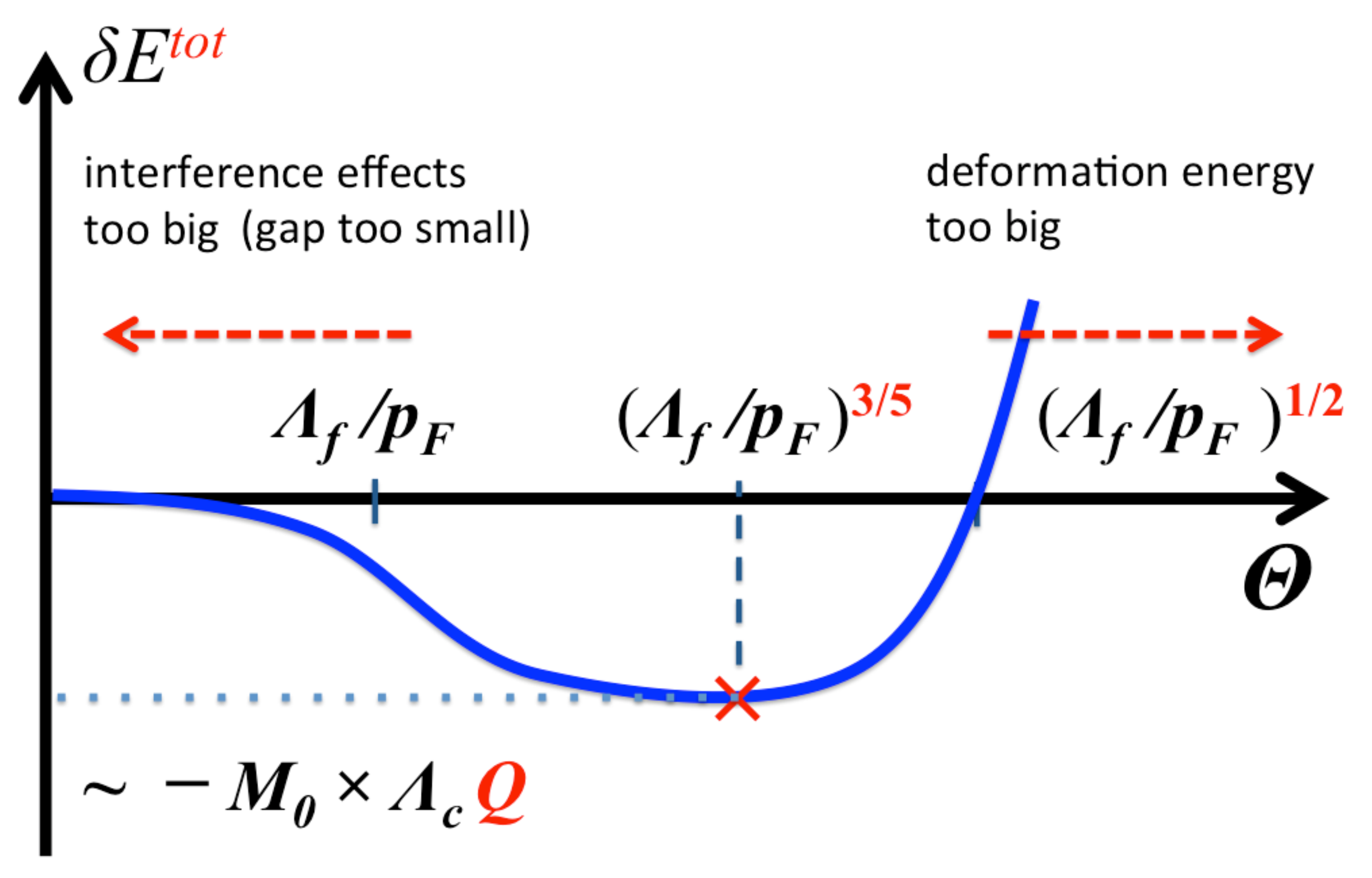} }
\end{center}
\vspace{0.0cm}
\caption{The schematic energy landscape as a function of $\Theta$.
  The region $\Theta \ll \lf/\pF$ is beyond the applicability of our
  analysis.
}
\label{fig:energylandscape}
\vspace{0.2cm}
\end{figure}

Now we can express the energy density as a function of $M$ and
$\Theta$.  In the domain $\lf/\pF \ll \Theta \ll 1$, it reads
\begin{align}
&\delta \calE(M,\Theta) 
\notag\\
&\;\;\sim \Nc \bigg(
\frac{ \lf^2 \, f(M_{\rm B}) }{\Theta}
- c_0\, M_0 \lf\, \pF
- c_2\, M_0 \lf\, \pF \Theta^2
+ c_4\, \pF^3 \Theta^4 + \cdots
\bigg)\,,
\end{align}
where coefficients $c_0, c_2, \cdots$ are positive, and we have
subtracted the free Fermi gas contribution.  The energy balance is
schematically illustrated in Fig.~\ref{fig:energylandscape}.  To
assure this expression by microscopic calculations is our goal in
later sections.

\begin{figure}[tb]
\vspace{0.0cm}
\begin{center}
\scalebox{1.0}[1.0] {
  \includegraphics[scale=.25]{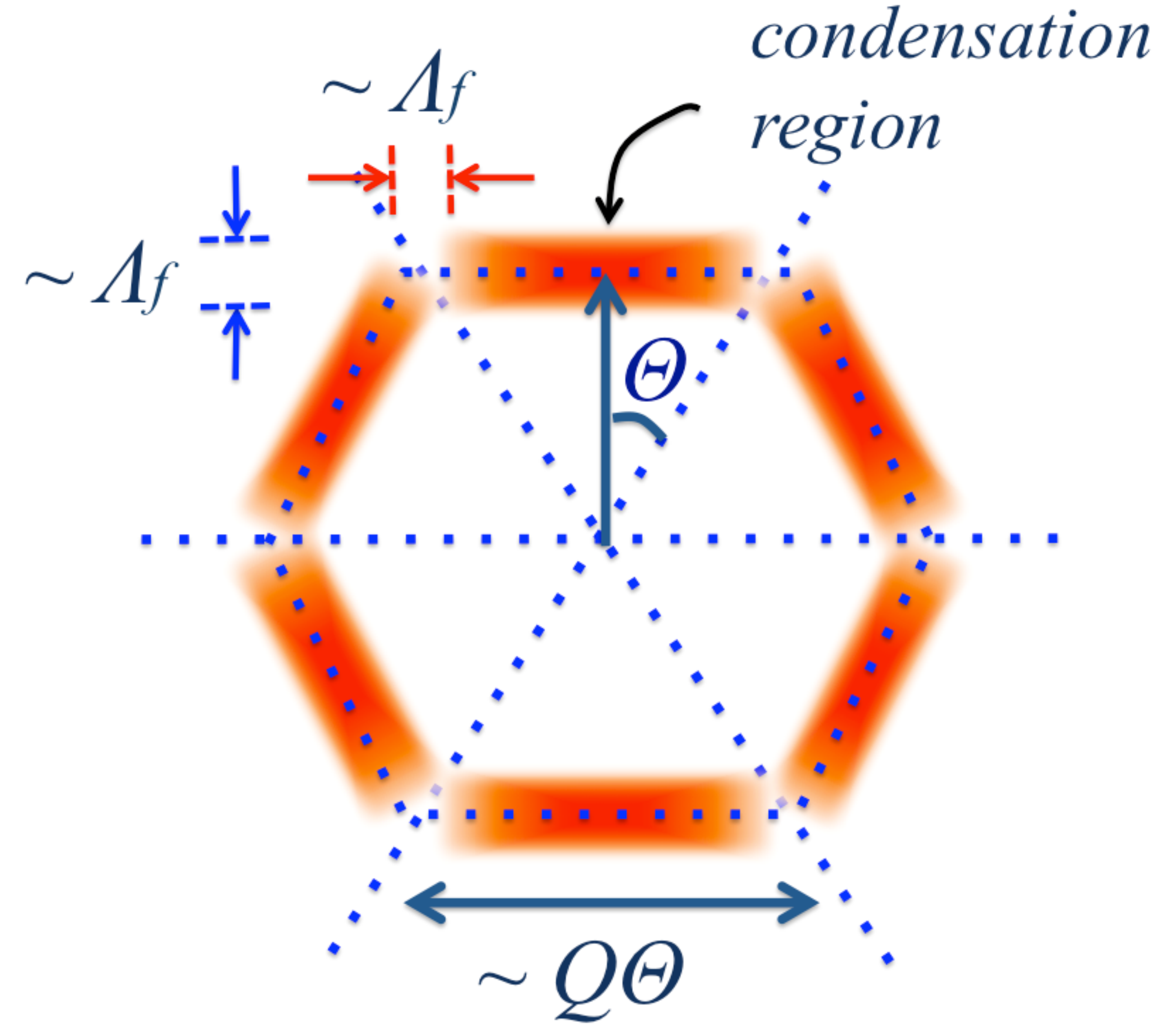} }
\end{center}
\vspace{0.0cm}
\caption{
The condensation region near the Fermi surface.  The thickness of the
region in the radial direction is $\sim \lf$.  In the boundary region
with the transverse size of $\sim \lf$, 
inter-patch interactions between
the nearest-neighbor chiral spirals destroy condensates one another,
reducing the energy gain from condensation effects.
}
\label{conregion}
\vspace{0.2cm}
\end{figure}

When $\lf/\pF < (\lf/\pF)^{1/2} < \Theta \ll 1$ is satisfied, the
$\pF^3 \Theta^4$ term dominates over other terms, and the deformation
energy supersedes the condensation energy.  Thus, there is an
upper bound of $\Theta$, and the energy minimum should lie in the
region, $\lf/\pF \ll \Theta \ll (\lf/\pF)^{1/2}$.  On the other hand,
the lower bound of $\Theta$ will be set by the patch-patch
interactions proportional to $1/\Theta$.  In the region of current
concern, we have
\begin{equation}
\frac{\partial\, \delta \calE(M,\Theta)}{\partial \Theta}
 \biggr|_{\Theta=\Theta_0}
 \!\sim\;
\Nc\bigg( -\frac{\lf^3}{\Theta_0^2} + 4 c_4\, \pF^4\, \Theta_0^3 \bigg)
 \;\sim\; 0 \,,
\end{equation}
for the energy minimum neglecting other terms.  Therefore we find
\begin{equation}
\Theta_0 \sim \bigg( \frac{\lf}{\pF} \bigg)^{\! 3/5} \;.
\end{equation}
As we promised, we can confirm in this way that
$\pF\Theta_0 \sim (\pF/\lf)^{2/5} \lf \gg \lf$,
which surely justifies Eq.~\eqref{eq:pfgg} {\it posteriori}.

The total energy is dominated by the condensation term
$\sim - \pF \lf^2$, which is independent of $\Theta_0$.  The leading
corrections come from the $1/\Theta_0$ term (patch-patch interactions)
and the $\Theta_0^4$ term (the deformation energy) which are
suppressed by $(\lf/\pF)^{2/5}$ compared to the leading contribution.
This implies that after solving the one-patch problem at the
mean-field level, other patch-patch contributions can be treated
within perturbation theory.  We will formulate our perturbation theory
within the domain of
$(\lf/\pF) \ll \Theta \ll (\lf/\pF)^{1/2}$. 

This paper is organized as follows.  In Sec.~\ref{secmodel}, we define
our model and describe consequences of form-factor effects in vacuum.
In Sec.~\ref{secdecomp}, the Fermi sea is decomposed into several
segments.  We formally separate the Lagrangian into the one-patch and
the patch-patch interactions.  In Sec.~\ref{seconepatch}, the
mean-field treatment for the one-patch problem is discussed.  We first
identify the dominant terms within one patch, and construct the mean
field for chiral spirals as well as the mean-field propagators for
quasi-particles.  In Sec.~\ref{formal}, we give a formal expression of
the perturbative expansion.  In Sec.~\ref{secpert}, we treat
corrections from subdominant terms in one patch.  It will be shown
that subdominant terms are suppressed by powers of $\Theta$ or
$\lf/Q$.  In Sec.~\ref{sec:interference}, the 
inter-patch interactions
at
the patch boundaries are discussed.  The size and sign of $1/\Theta$
terms are estimated in both perturbative and non-perturbative
manners.  In Sec.~\ref{secdiscussion}, we argue possible impacts of
several corrections ignored in this paper.  We also review other works
and attempt to place this work in perspective.
A coordinate space structure of the interweaving chiral spirals
is also discussed, leaving several interesting questions open.
Section~\ref{secsummary} is devoted to a summary and possible future
directions.

\section{A Model: The Four-Fermi Interaction with Form-factor Effects}
\label{secmodel}
In this section, we introduce explicit form factors into the NJL
model, so that the interaction is cutoff when the momentum transfer
becomes large.  This makes the NJL model non-local.  (For other
attempts to introduce non-locality which emphasize different aspects
from ours, see Ref.~\cite{Buballa:1992sz}.)  We will see that effects
of such a cutoff determine the momentum domains where condensation
phenomena are relevant.  This aspect is particularly important when we
consider the very large Fermi sea in which condensation phenomena
occur as Fermi surface effects, rather than the vacuum effects.

\subsection{Form factor effects}

Let us consider the scalar-scalar type of the four-Fermi interaction,
\begin{equation}
\int \rmd^3x \, \big( \bar{\psi} \psi(x) \big)^2
=
\int \rmd x_0 \! \int_{q,p,k}
\big( \bar{\psi}(\vp + \vq ) \psi(\vp) \big)
\big( \bar{\psi}(\vk ) \psi(\vk+ \vq ) \big)\,,
\label{form1}
\end{equation}
where we define a shorthand notation,
\begin{equation}
 \int_{q,p,k} \equiv \int \frac{\rmd\vq~ \rmd\vp~ \rmd\vk~}
 {(2\pi)^6}\,,
\end{equation}
and we did not explicitly write the coupling constant and
$x_0$ dependence of fermion fields.  Since the four-Fermi interaction
is not renormalizable beyond (1+1) dimensions, we need introduce some
UV cutoff.

We regularize the UV interaction by including form-factor effects,
\begin{equation}
\hspace{-0.5cm}
\int \rmd^3x \, \big( \bar{\psi} \psi(x) \big)^2 \rightarrow
\int \rmd x_0 \! \int_{q,p,k}
\big( \bar{\psi}(\vp + \vq ) \psi(\vp) \big)
\big( \bar{\psi}(\vk ) \psi(\vk+ \vq ) \big) 
\,\theta_{p,k} \;,
\label{form2}
\end{equation}
where
\begin{equation}
\theta_{p,k} \equiv \theta\big( \lf^2 - (\vp - \vk)^2 \big)\,.
\end{equation}
This mimics the form-factor effects in large-$\Nc$
QCD\footnote{
Here we put the cutoff on the spatial-momentum $\vec{p}^2$, not on the
Euclidean momentum $p_E^2$.  So, results in this work are connected to
those of large-$\Nc$ QCD in the Coulomb gauge, in which the dominant
non-perturbative part is of the instantaneous type
\cite{Gribov:1977wm}.  An alternative choice would be to introduce a
cutoff on $p_E^2$ keeping manifest Lorentz invariance.  Such a
treatment should mimic, for instance, Landau-gauge results in
Euclidean space.  However we do not know their Minkowskian behavior,
so we have to compute quantities in Euclidean space.  Then the price
we have to pay is that a simple physical intuition does not
necessarily work, especially at finite density.
},
and removes the UV divergences associated with interacting processes.

\begin{figure}[tb]
\vspace{0.2cm}
\begin{center}
\scalebox{3.0}[1.0] {
  \includegraphics[scale=.12]{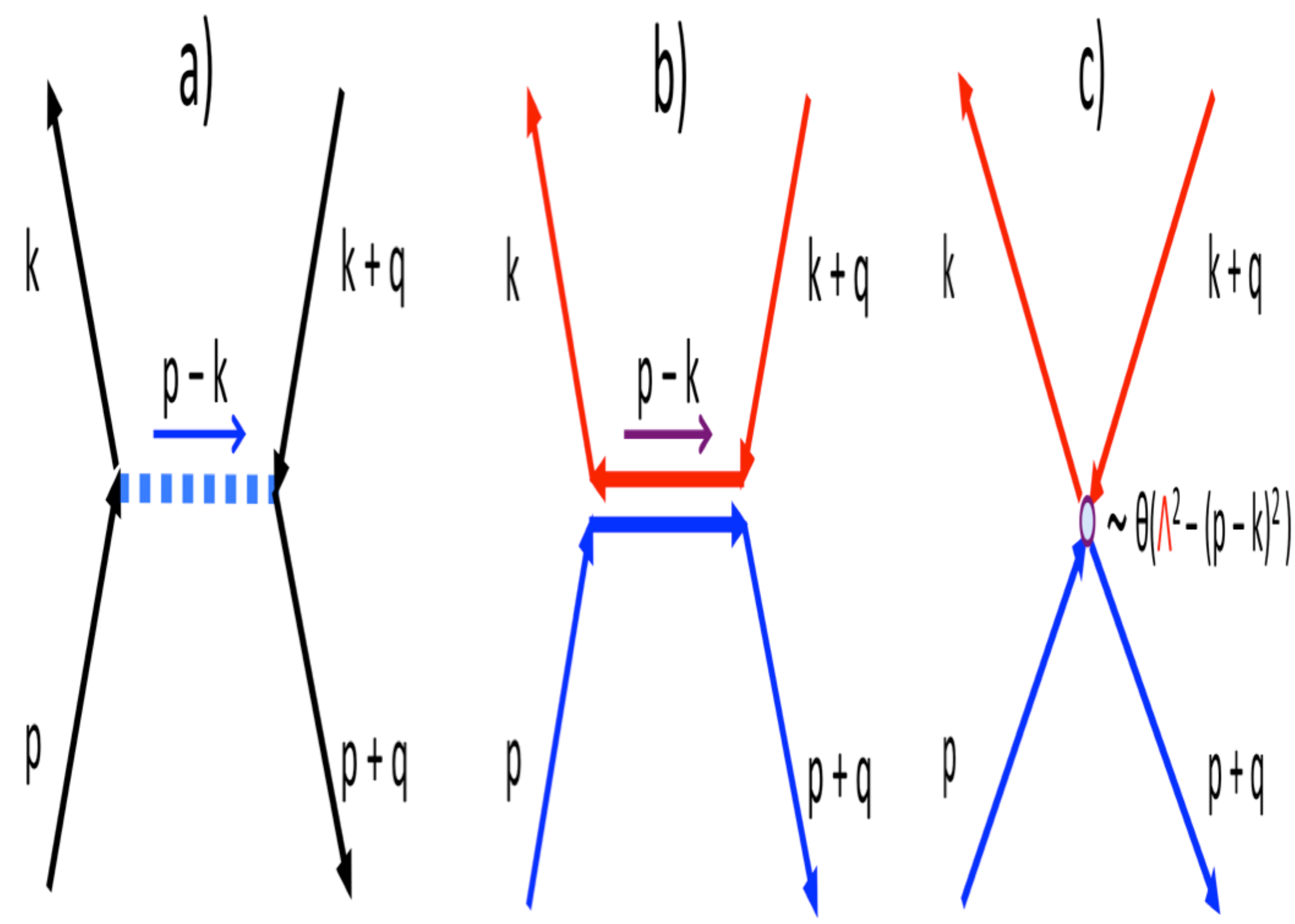} }
\end{center}
\vspace{0.2cm}
\caption{
(a) The non-perturbative gluon exchange which is supposed to damp
  quickly in the UV region.
(b) The color line representation of the one-gluon exchange.
(c) Our effective four-Fermi interaction including form factor
  effects.
}
\label{figform}
\vspace{0.2cm}
\end{figure}

The large-$\Nc$ QCD is mimicked as follows.  The one-gluon exchange
including non-perturbative effects are shown in Fig.~\ref{figform}(a).
Its strength damps as the momentum transfer becomes large.  We roughly
take into account this property by introducing a step function,
$\theta\big( \lf^2 - (\vp-\vk)^2\big)$, keeping the interaction
strength constant.  In QCD, the cutoff scale $\lf$ should be taken to
be of the order of $\lqcd$.

In Fig.~\ref{figform}(b), we show the color line representation to
illustrate how the one-gluon exchange 
interaction should be contracted into a four-Fermi type interaction.
Taking into account features in Figs.~\ref{figform}(a) and
\ref{figform}(b), we arrive at a simple model described in
Eq.~(\ref{form2}) and Fig.~\ref{figform}(c).

A relevant consequence of our choice of the form factor is that the
coupling between fermion fields and the condensate becomes strongly
momentum dependent.  In particular, fermions decouple from the
condensation effects if they belong to domains with a momentum
difference of the order of $\lf$.

Let us first see this property in the vacuum case by investigating the
Schwinger-Dyson equation for quasi-particles.  After picking up the
pole, we have\footnote{
We ignored the Lorentz vector self-energy for the sake of simplicity.
This simplification should not alter our basic statements in this
paper.}
\begin{equation}
\Sigma_m(\vp) 
= \int \! \frac{ d \vk }{ (2\pi)^2 } ~
 \frac{\Sigma_m( \vk )} {2 \epsilon( \vk ) } ~
\theta_{p,k}\,,
\label{SDeq1}
\end{equation}
where $\Sigma_m(\vk) $ is the self-energy and
$\epsilon(\vk ) = \sqrt{ \vk^2 + \Sigma_m^2(\vk) }$
is the quasi-particle energy.  For its diagrammatic expression, see
Fig.~\ref{fig:SDeq}\footnote{
The Fock-type gluon exchange, which is dominant in large-$\Nc$ QCD,
corresponds to the Hartree term of the four-Fermi interaction model,
so we can apply most of techniques used in the NJL model
calculations.
}.
When $|\vec{p}|$ is very large, $|\vec{k}|$ must be as large because
of the form factor.  In the asymptotic region
$|\vp| \gg \Sigma(\vp )$, the Schwinger-Dyson equation looks like
\begin{equation}
\Sigma_m ( \vp)
\sim \frac{1}{2 \epsilon( \vp) } 
\int \! \frac{ \rmd \vk }{ (2\pi)^2 } ~
 \Sigma_m ( \vk ) ~
\theta_{p,k}\,,
\end{equation}
from which one can show that $\Sigma_m (\vp)$ damps faster than
$1/\epsilon( \vp) \sim 1/|\vp|$ in the UV region, meaning that the
scattering between quasi-particles and the mean-field condensates
diminishes for large momentum.

This UV behavior tremendously simplifies considerations on the
energy benefit from condensation effects.  Due to decoupling from
condensation effects in the UV region, the normalized energy density,
in which the energy density without condensation are subtracted out,
is dominated by the IR contributions.  Also, the couplings between the
IR and the UV regions are allowed only in the limited region
constrained by the form factor effects.  
Thus, 
we can proceed with our calculations
independently from details of the UV physics.
 
At finite density, the same argument applies if we replace the Dirac
sea with the Fermi sea, namely, the excitation energy $\epsilon(\vp)$
is measured from the Fermi surface instead of the vacuum.  Then it
follows that dominant contributions to the condensation come from
fermions near the Fermi surface rather than those near vacuum.

\subsection{Bosonization of the Model}
Following usual treatments, let us introduce the auxiliary fields and
formally rewrite the action.  Although this formal treatment 
is not mandatory, it would be practically convenient to use several
relations derived in this framework.

Since our four-Fermi interaction is not of the separable type, we have
to modify conventional treatments slightly as
\begin{equation}
-\int \rmd x_0 \! \int \! \frac{\rmd \vq}{ (2\pi)^2 } ~ \Phi(\vq)\,
\Phi(-\vq)  
\rightarrow
-\int \rmd x_0 
\int_{q,p,k}
\Phi(\vq;\vp)
~ \theta_{p,k}
~ \Phi(-\vq;\vk)\,.
\end{equation}
Here we did not write $x_0$ dependence of the auxiliary boson field,
$\Phi$, explicitly for notational simplicity.  Next we shift the boson
fields to generate a four-Fermi interaction,
\begin{equation}
\Phi(\vq;\vp) 
\rightarrow
\Phi(\vq;\vp) + \bar{\psi}(\vp+\vq) \psi(\vp)\,, 
\end{equation}
then, by adding these trivial terms in $\Phi$, we can eliminate the
four-Fermi interactions out from the original action to obtain the
Yukawa-type model.  The form of the vertex is
\begin{equation}
- \int \rmd x_0 \! \int_{q,p,k} 
\Phi(\vq; \vp)
~ \theta_{p,k}~
\bar{\psi}(\vk) \psi(\vk + \vq)\,. 
\label{massvertex}
\end{equation}
We can obtain the gap equation from
$0=\int {\mathcal D} \Phi  {\mathcal D} \Psi {\mathcal D}
\bar{\Psi} ~ \frac{\delta}{\delta \Phi(\vq;\vp)} \rme^{iS}$ and find,
%
\begin{equation}
\int \frac{\rmd\vk}{(2\pi)^2}~~
\theta_{p,k} \; \big\{
\la \Phi(-\vq; \vk) \ra
+ 
\la \bar{\psi}(\vk) \psi(\vk + \vq) \ra
\big\}
=0\,.
\end{equation}
Since the equation should hold at arbitrary $\vp$, we conclude that
$\la \Phi(-\vq; \vk) \ra 
= - \la \bar{\psi}(\vk) \psi(\vk + \vq) \ra ~(\ge0)$, 
which is independent of $x_0$.

The relation between the mass self-energy and $\Phi$ 
can be read off 
from the coefficient of $\bar{\psi} \psi$ in Eq.~(\ref{massvertex}),
i.e.
\begin{equation}
\Sigma_m(\vq;\vk)
= 
\int \frac{\rmd\vp}{(2\pi)^2}
~ \Phi(\vq; \vp)
~ \theta_{p,k} \,.
\end{equation}
In the stationary phase approximation of the $\Phi$-integral (which is
rigorous in the large-$\Nc$ limit), we can replace $\Phi$ with
$\la \Phi \ra$.  One can check that 
substituting the gap equation for $\Phi$ and taking $\vq =\vec{0}$
(for uniform condensates)
reproduce the Schwinger-Dyson equation in Eq.~(\ref{SDeq1}).

Instead of computing self-consistent solutions precisely, let us
consider what kind of Ansatz would capture basic properties of
stationary solutions of $\Phi(\vq;\vp)$.  We will apply such arguments
to more complicated case at finite density where condensation effects
are important near the Fermi surface.

In the vacuum case, we already know qualitative behaviors of
$\Sigma_m(\vq; \vp)$ at high momenta from the previous arguments.  Its
value damps as $|\vp| \rightarrow \infty$, leading to damping of
$\Phi(\vq;\vp)$, as seen in the gap equation.  If we denote the
characteristic scale of damping as $\lc$, presumably the simplest
Ansatz would be
\begin{equation}
\Phi(\vq;\vp) 
= \Delta \cdot \theta ( \lc^2 - \vp^{~2} )\; \delta(\vq)\,.
\end{equation}
See also Fig.~\ref{fig:density}.  In principle, not only $\Delta$, but
also $\lc$ should be treated as variational parameters as
functions of $\lf$ that is an intrinsic scale in our model.

Perhaps it might be helpful to mention the relationship between this
Ansatz and usual spatial-momentum cutoff scheme.  In our model,
conventional treatment of the NJL model is recovered by taking
$\lf \rightarrow \infty$ with
$\lc$ kept fixed to be the NJL-model cutoff
$\Lambda_{{\rm NJL}} \sim 600\:\text{MeV}$ which semi-quantitatively
describes low-energy properties of the hadron phenomenology.  Such
successes of the NJL model imply that once we correctly cutoff
{\it the domain of condensation}, we should be able to express the
low-energy dynamics quantitatively.  In our model, this cutoff scale
should be dynamically determined by adjusting a value of $\lf$.

Adopting this interpretation, the effective cutoff of the NJL model
should be measured from the Fermi surface, not from zero momentum,
because condensation effects mainly appear near the Fermi surface.
This leads to a picture such that chiral symmetry is restored deeply
inside of the Fermi sea but is broken near the Fermi surface.

\subsection{Spinor Representations in (2+1) Dimensions}
\label{Spinors}
In this work we discuss (2+1) dimensional theory
instead of (3+1) dimensional one
because considerations for 
shapes of the Fermi surface are much simpler.
On the other hand, in (2+1) dimensions,
there is no chirality in a strict sense.
Although this fact does not modify our main considerations,
we shall give a brief remark on this special properties 
in (2+1) dimensions.

A spinorial representation of the Lorentz group $SO(2,1)$
is provided by two component spinors,
with $2\times 2$ representation of
the Dirac algebra which is given by the Pauli matrices
\begin{equation}
\gamma_0 = \sigma_2~,~~
\gamma_1 = i \sigma_3~,~~
\gamma_2 = i \sigma_1~.
\end{equation}
There is no other $2\times 2$ matrix anticommuting
with these $\gamma_\mu$
so that $\gamma_5$ can not be defined.

Therefore we use a four component spinor
for which $\gamma_5$ can be defined.
To get some feeling,
one can imagine that fermions in (3+1) dimensions
are restricted within (2+1) dimensional space
by imposing some external condition
as done for the Kaluza-Klein reduction.
Then the $\gamma$ matrices for four component spinors
can be taken in the same way as
(3+1) dimensional $\gamma$ matrices.

We expect that this prescription is the easiest way to directly 
convert our (2+1) dimensional manipulations
into higher dimensional ones.
For further discussions on (2+1) dimensional chirality, 
see Ref. \cite{Appelquist:1986fd} for instance.

\section{Decomposition of the Lagrangian into
Multiple Patch Domains}
\label{secdecomp}

In this section, we decompose the NJL Lagrangian into different
segments.  For the sake of simplicity, we begin with the Lagrangian
with one flavor,
\begin{equation}
{\cal L}
=
\bar{\psi}\,  \rmi\, \Slash{\partial}  \psi 
  + \frac{G}{\Nc} 
 \big( (\bar{\psi} \psi)^2 + (\bar{\psi}\, \rmi
 \gamma_5 \psi)^2 \big)\,,
\end{equation}
where we explicitly factor out the $\Nc$ dependence of the
interaction, so that $G =O(\Nc^0)$.  
In the (2+1)-dimensional system, $G$ is dimensionful,
and we take it to be $\sim \lf^{-1}$.
Our Lagrangian has continuous
chiral symmetry, $U(1)_{\rm L} \times U(1)_{\rm R}$.  In this work we
ignore the $U(1)_{\rm A}$ problem that is of $O(1/\Nc)$ and thus
negligible.
For the moment, we will not write the form factor explicitly for
notational simplicity.

We introduce the unit vectors $\vec{n}_{i}$ 
which point to the
center of the $i$-th patch, and the unit vector $\vec{n}_{i\perp}$
which is orthogonal to $\vec{n}_{i}$. 
We project out the spatial
components of vectors generically as
\begin{equation}
 p_{i \para} = \vec{n}_i \cdot \vec{p}\,,
\qquad  p_{i \perp} = \vec{n}_{i\perp} \cdot \vec{p}\,.
\end{equation}
With this definition, fermion fields can be decomposed into $\Np$
momentum domains,
\begin{align}
\hspace{-0.5cm}
\psi(x) 
&= \int \!\frac{\rmd^3 p}{(2\pi)^3}~ \tilde{\psi}(p)\,
\rme^{-\rmi p\cdot x} 
\notag \\
&= \frac{1}{(2\pi)^3} 
\sum_{i=1}^{\Np} 
\int^{\infty}_{-\infty} \rmd p_0
\int_{-\infty}^{\infty} \rmd p_{i \para} 
\int_{-p_{i\para} \tan \Theta }^{p_{i\para} \tan \Theta} \rmd p_{i\perp}\;
\tilde{\psi}(p) \, \rme^{-\rmi p\cdot x}
\notag\\
&\equiv
\sum_{i=1}^{\Np} \psi_i(x) \,.
\end{align}
It is worth mentioning here that our formulation has potential
relevance to analytic approaches based on the high-density effective
theory \cite{Hong:1998tn}.
A single patch includes a set of two domains with $p_{i \para} >0$
and $p_{i \para} <0$ for the positive energy states, in a similar way
as the (1+1)-dimensional problem.  Thus, the angle for one patch is
$2\times 2\Theta$, and $\Theta$ should satisfy  
\begin{equation}
 4\Theta \, \Np = 2 \pi\,.
\end{equation}
Let us first decompose the free part of the Lagrangian.  Since it is
diagonal in momentum space, we have
\begin{equation}
{\cal L}^{\rm kin}
=
\sum_i
 \bar{\psi}_i \, i \, \Slash{\partial} \psi_i
\equiv
\sum_i {\cal L}^{\rm kin}_i\,.
\end{equation}
The decomposition of four-Fermi interactions ${\cal L}^{\rm int}$ are much
more cumbersome, since the interaction terms combine different
domains.  Explicitly writing, it follows
\begin{align}
{\cal L}^{\rm int}&=
\frac{G}{\Nc} 
\big(  (\bar{\psi} \psi)^2 +  (\bar{\psi}\, \rmi \, \gamma_5
 \psi)^2 \big) \notag\\
&=
\frac{G}{\Nc} 
 \sum_{i,j,k,l} \big( 
(\bar{\psi}_i \psi_j)(\bar{\psi}_k \psi_l)
+
(\bar{\psi}_i \,\rmi \, \gamma_5 \psi_j)
(\bar{\psi}_k \,\rmi \, \gamma_5 \psi_l) 
\big)\,.
\end{align}
It can be decomposed into four types of interaction terms,
\begin{align}
\sum_i {\cal L}_i^{\rm int}
&=
\frac{G}{\Nc}
\sum_i \big( (\bar{\psi}_i \psi_i)^2 + (\bar{\psi}_i
 \,\rmi \gamma_5
\psi_i)^2 \big) \,,
\notag \\
\sum_{i\neq j} {\cal L}_{i,j}^{\rm  int}
&= \frac{G}{\Nc}  \sum_{i\neq j} \big( 
(\bar{\psi}_i \psi_i)(\bar{\psi}_j \psi_j) 
+ (\bar{\psi}_i \, \rmi \gamma_5 \psi_i)
(\bar{\psi}_j \, \rmi \gamma_5 \psi_j)
\big) \,,
\notag \\
\sum_{i, j\neq k} {\cal L}_{i,jk}^{\rm int}
&= \frac{2G}{\Nc} \sum_{i, j\neq k} 
\big( (\bar{\psi}_i \psi_i)(\bar{\psi}_j \psi_k) 
+ (\bar{\psi}_i \,\rmi \gamma_5 \psi_i)
(\bar{\psi}_j \,\rmi \gamma_5 \psi_k)
\big) \,,
\notag \\
 \sum_{i\neq j, k\neq l} {\cal L}_{ij,kl}^{\rm int}
&= \frac{G}{\Nc} \sum_{i\neq j, k\neq l} 
\big( (\bar{\psi}_i \psi_j)(\bar{\psi}_k \psi_l)
+ (\bar{\psi}_i \,\rmi \gamma_5 \psi_j)
(\bar{\psi}_k \,\rmi \gamma_5\psi_l)
\big) \,.
\end{align}
Except the first line, interaction terms involve fermions belonging to
different patches.  Now we have the Lagrangian separated into
one-patch and patch-patch interactions,
${\cal L} = \sum_i {\cal L}^{\text{1-patch}}_i + \Delta {\cal L}^{\rm int}$,
where
\begin{equation}
{\cal L}^{\text{1-patch}}_i = {\cal L}^{\rm kin}_i + {\cal L}^{\rm int}_i \,,
\quad
\Delta {\cal L}^{\rm int}
= \sum_{i\neq j} {\cal L}_{i,j}^{\rm int}
+ \sum_{i, j\neq k} {\cal L}_{i,jk}^{\rm int}
+ \sum_{i\neq j, k\neq l} {\cal L}_{ij,kl}^{\rm int} \,.
\end{equation}
The one-patch Lagrangian will play a dominant role for condensation
effects.  We will first solve the one-patch problem, and then include
contributions from other patches perturbatively.

\section{The One-Patch Problem at the Mean Field Level}
\label{seconepatch}

In this section, we provide a formal treatment of the one-patch
Lagrangian.  We consider sufficiently high density for which the
interaction scale $\sim \lf$ is much larger than the transverse
kinetic energy near the Fermi surface which is suppressed as
$\sim \vec{p}_\perp^{~2}/Q$ where $Q \sim \pF$.  Then the fermion
dispersion relation is robust and does not rely on the transverse
momentum;  corrections from the higher spatial dimensions than the
(1+1)-dimensional problem are generally suppressed by extra powers of
$1/Q$.  Therefore, the gap equation for the mean field effectively
becomes (1+1) dimensional.  We here analyze this quasi
(1+1)-dimensional problem in detail.

We first summarize some convenient notations used in (1+1)
dimensions.  They are useful to identify the dominant and subdominant
terms for the chiral spiral formation.  Then we bosonize the dominant
part of the four-Fermi interactions, and construct the mean-field
Lagrangian for the chiral spirals.  We prepare the mean-field
propagator and write the gap equation down.  The results in this
section are the basis for the perturbation theory, which treats
subdominant terms ignored at the mean-field level.

\subsection{Preliminaries}
In (1+1)-dimensional models, the chirality (i.e.\ the eigenvalue of
$\gamma_5$) characterizes the moving directions of particles.  In
higher dimensions, the corresponding $\gamma$-matrix is not
$\gamma_5$, but $\Gamma_{i5} \equiv \gamma_0 \gamma_{i\para}$ for
particles moving to the $x_{i\para}$-direction.  We define
\begin{equation}
\psi_{i \pm}
\equiv
\frac{1 \pm \Gamma_{i5} }{2} \psi_i \,.
\end{equation}
This $\Gamma_{i5}$ satisfies the following algebraic relations:
\begin{equation}
(\Gamma_{i5})^2=1\,,\quad
\{\Gamma_{i5}, \gamma_{i 0} \}=0\,,\quad 
\{\Gamma_{i5}, \gamma_{i \para} \}=0\,,\quad
[\Gamma_{i5}, \gamma_{i \perp} ]=0\,.
\end{equation}
%
The free Lagrangian can be decomposed into two pieces.  The
longitudinal part is defined with $(+,+)$ or $(-,-)$ 
combinations\footnote{
Our definition of $\bar{\psi} = \psi^\dag \gamma^0$.
Our metric is $g_{\mu \nu} = g^{\mu \nu} = {\rm diag}(1,-1,-1)$.};
\begin{equation}
{\cal L}^{\rm kin}_{i\para}
=
\psi^\dag_{i+} 
 \,\rmi(\partial_0 - \partial_{i\para})
\psi_{i+}
+
\psi^\dag_{i-} 
 \,\rmi(\partial_0 + \partial_{i\para}) 
\psi_{i-} \,,
\end{equation}
and the part made of $(+,-)$ combinations is
\begin{equation}
{\cal L}^{\rm kin}_{i\perp}
= 
\bar{\psi}_{i+} \,\rmi\,\Slash{\partial}_\perp \psi_{i-}
+
\bar{\psi}_{i-}  \,\rmi\,\Slash{\partial}_\perp  \psi_{i+} \,.
\end{equation}
Below, we will drop off the index $i$ as far as no confusion arises.

At finite density with the Fermi momentum $Q$, it is natural to
measure momenta of fermions from the Fermi surface.  Accordingly, we
take fields with shifted momenta\footnote{
We will use lower index expressions for momenta,
and $Q$ should be interpreted as the lower component.
And when we use the vector, that means the lower index components.
For instance,
$\vq = (q_1, q_2)$ and $\vx = (x_1, x_2)$.},
\begin{equation}
\psi_\pm (x)
= \rme^{\rmi\, Q x_\para \Gamma_5} \psi'_\pm (x) 
= \rme^{\pm \rmi\, Q  x_\para} \psi'_\pm (x) 
\end{equation}
or
\begin{equation}
\psi'_\pm ( \delta \vp) 
= \psi_\pm (\delta p_\para \pm Q, p_\perp) \,,
\qquad \big(\delta \vp = (\delta p_\para, p_\perp) \big)\,,
\end{equation}
in momentum space.  We use the notation $\delta \vp$ to emphasize that
momenta of $\psi'$ field are measured from the Fermi surface.  Using
the $\psi'$ field, one can easily identify dominant and subdominant
terms at large density.

In the $\psi'$-representation, the longitudinal part becomes\footnote{
In the grand canonical ensemble the basis $\psi'$ with $Q=\mu_q$
eliminates the chemical potential term reflecting that in the
$\psi'$-representation we can deal with dynamics near the Fermi
surface as in vacuum.  This simple logic is not directly applicable in
the canonical ensemble since the density constraint does not
explicitly appear at the Lagrangian level.
}
\begin{equation}
{\cal L}^{\rm kin}_{\para}
\rightarrow
\psi'^\dag_{+} 
\big[ \rmi(\partial_0 - \partial_{\para}) - Q \big]
\psi'_{+}
+
\psi'^\dag_{-} 
\big[ \rmi(\partial_0 + \partial_{\para}) - Q \big]
\psi'_{-}\,,
\end{equation}
and the transverse kinetic term and the mass term acquire the
oscillating factors,
\begin{equation}
{\cal L}^{\rm kin}_{\perp}
\rightarrow 
\bar{\psi}'_{+}  \,\rmi\,\Slash{\partial}_\perp \psi'_{-}\  
\rme^{-2\,\rmi Q  x_\para} 
+
\bar{\psi}'_{-} \,\rmi\,\Slash{\partial}_\perp \psi'_{+}\ 
\rme^{2\,\rmi Q x_\para}\,. 
\end{equation}
Such oscillatory terms are suppressed near the Fermi surface by powers
of $1/Q$.  In the free theory, in fact, the excitation energy at
$|\delta p_\para|\ll Q$ is
\begin{equation}
\epsilon^{{\rm free}}( \delta \vp )
= 
\sqrt{ (Q + \delta p_\para)^2 + p_\perp^2}  - Q
= |\delta p_\para| 
+ \frac{ \delta p_\para^2 + p_\perp^2 }{2 Q} +\cdots \,.
\end{equation}
%
Terms with oscillating factors define what we call ``subdominant''
terms.

The four-Fermi interactions can be also separated into dominant terms
and subdominant terms,
\begin{align}
(\bar{\psi} \psi)^2 
&= \frac{1}{2} \big( (\bar{\psi} \psi)^2 
+ (\bar{\psi}\,\rmi\, \Gamma_5 \psi)^2 \big)
+
\frac{1}{2} \big( (\bar{\psi} \psi)^2 
- (\bar{\psi}\,\rmi\, \Gamma_5 \psi)^2 \big) \notag \\
&=
2\,(\bar{\psi}_+ \psi_- ) (\bar{\psi}_- \psi_+ ) 
+ \big(
 (\bar{\psi}_- \psi_+ )^2 
+ (\bar{\psi}_+ \psi_- )^2 \big)
\notag \\
&\longrightarrow~
2\,(\bar{\psi}'_+ \psi'_- ) (\bar{\psi}'_- \psi'_+ ) 
+ (\bar{\psi}'_- \psi'_+ )^2 \rme^{4\,\rmi Q x_\para}  
+ (\bar{\psi}'_+ \psi'_- )^2 \rme^{-4\,\rmi Q x_\para} \,.  
\label{eq4fermi'1}
\end{align}
The first term corresponds to the continuous symmetric part, which
becomes IR dominant at high density.  We will apply the mean-field
Ansatz to dominant terms, while subdominant terms are treated as
perturbation.  In this treatment, a gap will be found only near the
Fermi surface.  The gaps will not open periodically in momentum space
because of the absence of different harmonics\footnote{
The current problem is different from
the problem of the Peierls instability with an external periodic
potential.
In our case the coupling between the mean field and particles
depend on the $\pm$ combination.  For instance, the mean field
$\la \bar{\psi}_+ \psi_- \ra = \Delta \rme^{-2\,\rmi Qx_\para}$ can
scatter the particles from the $+$ region to the $-$ region, but
cannot scatter from the $-$ region to the $+$ region.  In this way
particles and holes are kept around the Fermi surface, without going
to the higher harmonic regions.}.

Finally, for later convenience, we write Eq.~(\ref{eq4fermi'1}) in
momentum space including the form factor explicitly.  For $(+-)(-+)$
combinations of the dominant part, we have
\begin{align}
&
\int_{q,p,k}
\big( \bar{\psi}_+( \vp+\vq ) \psi_-( \vp ) \big)~
\big( \bar{\psi}_-( \vk) \psi_+( \vk+\vq) \big)~ 
\theta_{p,k}
\notag \\
=
& \int_{q, \delta p, \delta k}
\big( \bar{\psi}'_+( \delta \vp+\vq -2Q \vec{n} ) \,
\psi'_-( \delta \vp ) \big)\,
\big( \bar{\psi}'_-( \delta \vk) \,
\psi'_+( \delta \vk+\vq - 2Q \vec{n} ) \big)\,
\theta_{\delta p, \delta k}\,.
\label{eq4fermi+-1} 
\end{align}
Note that if $\vq \simeq  2Q \vec{n}$, all fields can be close to the
Fermi surface simultaneously when
$\delta \vp \sim \delta \vk \sim \vec{0}$.  This is the reason why the
configuration with $\vq \sim 2Q\vec{n}$ becomes dominant in the path
integral.  For this reason, we should choose the wave vector of the
chiral spirals to be $\vq = 2Q\vec{n}$ in the high-density limit.
This expression also indicates that the exciton-type condensation
(i.e.\ homogeneous chiral condensation) with $\vq = \vec{0}$ is
not favored energetically.

On the other hand, for $(+-)(+-)$ combinations of the subdominant
part, we have
\begin{align}
&
\int_{q,p,k}
\big( \bar{\psi}_+( \vp+\vq ) \psi_-( \vp ) \big)~
\big( \bar{\psi}_+( \vk) \psi_- ( \vk+\vq) \big)~ 
\theta_{p,k}
\notag \\
=
&
\int_{q, \delta p, \delta k}
\big( \bar{\psi}'_+( \delta \vp+\vq -2Q \vec{n} ) 
\psi'_-( \delta \vp ) \big)~
\big( \bar{\psi}'_+ ( \delta \vk) 
\psi'_- ( \delta \vk+\vq + 2Q \vec{n} ) \big)~ 
\notag \\
& \qquad
~\times 
~\theta 
\big( 
\lf^2 - ( \delta \vp - \delta \vk - 2Q \vec{n} )^2
\big)\,.
\label{eq4fermi+-2} 
\end{align}
In contrast to the dominant terms, subdominant interactions require
that at least one fermion must go far away from the Fermi surface
regardless of any $\vec{q}$.  The propagation of such a fermion serves
the $1/Q$ suppression in the quantum corrections.

\subsection{Formal Treatment: Bosonization}

Now let us introduce the mean-field Ansatz for one patch.  The
auxiliary-field method may be applied as before.  There are two slight
modifications on the standard approach.  One is that our boson field
is introduced as a complex field since we need to eliminate quark
bilinears, $\bar{\psi}_+ \psi_-$ and $\bar{\psi}_- \psi_+$, which have
complex phase factors that are opposite to each other.  Another is
that boson fields are used to be replaced with only dominant
four-Fermi interactions.

Inserting an identity to the original partition function, we introduce
the bosonic terms,
\begin{equation}
{\cal S}_\Phi = - 
\frac{\Nc}{2G}~ \int \rmd x_0 
\int_{p,q,k} \Phi^\dag( \vec{q}; \vp, x_0)
~\theta_{p,k}~
\Phi (\vec{q}; \vec{k}, x_0)\,,
\label{alboson}
\end{equation}
to the original action. 
For the moment, we will explicitly write the $x_0$ coordinate. 
We replace the dominant four-Fermi
interaction with a Yukawa-type vertex by shifting the boson fields,
\begin{equation}
\begin{split}
\Phi^\dag ( \vec{q}; \vec{p}, x_0) 
~&\longrightarrow ~
\Phi^\dag ( \vec{q};  \vec{p}, x_0)  
+ \frac{2G}{\Nc} ~\bar{\psi}_+ ( \vp + \vq, x_0) 
~ \psi_- (\vp, x_0)\,,
\\
\Phi  (\vec{q}; \vec{k}, x_0) 
~&\longrightarrow ~
\Phi ( \vec{q}; \vec{k}, x_0  )  
+ \frac{2G}{\Nc} ~\bar{\psi}_- ( \vk, x_0 ) 
 ~ \psi_+ ( \vk + \vq, x_0 )\,,
\end{split}
\end{equation}
then the Yukawa vertex is
\begin{align}
\hspace{-1em}
{\cal S}_{\Phi,\psi} =
& 
- \int \rmd x_0 \int_{p,q} \Big[~\int_k
\theta_{p,k}~
\Phi (\vec{q}; \vec{k}, x_0) ~\Big]
~\bar{\psi}_+ (\vp+\vq, x_0 ) 
 ~ \psi_- (\vp, x_0 ) \notag \\
& \quad
- \int \rmd x_0 \int_{p,q} \Big[~ \int_k
\theta_{p,k}~
\Phi^\dag (\vec{q}; \vec{k}, x_0) ~\Big]
~\bar{\psi}_- (\vp, x_0 ) 
 ~ \psi_+ (\vp + \vq, x_0  )\,,
\end{align}
where terms inside of $[\cdots]$ in the first and second terms are the
self-energy,
$\Sigma_m^\dag (-\vec{q}; \vec{p})$ and
$\Sigma_m(\vec{q}; \vec{p})$, respectively.
And the equation motion is
\begin{align}
&\la \Phi (\vq;\vk) \ra 
= - \frac{2G}{\Nc} ~\la \bar{\psi}_- (\vk, x_0 ) ~ \psi_+ (\vk + \vq, x_0 )
\ra\,, \notag \\
&\la \Phi^\dag (\vq;\vk) \ra 
= - \frac{2G}{\Nc} ~\la \bar{\psi}_+ (\vk+\vq, x_0 ) ~ \psi_- (\vk, x_0 ) \ra\,.
\end{align}
The RHS will appear to be $x_0$ independent,
so we will not have to write the $x_0$ coordinate in
$\la \Phi (\vq;\vk) \ra$ and $\la \Phi^\dag (\vq;\vk) \ra$
anymore.
Below we will not explicitly write $x_0$ dependence
of fermion fields for notational simplicity.
\subsection{Mean Field for Chiral Spirals and the Quasi-Particle Spectrum}
According to the arguments around Eq.~(\ref{eq4fermi+-1}), in the
high-density limit the Ansatz for chiral spirals in the $i$-th patch
is\footnote{
In our definition of $\Phi^\dag(\vq;\vk)$ in Eq.(\ref{alboson}),
$\Phi^\dag(\vq;\vk)$ actually carries momentum $-\vq$.}
\begin{equation}
\Phi_{0} (\vec{q};\vec{k})
= (2\pi)^2 \delta ( \vec{q} - 2Q \vec{n}_i ) \, \Delta (\vec{k})\,, 
\qquad
\Phi_{0}^\dag (\vec{q};\vec{k})
= (2\pi)^2 \delta ( \vec{q} - 2Q \vec{n}_i ) \, \Delta (\vec{k})\,, 
\end{equation}
where $\Delta$ is a real field which characterizes the magnitude of
the condensate.  We will give what would happen if we chose different
$\vq$ in Appendix~\ref{vqneq2Q}.  The Fourier transformation with
respect to the total momentum of the boson fields gives the
expression,
\begin{equation}
\Phi_{0} (\vx;\vec{k}) = \Delta(\vec{k}) ~\rme^{2\,\rmi Q x_\para}\,, 
\qquad
\Phi^\dag_{0} (\vx;\vec{k}) 
= \Delta(\vec{k}) ~\rme^{-2\,\rmi Q  x_\para}\,, 
\end{equation}
from which we can see that the complex nature of the boson fields is
taken into account in the phase factor.

For later convenience, let us define the mass gap function,
\begin{equation}
M (\vec{p}) 
\equiv 
\int_k~ \theta_{p,k}~ \Delta (\vec{k}) \,,
\end{equation}
which will be determined self-consistently after constructing the
mean-field propagator.  Using this shorthand notation and shifting
momentum $\vec{p} \rightarrow \vec{p} + Q \vec{n}_i$, 
the mass vertex
becomes
\begin{align}
& {\cal S}_{\Phi,\psi} \rightarrow 
{\cal S}_M  =
- \int \rmd x_0 \int_p  \big(\,
 M(\vec{p}-Q\vec{n}_i )~ 
\bar{\psi}_+ (\vec{p} + Q \vec{n}_i )   \psi_- (\vec{p} - Q \vec{n}_i) 
\notag \\
& \qquad\qquad\qquad
+
M (\vec{p} - Q \vec{n}_i ) ~
\bar{\psi}_- (\vec{p} - Q \vec{n}_i )   \psi_+ (\vec{p} + Q \vec{n}_i ) 
~\big) \notag \\
&\; =
- \int \rmd x_0 \! \int_{\delta p}  
\big(\,
 M' (\delta\vec{p} )~ 
\bar{\psi}'_+ ( \delta \vec{p} )   \psi'_- ( \delta \vec{p} )
+
 M' (\delta\vec{p} )~ 
\bar{\psi}'_- ( \delta \vec{p} )   \psi'_+ ( \delta \vec{p} ) 
~\big) \,.
\end{align}
In the last line of the above equation, we have replaced the loop
momentum $\vp$ with $\delta \vp$ and have defined
$M'(\delta \vec{p}) \equiv M(\delta \vec{p} - Q \vec{n}_i)$. 

Here it is very important to notice that although our inhomogeneous
condensate breaks the translational invariance, momenta measured from
the Fermi surface, $\delta \vec{p}$, is a conserved quantity.  This is
true as far as we include only dominant terms.  The conservation of
$\delta \vec{p}$ is violated by corrections such as transverse kinetic
terms or subdominant parts of four-Fermi interactions,  but they are
subleading effects suppressed by powers of $1/Q$.

At leading order, thanks to the conservation of $\delta \vec{p}$, the
eigenvalue problem is diagonal in momentum space.  Then we can
formally derive the mean-field spectrum of quasi-particles using the
mass function $M'(\delta\vec{p})$.  The eigenvalue problem for the
longitudinal plus mass terms is\footnote{
When we write $2\times 2$ matrix expressions,
that means that each element is proportional to
the $2\times 2$ identitiy matrix, ${\bf 1}_{2\times 2}$.}
\begin{equation}
\Psi'^\dag
\begin{pmatrix}
\delta p_\para + Q ~&~ M'(\delta \vec{p})  \\
M' (\delta \vec{p})  ~&~ - \delta p_\para + Q
\end{pmatrix}
\Psi'(\delta \vec{p} )
= E_{{\rm MF}}  (\delta \vec{p} )~ 
\big( \Psi'^\dag \Psi'( \delta \vec{p} ) \big) \,,
\end{equation}
where a 
notation,
$\Psi' (\delta p_\para) 
= \big(\psi'_{+}(\delta \vec{p}),
 \psi'_{-}(\delta \vec{p}  ) \big)^T$,
is introduced.
The eigenvalue has upper and lower branches,
\begin{equation}
E_{ {\rm MF} } (\delta \vec{p})
= Q \pm \omega(\delta \vec{p} )\,,
\qquad \big(~ \omega(\delta \vec{p} ) 
= \sqrt{ \delta p_\para^2 + M' (\delta \vec{p} )^2  } ~\big) \,,
\end{equation}
where the gap opens at $|p_\para|=Q$, and its influence survives up to
distance $\sim \lc$ from the Fermi surface.  (For the case that the
chiral spiral has a wave vector $Q'\neq Q$, see
Appendix~\ref{vqneq2Q}.)
Since the energy level for particles with a mass gap is pushed
down as compared to free particles,
the energy gain inside of one patch is
$\sim M \times \text{(phase space)} = M \times \lc Q \tan \Theta$.
Also, note that the phase space does not change before and after the
formation of the mass gap, so that the Fermi volume conservation is
automatically satisfied.

It is important to specify the relation between these upper and lower
branches and the respective momentum regions.  They can be summarized
as
\begin{equation}
E_{\nearrow}
= Q + \omega(\delta p_\para)\quad (p_\para >Q)\,, \qquad 
E_{\swarrow} 
= Q - \omega(\delta p_\para)\quad (p_\para <Q)\,,
\end{equation}
for states moving to the $+$ direction, and
\begin{equation}
E_{\nwarrow} 
= Q + \omega(\delta p_\para)\quad  (p_\para < -Q) \,, \qquad 
E_{\searrow} 
= Q - \omega(\delta p_\para)\quad  (p_\para > -Q) \,,
\end{equation}
for states moving to the $-$ direction.  See
Fig.~\ref{fig:singleparticleE} for a graphical summary.

In Appendix~\ref{vqneq2Q}, we repeat calculations for
$\vq \neq 2Q\vec{n}$, and explain that at sufficiently high density,
the choice $\vq = 2Q\vec{n}$ is the best way to minimize the
single-particle contributions with the fixed particle number
constraint.
%
\begin{figure}[tb]
\vspace{0.0cm}
\begin{center}
\scalebox{0.6}[0.6] {
  \includegraphics[scale=.30]{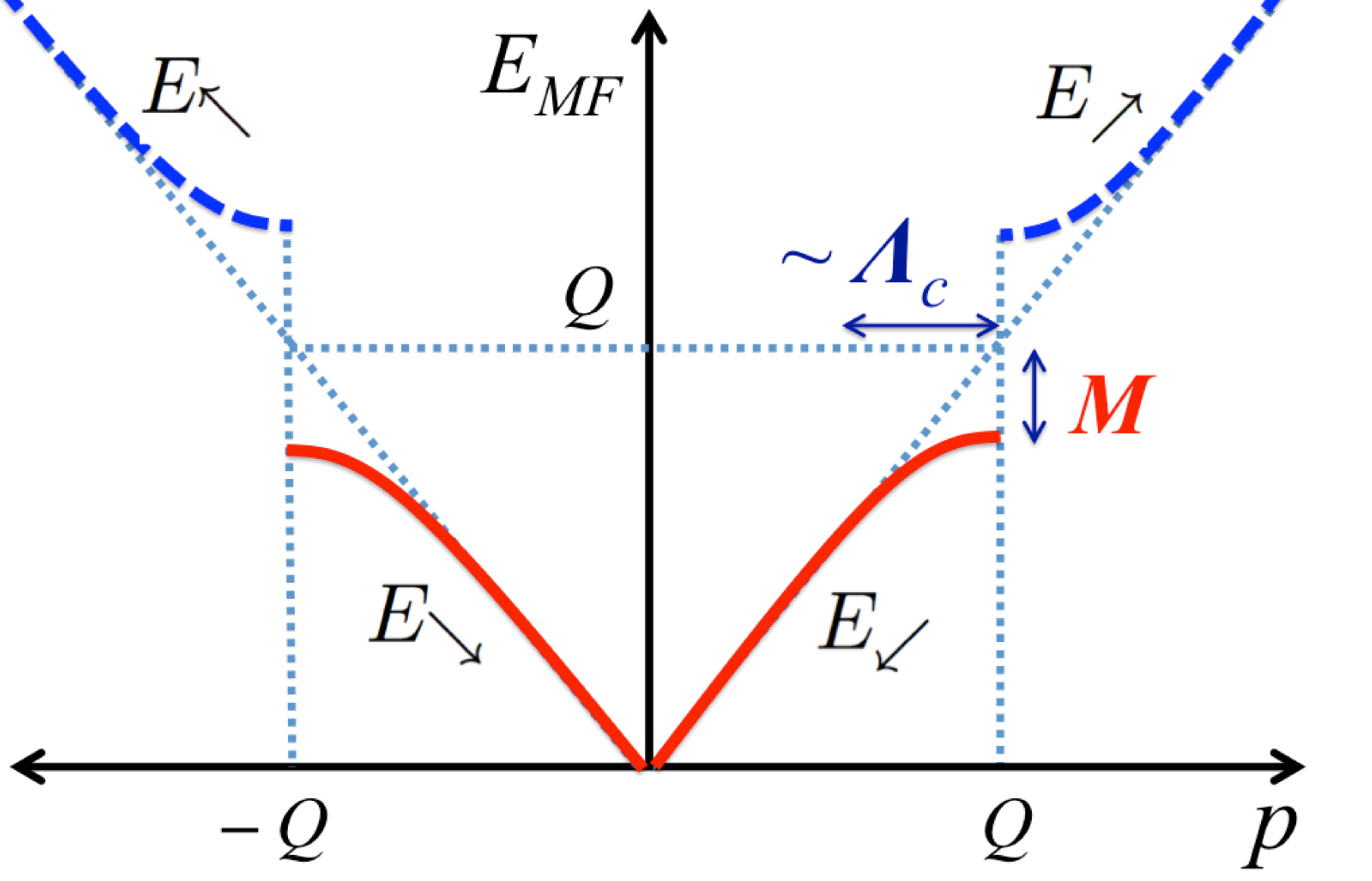} }
\hspace{0.3cm}
\scalebox{0.6}[0.6] {
  \includegraphics[scale=.30]{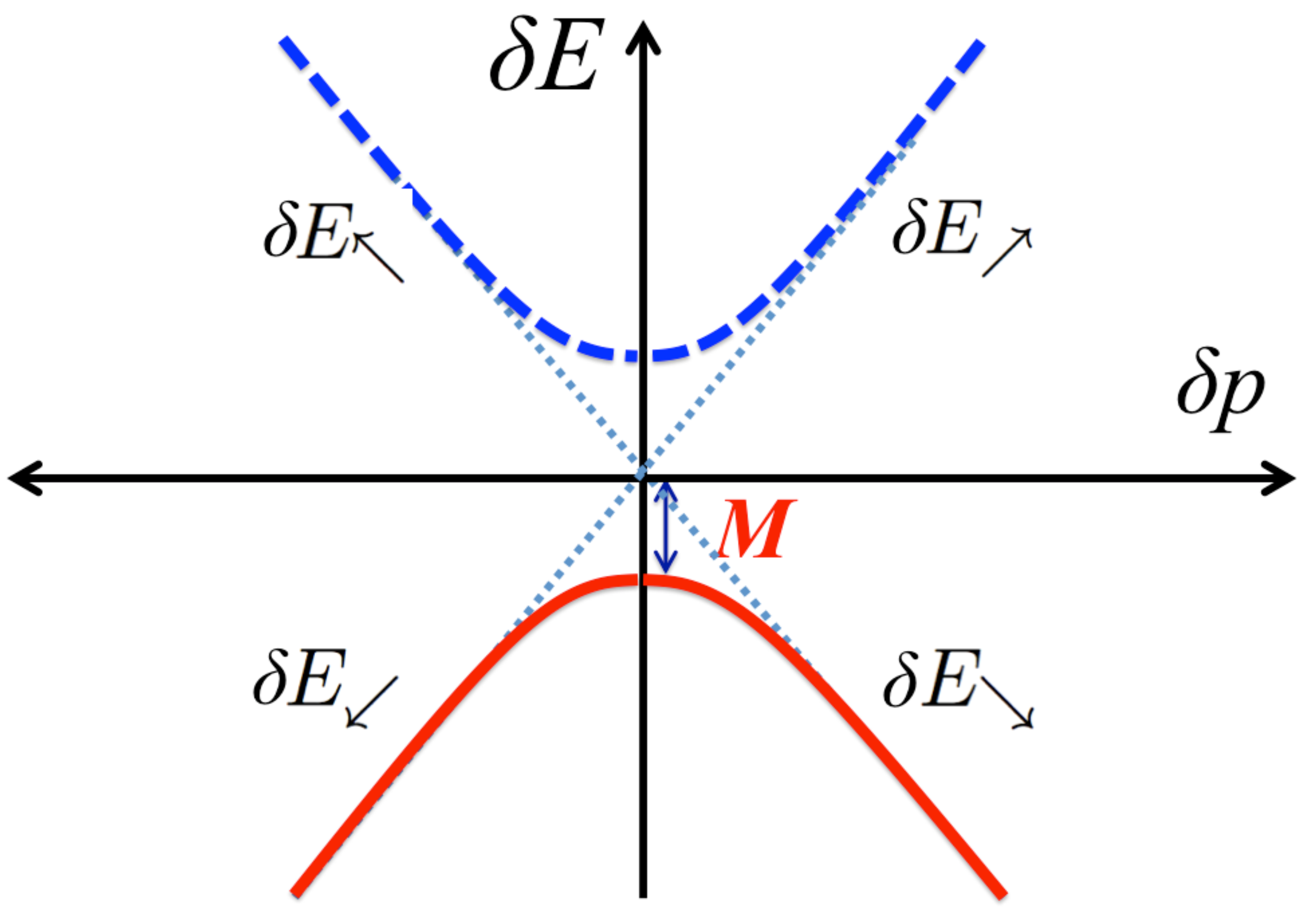} }
\end{center}
\vspace{0.2cm}
\caption{
The mean-field single-particle dispersion in the presence of the
chiral spirals.  The particle orbit is occupied up to
$|p_\para| \le Q$.  The Fermi volume conservation is satisfied before
and after the formation of chiral spirals.\ \
(Left)
The energy $E_{ {\rm MF} }$ in the $\psi$-representation.  The gap
opens at the edge of the Fermi surface.\ \
(Right)
The energy $\delta E = E_{{\rm MF}} - Q$ in the $\psi'$-representation.
}
\label{fig:singleparticleE}
\vspace{0.2cm}
\end{figure}
%

\subsection{The Mean-Field Propagator in the Canonical Ensemble}
Let us construct the mean-field propagator for the quasi-particle.  We
are working in the canonical ensemble with a fixed shape of the Fermi
sea, i.e.\ one wedge, filled with quasi-particles up to
$|p_\para| \le Q$.   This can be restated as the condition that all
occupied states must take the lower energy branch of the mean-field
spectra, $E=Q-\omega$.  To take into account such information in the
propagator, we have only to take a proper $\rmi \epsilon$
prescription as we usually do in the vacuum case.

The dominant fermion bilinear part of the action (longitudinal plus
mass terms) is
\begin{equation}
\hspace{-1em}
{\cal S}_{\psi'}^{\para+M} \!=\! 
\int \!\rmd p_0 \int_{\delta p} \bar{\Psi}' (p_0, \delta \vec{p}) ~\big(\,
(p_0-Q) \gamma^0 + \delta p_\para \gamma^\para
- M' (\delta \vec{p}) 
\,\big) ~\Psi'(p_0, \delta \vec{p}) \,.
\end{equation}
Note that $Q$ couples $\gamma^0$, not to $\gamma^\para$, as a
consequence that we measure momenta of $\psi_\pm$ in an opposite way.
Defining $\delta p_0 = p_0 -Q$, our mean-field propagator in the
region of condensation is,
\begin{equation}
{\cal S}_{{\rm MF}}(\delta p) 
= \rmi ~\frac{ \delta p_0 \gamma^0 +\delta p_\para \gamma^\para 
 + M'( \delta \vec{p}) }
{~(\delta p_0)^2 - (\delta p_\para)^2  - M'^2 (\delta \vec{p} )
 + \rmi \epsilon~} 
\cdot
\theta_\perp (\delta \vec{p} ) \,,
\end{equation}
and out of the condensation region, it is just a free quark
propagator.  Here $\rmi\epsilon$ is introduced as in the vacuum case
in order to classify the propagations of
particles in the upper and the lower energy branches.
The last function $\theta_\perp$ takes into account
the phase space restriction of one wedge,
\begin{align}
\theta_\perp(\delta \vec{p} )
~&\equiv~
\theta \big( |Q + \delta p_\para| \tan\Theta - |p_\perp| \big) ~ 
\theta \big( |Q - \delta p_\para| \tan\Theta - | p_\perp | \big)  
\notag \\
~&\simeq~
\theta \big( Q \tan \Theta - |p_\perp| \big) 
\qquad
(~ {\rm for}~ Q\gg
|\delta p_\para| ~) \,.
\end{align}
This means that both quasi-particles and holes must be within one
wedge to create chiral spirals.  For the most part, we will assume the
approximate expression in the second line, provided that $Q$ is
sufficiently large.

In the following, we will frequently use the component expression
which is defined as
\begin{equation}
{\cal S}_{{\rm MF}} (\delta p)
= 
\begin{pmatrix}
~S_{++} ~ &~ S_{+-} ~  \\
~S_{-+} ~ & ~ S_{--} ~ 
\end{pmatrix}
= 
\begin{pmatrix}
\displaystyle
\frac{1+\Gamma_5}{2}~{\cal S}_{{\rm MF}} ~\frac{1-\Gamma_5}{2}  ~
&~~
\displaystyle
\frac{1+\Gamma_5}{2}~{\cal S}_{{\rm MF}}~ \frac{1+\Gamma_5}{2}  ~  \\[1.6ex]
\displaystyle
\frac{1-\Gamma_5}{2}~{\cal S}_{{\rm MF}} ~\frac{1-\Gamma_5}{2}  ~
&~~
\displaystyle \frac{1-\Gamma_5}{2}~{\cal S}_{{\rm MF}} ~\frac{1+\Gamma_5}{2}  ~
\end{pmatrix}\,.
\end{equation}
Some useful component expression is given also in
Appendix~\ref{apppro}.

\subsection{The Gap Equation and Its (1+1) Dimensional Character}

Since we already have the formal expression of the propagator, we can now
write the gap equation explicitly.  The equation is (with the trace
for the Dirac indices\footnote{
Here we remind readers that 
we are using four component spinors,
so the trace gives a factor $4$.}),
\begin{align}
\hspace{-2ex}
M'(\delta \vec{p}) 
&=  2 G \int \frac{\rmd \delta k_0}{2\pi} 
 \int_{\delta k} ~\theta_{\delta p,\delta k}~
\tr \bigg[
\frac{ \rmi M'( \delta \vec{k}) }
{~(\delta k_0)^2 - (\delta k_\para)^2 - M'^2 (\delta \vec{k}) + 
 \rmi\epsilon~} \bigg]
\,
\theta_\perp (\delta \vec{k} ) \notag \\
&= 
4 G \int \frac{\rmd \delta k_\para\, \rmd \delta k_\perp}{(2\pi)^2} 
\frac{M'(\delta \vec{k}) }
 {~\sqrt{ \delta k_\para^2 +  M'^2(\delta \vec{k})~}~} 
~\theta_{\delta p,\delta k} ~\theta_\perp(\delta \vec{k}) \,.
\end{align}
When we treat the self-consistent equation, we have to distinguish two
situations depending on the fermion momenta.  One case is that
$p_\perp$ resides sufficiently far from the patch boundary.  The other
case is that $p_\perp$ is so close to the boundary that its mass gap
is affected by other patches.

In this section, we focus only on the former case without the boundary
effects and postpone the discussion on the latter case to the 
later sections, which actually goes beyond the one-patch problem.

For $\vec{p}_\perp$ sufficiently far from the patch boundary, the
restriction $\theta_\perp (\vec{ \delta k} )$ is automatically
satisfied by the other restriction $\theta_{\delta p, \delta k}$, so the former does
not play any essential role.  Then the gap equation has
(1+1)-dimensional solutions.  Indeed, we can find the mass gap
function independent of $p_\perp$,
\begin{equation}
M'_0(\delta \vec{p}) = M'_0 (\delta p_\para) \,.
\end{equation}
When we look for such a solution, we can factorize the integral over
$k_\perp$, which gives the (1+1)-dimensional gap equation,
\begin{align}
\hspace{-0.5cm}
M'_0 (\delta p_\para) 
&= 4 G \int \frac{\rmd \delta k_\para}{2\pi}  
\frac{M'_0 (\delta k_\para ) }
 {~\sqrt{ \delta k_\para^2 +  M_0^{\prime~ 2} (\delta k_\para ) }~}
\int \frac{\rmd \delta k_\perp}{2\pi} 
~\theta_{\delta p,\delta k} \notag \\ 
&=
\frac{4 G}{\pi} 
\int_{\delta p_\para - \Lambda_f}^{ \delta p_\para + \Lambda_f}
\frac{\rmd \delta k_\para}{2\pi}  
\frac{M'_0 (\delta k_\para ) }
 {\sqrt{ \delta k_\para^2 +  M_0^{\prime ~2} (\delta k_\para ) }}
\cdot
\sqrt{ \lf^2 - (\delta p_\para - \delta k_\para)^2 }\,,
\end{align}
where the $p_\perp$ dependence disappears from the RHS as it should
be.

Let us note that this factorization could not yield the
$p_\perp$-independent solutions if we did not ignore corrections from
the transverse kinetic terms $\sim p_\perp^2/Q$ or
$\theta_\perp(\delta \vec{k})$.  If one of such terms become relevant,
then the $p_\perp$ dependence of the RHS does not disappear in the
above manipulations.  This means that the (1+1)-dimensional mass
function can be obtained only if $Q$ is sufficiently large and
$p_\perp$ is not too close to the one-patch boundary.

Now let us show that the non-trivial gap arises from the IR effects
near the Fermi surface.  We shall consider the case with
$\delta p_\para =0$.  If we separate the integral region of the RHS
into regions below and above a certain scale $c \lf$ such that
the momentum dependence in $M_0'$ can be ignored below $c \lf$.
Assuming that $c \lf < \lf$, we have
\begin{align}
M'_0 (0) 
&\simeq
\frac{8 G \lf}{\pi} 
\int_{ 0 }^{ c \lf }
\frac{\rmd \delta k_\para}{2\pi}  
\frac{M'_0 (0) }
 {~\sqrt{ \delta k_\para^2 +  M_0^{\prime~ 2}(0) }~}
+ \text{(finite positive terms)}
\notag \\
&= M'_0(0) \cdot
\frac{4 G \lf}{\pi^2}  
\ln \bigg( \frac{ c \lf }{M'_0(0)} \bigg)
+ \text{(finite positive terms)} \,,
\end{align}
where the logarithmic term comes from the (1+1)-dimensional character
of the equation, and expresses the IR effects.

In the same way as the Cooper instability in superconductivity, we can
find the solution of the gap equation from the IR structure of the
equation.  Indeed, if we take $M_0'(0)$ too small,
the logarithmic part is divergingly large and the RHS exceeds the LHS much.
Thus $M'_0(0)$ must be taken substantially large until the IR
contributions are tempered to be the same order of the LHS.

We point out that one oversimplification in the above expression is
related to our approximation to ignore the transverse kinetic terms.
If we recover those kinetic terms, the logarithmic part must be
modified effectively as
\begin{equation}
 \ln \bigg( \frac{ c \lf }{M'_0(0)} \bigg) 
~ \rightarrow  ~\ln \bigg( \frac{ c \lf }{M'_0(0) +
  p_\perp^2/Q} \bigg) \,,
\end{equation}
which tempers the growth in the IR region.  Thus, the gap solution
might not be found unless the density is sufficiently large.

Finally, as a solution of the gap equation, $M_0'$ is parametrically
given as
\begin{equation}
M_0'  \sim \lf ~ \rme^{ -C/G\lf} \,,
\end{equation}
where $C$ is some number.
Within our approximation,
the size of the gap should be at least larger 
than that in vacuum because of 
larger phase space for low energy excitations
which contribute to the formation of the gap.

\section{Perturbation Theory with Chiral Spiral Mean Fields}
\label{formal}

In this section we develop systematic computation of the corrections
from subdominant terms up to the leading order of the $1/\Nc$ expansion.

Using the stationary phase approximation at large $\Nc$, the fermionic
partition function under chiral spiral backgrounds for fixed $\Theta$
is
\begin{align}
Z_\psi [\la \Phi \ra,\Theta]
&= \int {\cal D}\psi' {\cal D}\bar{\psi'} 
~\rme^{\rmi  ({\cal S}_{{\rm MF}}
+ \Delta {\cal S}
)} \notag\\
&=
Z_{{\rm MF}} ~ \big\la  1 + \rmi\,  \Delta {\cal S}  
+ \frac{\rmi^2}{2!}  (\Delta {\cal S})^2 + \cdots
\big\ra_{{\rm MF}}\,, 
\end{align}
where $\la \cdots \ra_{{\rm MF}}$ is the expectation value when we use
the mean-field weight, $\rme^{\rmi {\cal S}_{{\rm MF} } }$, in the
path integral.  The action is made of
\begin{align}
{\cal S}_{{\rm MF}} 
&=
\sum_i \big( {\cal S}_i^{\para + M} + {\cal S}_i^{\Phi} \big), 
\notag \\
\Delta {\cal S} 
&=
\sum_i \big( {\cal S}_i^{\perp} + {\cal S}_i^{\rm sub. int}\big)
+
\sum_{i \neq j} {\cal S}_{i,j} 
+
\sum_{i,j \neq k} {\cal S}_{i,jk}
+
\sum_{i \neq j,k \neq l} {\cal S}_{ij,kl} \,,
\end{align}
where ${\cal S}_{{\rm MF}}$ is the mean-field action which is the
(uncorrelated) sum of one-patch actions.
The ${\cal S}_i^\perp$ and ${\cal S}_i^{\rm sub.int}$ are 
subdominant
terms inside of the $i$-th patch, which  were not treated in the last
section.  Finally ${\cal S}_{i,j},~\cdots$ describe the interactions
among different patches.

The energy density functional is given by
${\cal E} [\la \Phi \ra, \Theta] 
= -\rmi \ln Z / {\cal V}_3
= {\cal E}_{{\rm MF}} + \Delta {\cal E}$
(with ${\cal V}_3$ being the (2+1)-dimensional space-time volume),
where
\begin{equation}
{\cal E}_{{\rm MF}}
= \frac{-\rmi}{ {\cal V}_3 } \ln Z_{{\rm MF}}\,, \qquad
\Delta {\cal E}
= \frac{-\rmi}{ {\cal V}_3 } \cdot 
\big\la \rmi\,\Delta {\cal S} + \frac{\rmi^2}{2!}  (\Delta {\cal S})^2 +
\cdots \big\ra_{{\rm MF}}^{{\rm conn.}} \,.
\end{equation}
We wish to measure the energy benefit from the chiral spiral
formations.  To do so, we need to subtract the energy of the trivial
configuration, and so we should compute,
\begin{equation}
\delta {\cal E}  [\la \Phi \ra, \Theta]
= {\cal E} [\la \Phi \ra, \Theta] - {\cal E}[0, 0] \,.
\end{equation}
When we apply the perturbative expansion soon later, it is convenient
to reorganize the above expression into
\begin{equation}
\delta {\cal E}  [\la \Phi \ra, \Theta]
= \big( {\cal E} [\la \Phi \ra, \Theta] - {\cal E}[0, \Theta] \big)
+ \big( {\cal E} [0, \Theta] - {\cal E}[0, 0] \big) \,.
\end{equation}
The first term expresses genuine condensation effects, while the
second term comes from the deformation energy, which was already
computed at the introduction and turned out to be
$\sim \pF^3 \Theta^4$.  Our task below is the computation of the
quantity inside of the first parentheses.

In the following, we will use the $\psi'$-representation.  The
advantage of doing this is that the momentum $\delta p$ in the
propagator is conserved, and we can express the propagator as a
function of relative distance in space-time, $x-y$.  Noting this fact,
many of diagrams can be easily cast away.  Let us recall that the
subdominant terms have oscillating factors multiplied to the fermion
fields.  Non-zero contributions remain only if the combination of
vertices has the oscillating factors in a form of
$\rme^{\pm \rmi Q(x-y)_\para},\,\rme^{\pm 2\,\rmi
  Q(x-y)_\para},\,\cdots$ 
because propagators are functions of $x-y$.  In other words, this is
simply a consequence of the momentum conservation in our shifted
variables.
 
For example, let us see the first-order expansion, that gives a
tadpole contribution, such as
\begin{equation}
\int \rmd^3 x \Big(
\big\la \bar{\psi}'_{+}  \rmi \Slash{\partial}_\perp
\psi'_{-}\big\ra_{{\rm MF}}~
\rme^{-2\rmi Q x_\para} 
+
\big\la \bar{\psi}'_{-} \rmi \Slash{\partial}_\perp
\psi'_{+}\big\ra_{{\rm MF}}~ 
\rme^{2\rmi Q x_\para} \Big) \,. 
\end{equation}
This is, however, a space-independent quantity times an oscillation
factor, and so its spatial integral vanishes for $Q\neq 0$.  This is
an example of the momentum conservation.  Non-zero correction terms
start to arise from the second order of the perturbative expansion.

In the following, we will first compute the corrections from one patch
which include transverse kinetic terms and subdominant four-Fermi
interactions in one patch.  Second, we will compute interaction terms
including patch-patch interactions.

%
\section{Corrections from One Patch}
\label{secpert}
We consider the perturbative corrections inside of the $i$-th patch.
For the moment we will omit the subscript $i$.  The sources of
second-order corrections are enumerated as follows: 
(1) (transverse terms)$^2$, (2) (four-Fermi interaction terms)$^2$, (3)
(transverse terms) $\times$ (four-Fermi interaction terms).
The last one, (3), cannot cancel oscillating factors out, and only (1)
and (2) contribute to the free energy.

\subsection{Product of Transverse Kinetic Terms}
\begin{figure}[tb]
\begin{center}
\scalebox{1.0}[1.0] {
  \includegraphics[scale=.20]{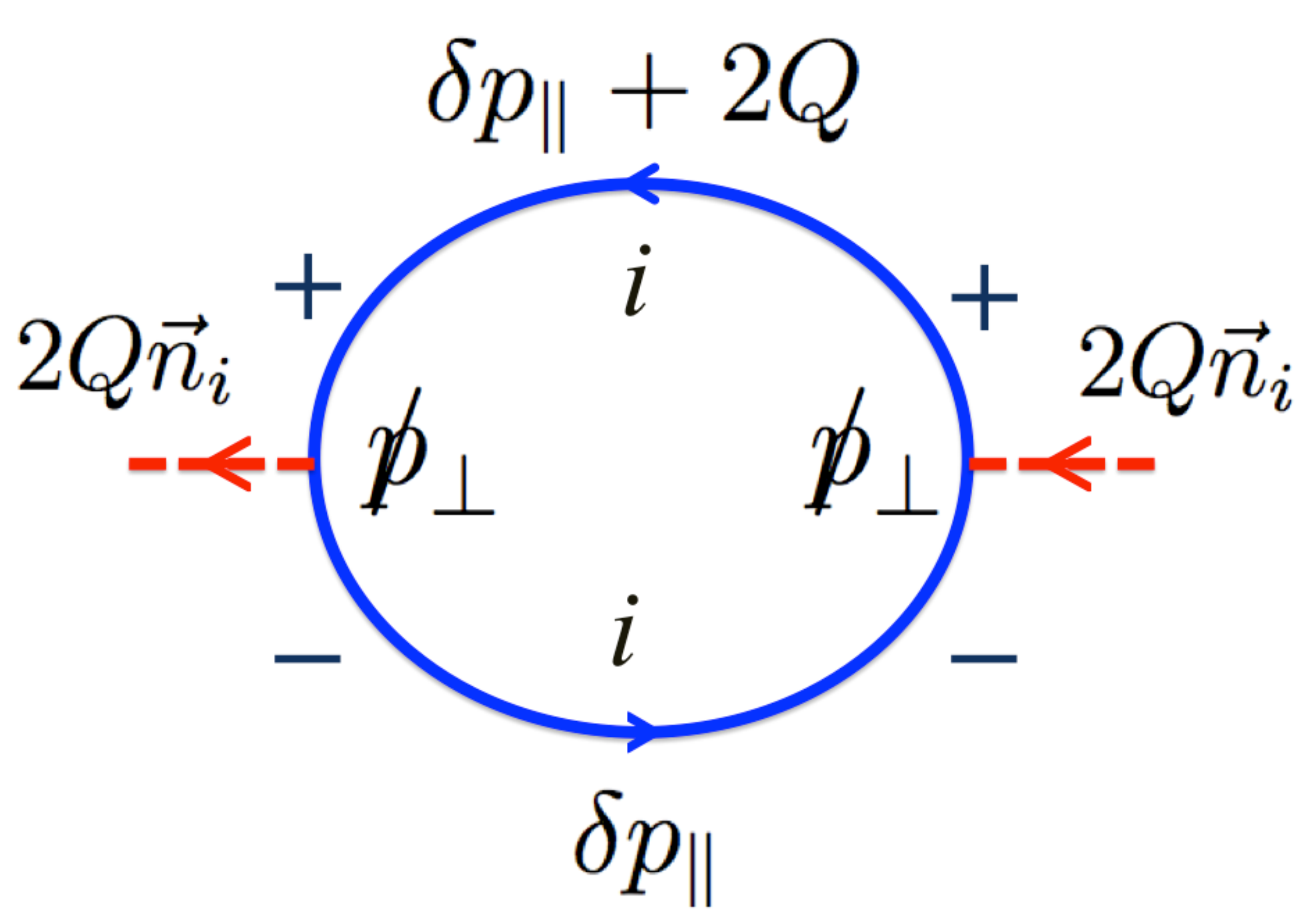} }
\end{center}
\caption{
A diagram for the transverse kinetic terms in the $i$-th patch.  Here
we used the $\psi'$-representation in which $\psi'$ fields effectively
feel incoming and outgoing momenta, $2Q\vec{n}_i$.  This makes at
least one fermion go far away from the Fermi surface.
}
\label{fig:transpert}
\vspace{0.3cm}
\end{figure}

The non-zero contributions from the product of transverse 
terms arise from $(+-)(-+)$ or $(-+)(+-)$ combinations, i.e.
\begin{align}
\hspace{-0.8cm}\Delta {\cal E}_{\rm trans}  &=
\frac{\rmi}{ { \cal V}_3 } 
\int \rmd^3 x\,\rmd^3 y~
\big\la 
\big[ \bar{\psi}'_{+}  \rmi \Slash{\partial}_\perp  \psi'_{-}(x) \big]
\big[ \bar{\psi}'_{-} \rmi \Slash{\partial}_\perp  \psi'_{+}(y) \big]
\big\ra_{{\rm MF}}~
\rme^{-2\rmi Q (x - y)_\para } 
\notag \\
&=
 \Nc 
\int \frac{\rmd^2 \delta p}{(2\pi)^2} ~ p_\perp^2 \,
\int \frac{\rmd\delta p_0}{2\pi} \notag\\
&\qquad\times (-\rmi)\, \tr \big[
~S_{--}(\delta p_0,\delta p_\para ,p_\perp)
~S_{++}(\delta p_0,\delta p_\para +2 Q,p_\perp)
~ \big]  \,.
\label{eqtrans1}
\end{align}

After the $p_0$ integration, we have
\begin{equation}
\Delta {\cal E}_{\rm trans} = 
 \Nc \int \frac{\rmd \delta p_\para}{2\pi} 
\int_{\theta_1(Q)}  \frac{\rmd p_\perp}{2\pi}
~ p_\perp^2 \cdot
\frac{ F( \delta p_\para,2Q) }
 {~\omega( \delta p_\para ) + \omega( \delta p_\para + 2Q) ~}
\,,
\label{eqtrans2}
\end{equation}
where the integral of the transverse momentum is restricted within one
patch,
\begin{equation}
\int_{\theta_1(Q)} \frac{\rmd p_\perp}{2\pi} 
\equiv  
~ \int \frac{\rmd p_\perp}{2\pi} 
~\theta_\perp( \delta p_\para , p_\perp)
~ \theta_\perp( \delta p_\para+2Q, p_\perp) \,,
\end{equation}
and we defined the function $F(\delta p_\para, 2Q)$ as
\begin{equation}
F(\delta p_\para, 2Q) 
= 1 + 
\frac{ \delta p_\para }{~ \omega( \delta p_\para)~}
 \frac{ \delta p_\para +2Q }{~ \omega( \delta p_\para +2Q) ~ }
\qquad
( 0 \le F \le 2) \,.
\label{F}
\end{equation}

Now let us analyze Eq.~(\ref{eqtrans2}).  The key point is that at
least one of the momenta, $\delta p_\para$ or $\delta p_\para +2Q$,
must be much larger than $\lc$.  If both of the momenta are much
bigger than $\lc$, then the mass gap does not exist, so that the
result is just reduced to the one in the free theory.  It is thus a
non-trivial case when one of the momenta is close to the Fermi
surface.  Let us consider such a situation in the following.

Supposing that $|\delta p_\para | < \lc$, the other momentum
$\delta p_\para + 2Q$ satisfies
$2Q-\lc < \delta p_\para +2Q < 2Q +\lc$.  Therefore we can use the
free fermion dispersion for
$\omega(\delta p_\para + 2Q) = \delta p_\para + 2Q$, and can apply an
approximate upper bound, $Q \tan\Theta$, for the $p_\perp$-integral.

In order to make physical interpretations, we should notice that
generically the perturbative expansion contains the trivial
contribution that is independent of the condensate, and this must be
subtracted.  Here we subtract such trivial contributions with the same
phase space as the non-trivial one.  Outside of the condensation
domain, two contributions cancel out.  After subtracting terms at
$M'=0$, the non-trivial contribution of Eq.~(\ref{eqtrans2}) is given
by
\begin{align}
\Delta {\cal E}_{\rm trans} &\simeq~
 \frac{\Nc}{\pi^2} \int_0^{\lc}
 \rmd \delta p_\para 
\int_0^{Q\tan \Theta} \! \rmd p_\perp\,
 \frac{p_\perp^2}{~2Q ~}
\bigg[ \bigg(1 + \frac{\delta p_\para}
{ \sqrt{ \delta p_\para^2 + M'^2(\delta p_\para) } }
\bigg)
- 2\bigg] \notag \\ 
&
\sim ~ 
-\Nc \cdot M'  Q^2 \tan^3 \Theta
\qquad(M'\sim \Lambda_f)\,,
\label{eqtrans3}
\end{align}
where $-2$ in the first line represents the subtraction of the trivial
contribution.  Note that the energy correction before performing the
integration is $\sim p_\perp^2/Q$, and so at the level of the
computation of the fermion dispersion relation, this terms were
negligible effects.

The sign of Eq.~(\ref{eqtrans3}) is negative, which means that this
term corresponds to an energy gain.  This term can be regarded as a
controllable correction only if it is sufficiently smaller than the
energy gain from the one-patch mean field,
$\Nc \cdot \lf^2 Q \tan\Theta $.  This requires the condition,
$\Theta \ll (\lf/Q)^{1/2}$, as already stated at the introduction.

Finally let us mention on how the perturbative analysis of the
correction from the transverse kinetic energy can be organized in a
systematic way.  If we go to higher order of the transverse kinetic
terms, we have terms of higher powers of $p_\perp^2/Q$ in the
integrand.  Computing the $p_\perp$-integration with the integral
region $\sim Q\tan \Theta$ results in terms of higher powers of
$\Theta \ll 1$, which appears as an expansion parameter in the
systematic computations.

\subsection{Product of Four-Fermi Interaction Terms}
\begin{figure}[tb]
\begin{center}
\scalebox{0.5}[0.5] {
  \includegraphics[scale=.35]{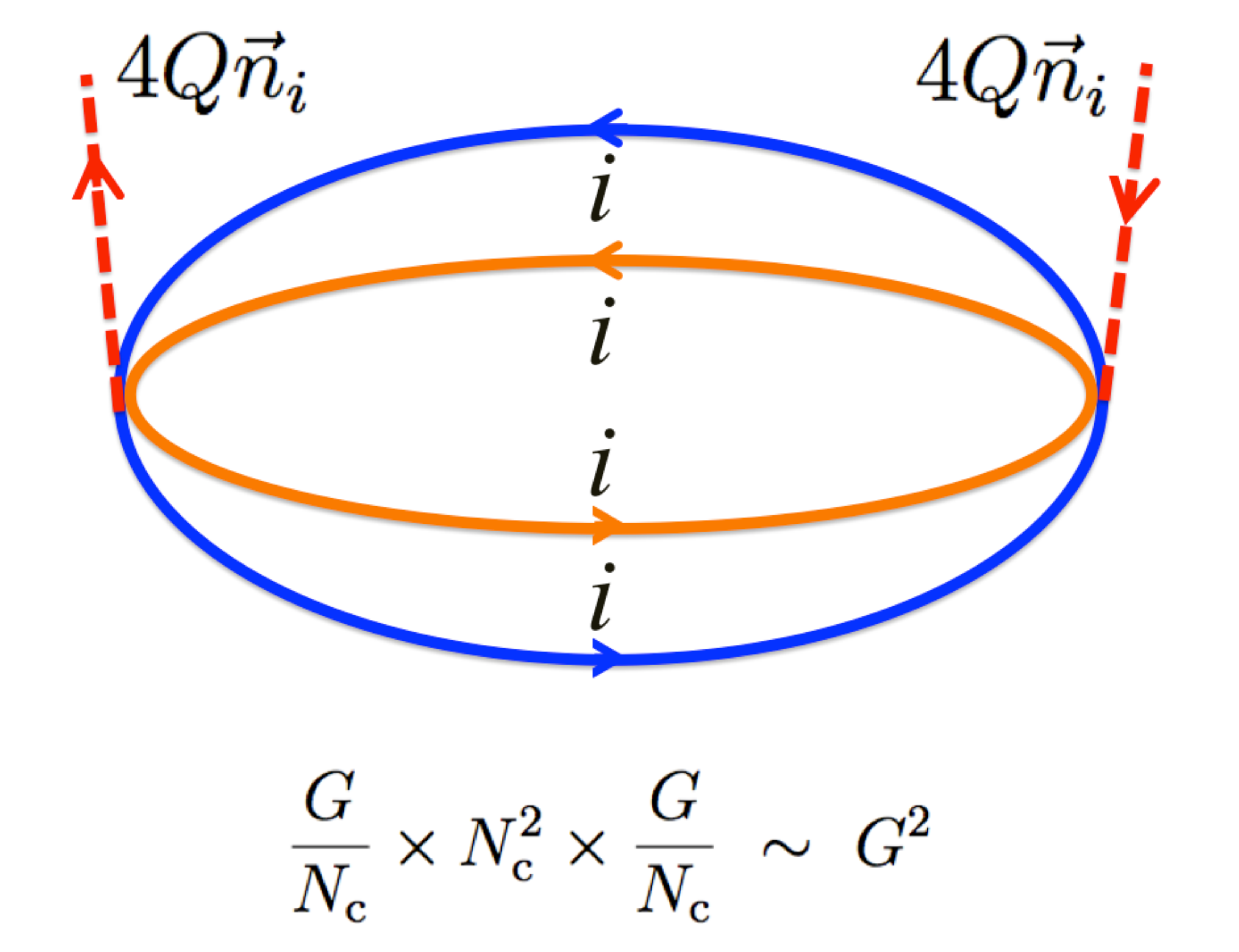}  }
\scalebox{0.5}[0.5] {
  \includegraphics[scale=.35]{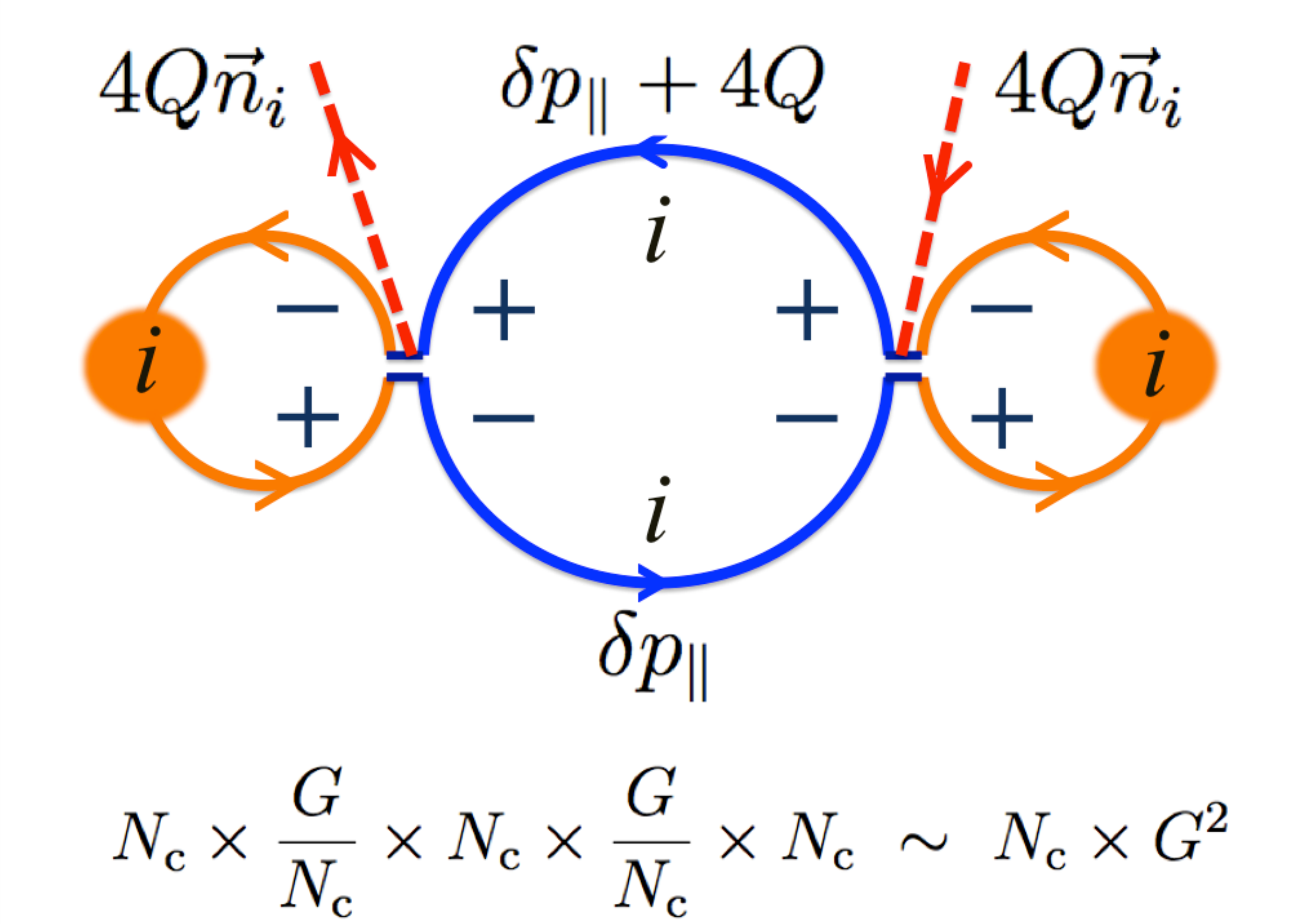}  }
\end{center}
\caption{
Contributions from the subdominant vertices.  The
$\psi'$-representation is used and all fields belong to the $i$-th
patch.\ \
(Left) The $O(\Nc^0)$ contribution which is ignored in this work.\ \
(Right) The $O(\Nc)$ contributions.  The diagram can be interpreted as
the condensate-condensate interaction mediated by particle-hole
excitations.  One of fermions in the internal loop must go far from
the Fermi surface.
}
\label{fermisphere2}
\vspace{0.2cm}
\end{figure}

We will compute the second-order perturbation of subdominant four-Fermi
interaction terms, which reads
\begin{equation}
\frac{\rmi}{ {\cal V}_3 } \cdot \frac{G^2}{\Nc^2}
\int \rmd^3 x\, \rmd^3 y~
\big\la 
 \big(\bar{\psi}'_{+} \psi'_{-}(x) \big)^2
\big( \bar{\psi}'_{-} \psi'_{+}(y) \big)^2
\big\ra_{ {\rm MF} }~
\rme^{-4\rmi Q (x - y)_\para}\,. 
\label{512eq4fermi1}
\end{equation}
Equation~(\ref{512eq4fermi1}) yields several distinct terms depending
on how fermion lines are contracted.  Here we introduce the hierarchy
for different contractions by applying the $1/\Nc$ expansion.  At
large $\Nc$, the dominant contributions come from
condensate-condensate interactions mediated by virtual quark-hole
exchange.  This situation is schematically described as
\begin{equation}
\hspace{-0.6cm} 
\sim ~G^2~
\frac{\la \bar{\psi}'_+ \psi'_- \ra_{ {\rm MF} } }{\Nc}
 ~ \bigg( (-\rmi) \Nc \!\int \rmd^3 z ~
\tr \big[ 
S_{++}(z) S_{--}(-z) 
\big] ~ \rme^{-4\rmi Q z_\para} \bigg) ~
\frac{\la \bar{\psi}'_- \psi'_+ \ra_{ {\rm MF} } }{\Nc}
\label{512eq4fermi2}
\end{equation}
with $z=x-y$ (see the right panel in Fig.~\ref{fermisphere2}).
Here the form factor is not explicitly written yet.  This contribution
is $O(\Nc)$ and positive, according to the previous calculations which
treated the integral in the bracket.  One fermion pair at $x$ and $y$
is contracted at one space-time point and becomes condensates.  The
remaining part is for the particle-hole propagation between $x$ and
$y$.  Loosely speaking, this part can be interpreted as the
propagation of meson-like objects with total momentum $4Q$ (though
ladder-type resummation is necessary to construct the meson
propagation in fact).

Let us note that the size of the condensate of $O(\Nc)$ compensates
for the suppression factor of $O(1/\Nc)$ in the intrinsic interaction
vertex.  Other contractions without the condensate cannot be
accompanied by a fermion loop of $O(\Nc)$ and are suppressed by
$1/\Nc$.

We should not take Eq.~(\ref{512eq4fermi2}) literally, however, since
the loop integral is UV divergent, which is regulated with the form
factor.  It should be mentioned that such apparent UV divergence
couples to the condensate, so the subtraction of the trivial
configuration without the condensate cannot regulate the UV behavior.

Including the form factor explicitly and carrying out the 
integral in Eq.~(\ref{512eq4fermi1}), we have
\begin{align}
& \hspace{-4ex}
\sim \Nc G^2
\int_{\delta p, \delta k, \delta l}
\theta 
\big( 
\lf^2 - ( \delta \vp - \delta \vk + 2Q \vec{n} )^2
\big)
~
\theta 
\big( 
\lf^2 - ( \delta \vp - \delta \vl + 2Q \vec{n} )^2
\big)~ \notag \\
&
\hspace{-3ex}
\times 
\bigg(
\int \!\rmd \delta k_0 
\frac{ \la \bar{\psi}'_+ \psi'_-( \delta k) \ra }{\Nc} 
\bigg)
\frac{ F( \delta p_\para, 4Q) }
 {~\omega( \delta p_\para ) + \omega( \delta p_\para + 4Q) ~}
\bigg(
\int \! \rmd\delta l_0 
\frac{ \la \bar{\psi}'_- \psi'_+ ( \delta l) \ra }{ \Nc}
\bigg) \,,
\label{eqrestriction}
\end{align}
or equivalently,
\begin{equation}
\sim 
\Nc
\int_{\delta p}
M'
\big( \delta \vp + 2Q \vec{n} \big)
~
\frac{ F( \delta p_\para, 4Q) }
 {~\omega( \delta p_\para ) + \omega( \delta p_\para + 4Q) ~}
~
M'
\big( \delta \vp + 2Q \vec{n} \big)\,.
\label{eqrestriction2}
\end{equation}
Here the phase-space restriction for the transverse momentum is not
explicitly written.  The condensate takes a finite value only if
$|\delta k_\para|,~|\delta  l_\para| < \lc$.  Once they are restricted
within such a domain, $\delta \vp~$ is also restricted around
$-2Q\vec{n}$, and so the integral over the phase space has a UV
cutoff.  The integrand itself is suppressed by $1/Q$.  Finally the
energy cost from this contribution can be estimated as
\begin{equation}
\hspace{-0.3cm}
 \text{(one-patch energy cost)}
\sim~ \Nc \, \frac{\Lambda_f^2}{Q} 
\cdot \lf Q \tan \Theta
~\sim~ 
\Nc \lf^3 \tan \Theta \,.
\end{equation}
%
Here, a phase-space factor
$\sim \lf Q\tan \Theta$ arises because one spatial momentum is not
restricted by the $\theta$ function constraint in
Eq.~(\ref{eqrestriction}).  As expected, this contribution is much
smaller than the one-patch mean-field contribution
$\sim \Nc \cdot \lf^2 Q\tan \Theta$, so can be safely ignored at large
density.

%
\section{Contributions from Patch Boundaries: 
Inter-Patch Effects} 
\label{sec:interference}
\begin{figure}[tb]
\vspace{0.2cm}
\begin{center}
\scalebox{1.0}[1.0] {
  \includegraphics[scale=.20]{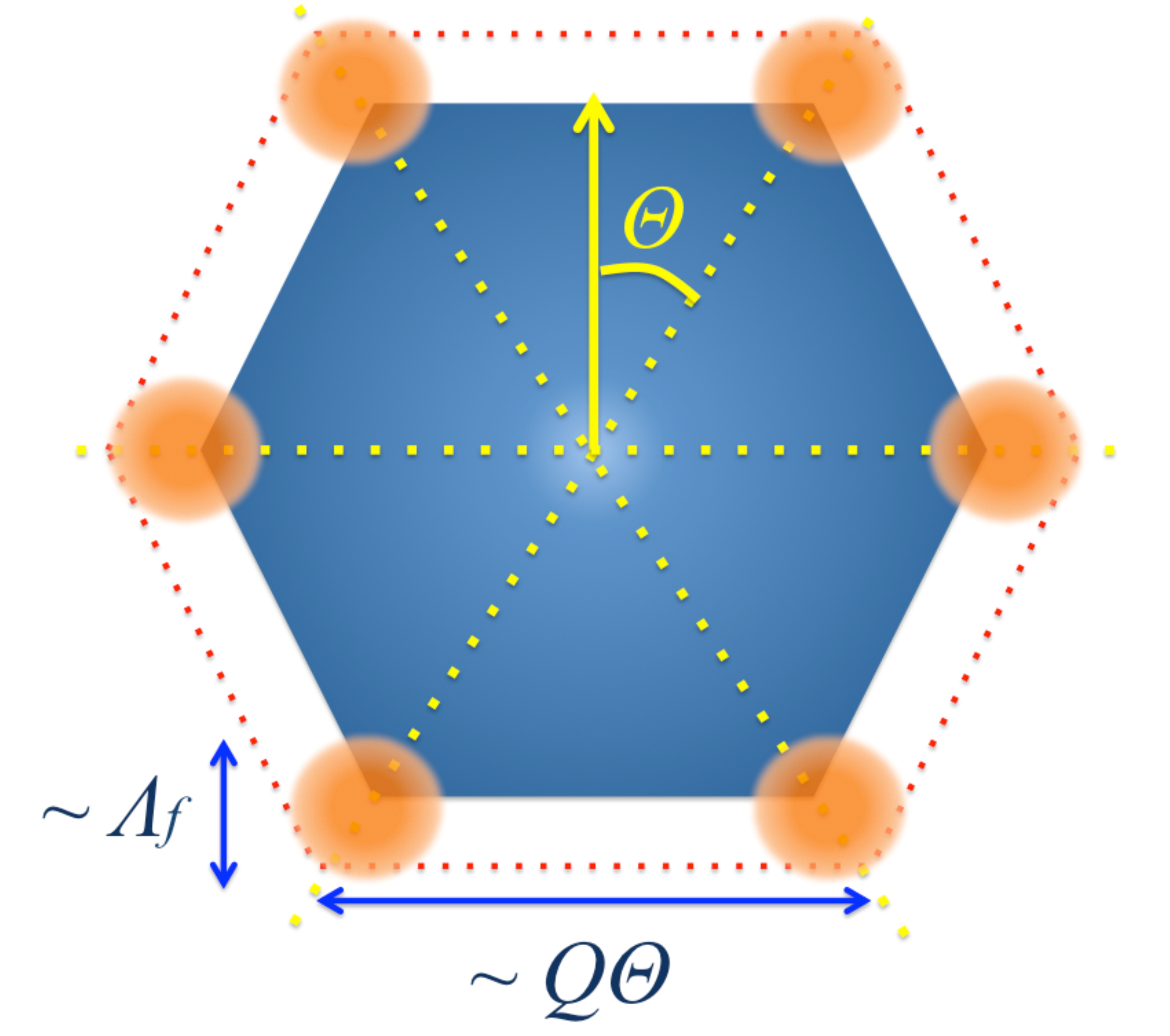} }
\end{center}
\vspace{0.2cm}
\caption{
The domain where the chiral spiral mean field in one patch
couples to particles (holes) in its nearest neighbor patches.
The size of such a domain is 
$\sim \Lambda_f^2$ due to form factor effects.
}
\label{fermisphere3}
\vspace{0.2cm}
\end{figure}

So far, we have ignored the interactions among fermions belonging to
different patches.  Due to the form-factor effects, such interactions
occur only near the boundaries of patches.  We will discuss the impact
of such effects.

The outline for our computational procedure is the following.
The condensation effects generate the following three types of the
energy contributions,
\begin{equation}
\hspace{-0.5cm}
{\cal E}_{{\rm cond.}}
\sim ~ N_p \, 
\big( {\cal E}^{\text{1-patch}}_{\rm inside}(M_0, \Theta)
+ {\cal E}^{\text{1-patch}}_{\rm B} (M_{\rm B},\Theta)
+ {\cal E}_{\rm int}^{\text{patches}} (M_{\rm B},\Theta) 
\big) \,,
\end{equation}
where ${\cal E}^{\text{1-patch}}_{\rm inside}$ and
${\cal E}^{\text{1-patch}}_{\rm B}$ are one-patch condensation energy
in the region far from and close to the boundaries, and
${\cal E}_{\rm int}^{\text{patches}}$ represents the patch-patch
interaction at the boundaries.  The subscript B means the boundary,
and $M_0$ and $M_{\rm B}$ can be considerably different.  The energy
density schematically behaves as
\begin{align}
{\cal E}_{{\rm cond.}}/\Nc
~&\sim~
\frac{1}{\Theta} ~
\big(
-M_0 \lf (Q\tan \Theta - \lf) 
-
M_{\rm B} \lf^2
+
 {\cal E}_{\rm int}^{\text{patches}} (M_{\rm B},\Theta)
\big)
\notag \\
&\sim~ 
- M_0 \lf Q 
~+~ 
\frac{1}{\Theta}
~
\big( \lf^2
( M_0 - M_{\rm B} )
+ 
{\cal E}_{\rm int}^{\text{patches}} ( M_{\rm B},\Theta )
\big) \,. 
\label{eq:conden}
\end{align}
We will show that ${\cal E}_{\rm int}^{\text{patches}}$ is positive
(at least at the level of the second-order perturbation theory).  Its
strength is determined by the size of $M_{\rm B}$, and vanishes when
$M_{\rm B} \rightarrow 0$.

Equation~(\ref{eq:conden}) can be understood in twofold ways.  If we
regard the incoherent sum of the one-patch actions as our unperturbed
action, $M_{\rm B}$ at the unperturbed level is $\sim M_0$, and
${\cal E}_{\rm int}^{\text{patches}}$ provides relatively large,
positive contributions.  This means that the 
inter-patch interactions
between different chiral spirals reduce the energy gain from creating
the condensates at the patch boundaries.  Instead, if we assume
$M_{\rm B}\ll M_0$, then the one-patch contribution is
$M_0 - M_{\rm B} \sim M_0 \sim \lf$, while
${\cal E}_{\rm int}^{\text{patches}}$ is negligible.  In both limiting
cases, the sign of $1/\Theta$ terms is positive and of $O(\lf^3)$.

The precise estimate of the $1/\Theta$ term would be given by
self-consistently solving the gap equation of $M_{\rm B}$ with
${\cal E}_{\rm int}^{\text{patches}}$.  This is beyond our scope in
this paper.  Instead, we will give several indicative discussions to
understand inter-patch interactions
at the patch boundaries, in both
perturbative and non-perturbative manners.

\subsection{Preliminaries}
\begin{figure}[tb]
\vspace{0.2cm}
\begin{center}
\scalebox{0.6}[0.6] {
  \includegraphics[scale=.35]{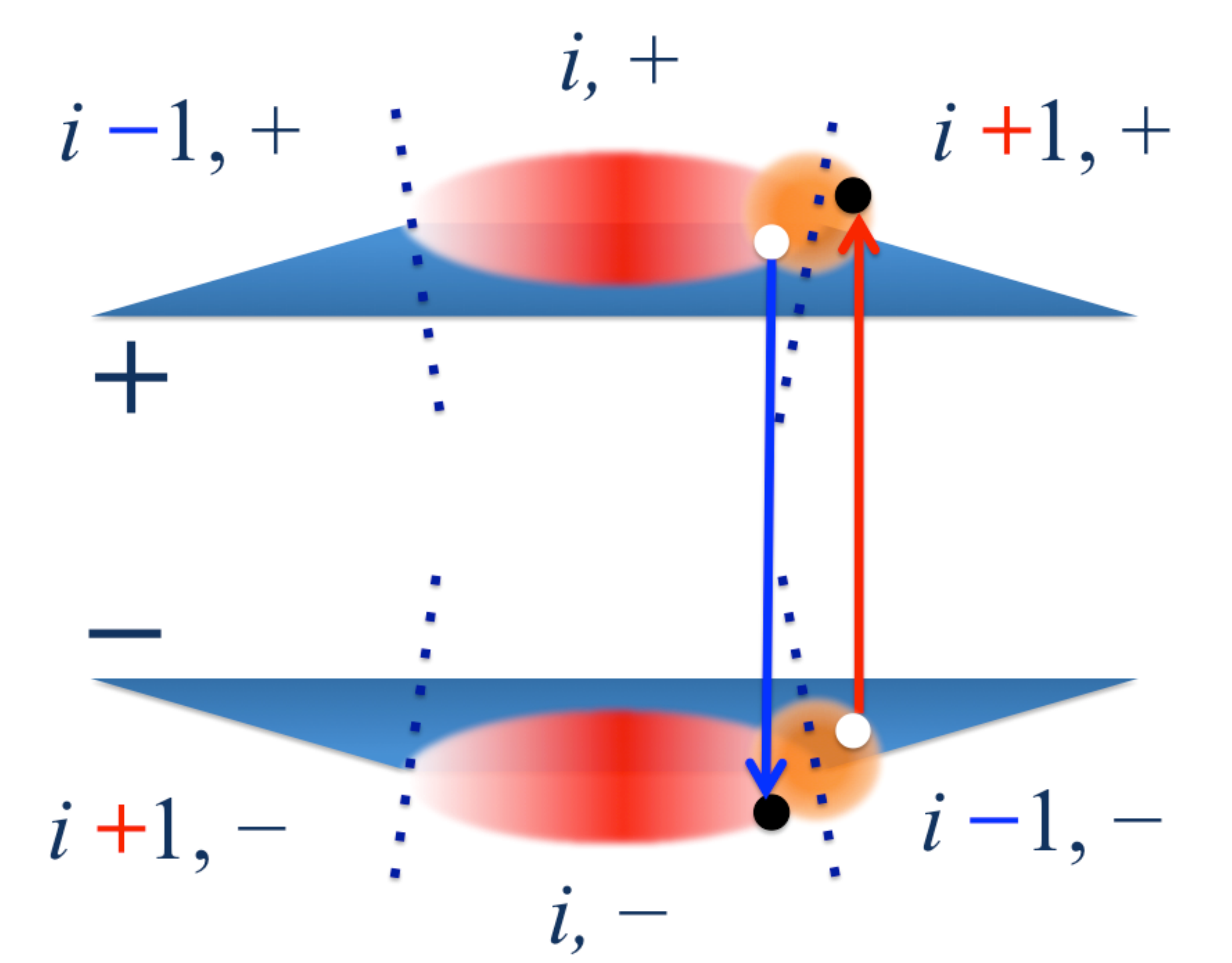} }
\scalebox{0.6}[0.6] {
  \includegraphics[scale=.35]{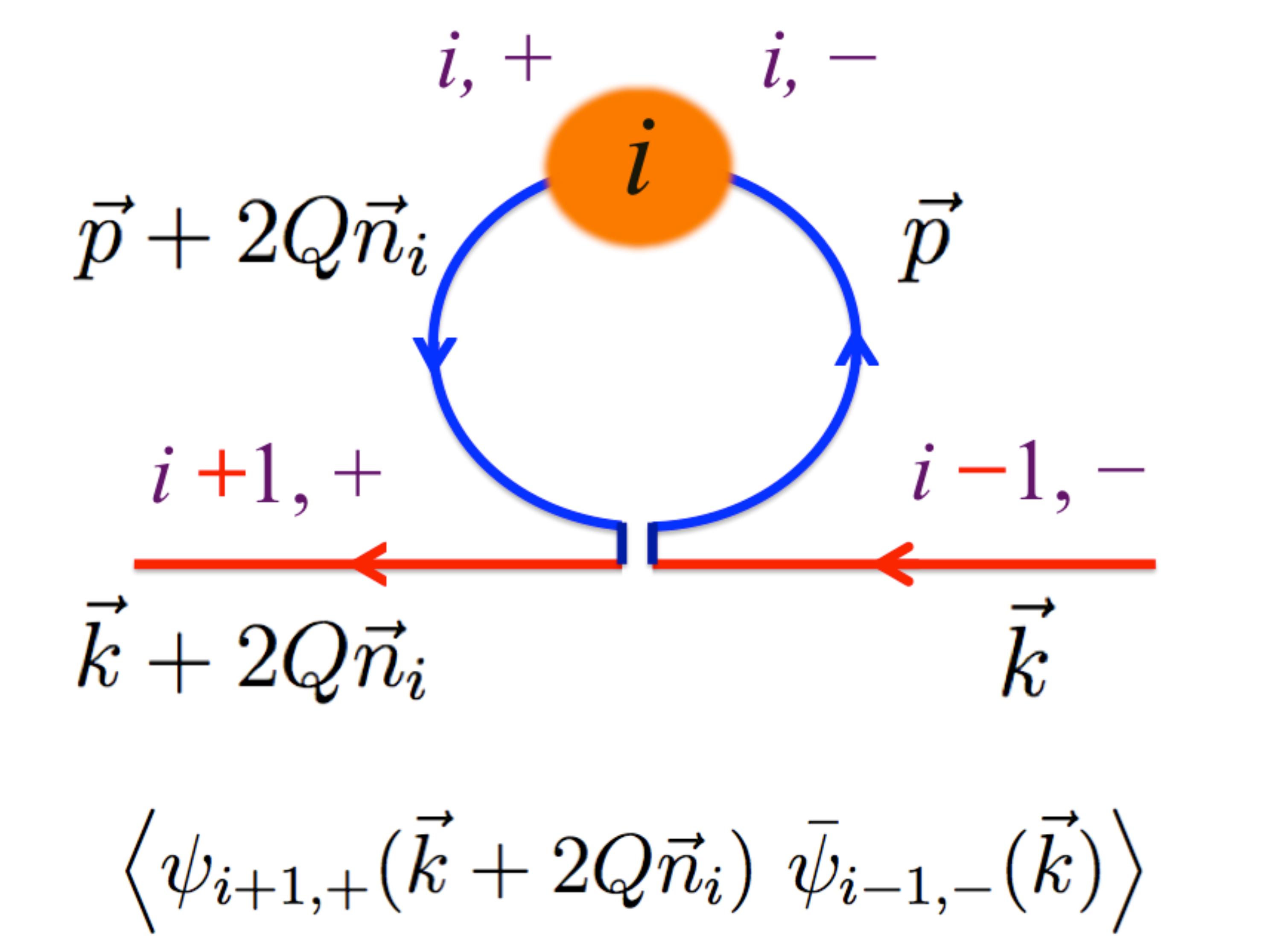} }
\end{center}
\vspace{0.2cm}
\caption{(Left)
An influence of the $i$-th chiral spiral mean-field at the patch
boundaries.  Only the Fermi surface close to $i$-th patch is shown.
The $i$-th chiral spiral can scatter a particle (hole) state in the
$(i+1)$-th patch into a hole (particle) state in the $(i-1)$-th
patch.  Such processes are possible only for a particle and a hole
close to the $i$-th patch boundary.\ \
(Right) A diagrammatic expression of the particle-condensate
scattering.   The vector $\vp$ and $\vk$ must be close each other.
}
\label{fig:patchint23}
\vspace{1.0cm}
\end{figure}

As we have already seen, at large $\Nc$, the dominant correction terms
come from condensate-condensate interactions mediated by virtual
quark-hole excitations.  Therefore we will take into account only such
terms.  Then we need consider only vertices in which two indices of
the patches are identical, such as terms in
${\cal L}^{\rm int}_{i,j}$ or ${\cal L}^{\rm int}_{i,jk}$.

Actually,  because of the form factor, only a particle and a hole in
$(i\pm1)$-th patches can directly couple to the $i$-th chiral-spiral
mean field.  Therefore we have only to consider the following types of
the vertices (see also Fig.~\ref{fig:patchint23}),
\begin{equation}
\int_{p,k} \big\la 
 \bar{\psi}_{i+} (\vp + 2Q\vec{n}_i ) ~ \psi_{i-}
(\vp) \big\ra 
 \,
 \bar{\psi}_{i-1,-} (\vk) ~ \psi_{i+1,+} (\vk + 2Q\vec{n}_i )
~ \theta_{p,k}\,,
\label{patchint1}
\end{equation}
where we have replaced a fermion bilinear in the $i$-th patch with the
chiral spiral mean field.  Using the shifted momentum variables,
$\delta \vp = \vp +Q \vec{n}_i$ and
$\delta \vk=\vk + Q \vec{n}_{i-1}$, and the $\psi'$-representation,
Eq.~(\ref{patchint1}) can be rewritten as
\begin{align}
& 
\int_{\delta p, \delta k} \!
\big\la 
 \bar{\psi}'_{i+} (\delta\vp ) ~ \psi'_{i-}(\delta \vp) 
\big\ra 
\,
\bar{\psi}'_{i-1,-} (\delta \vk) 
~ \psi'_{i+1,+} 
\big(\, \delta \vk + Q(2\vec{n}_i - \vec{n}_{i-1} -
\vec{n}_{i+1} ) \, \big) 
\notag \\
& \qquad\qquad
\times 
\theta \big(\,
\lf^2 
- (\delta \vp - \delta \vk - Q \vec{n}_i + Q \vec{n}_{i-1}   )^2 
\,\big) ~ .
\label{patchint2}
\end{align}
In perturbative computations, it is necessary to decompose momenta of
the fermion field into the longitudinal and transverse components.
Here let us briefly summarize necessary ingredients for the
computations.  Note that
\begin{equation}
\vec{n}_{i\pm1}
 = \cos 2\Theta ~\vec{n}_i \mp \sin 2\Theta ~\vec{n}_{i\perp} \,,
\qquad
\vec{n}_{i\pm1, \perp} 
= \pm \sin 2\Theta ~\vec{n}_i + \cos 2\Theta ~\vec{n}_{i\perp} \,.
\end{equation}
If we write
$\delta \vk = \delta k_{\para} \vec{n}_{i-1} + k_{\perp} \vec{n}_{i-1,\perp}$,
then the momentum in $\psi'_{i+1,+}$ can be decomposed into the
$\vec{n}_{i+1}$ and the $\vec{n}_{i+1,\perp}$ directions,
\begin{align}
&
\delta \vk + Q(2\vec{n}_i - \vec{n}_{i-1} -\vec{n}_{i+1} )
\notag \\
&
= ~
\big[ ~
\delta k_{\para} \cos 4\Theta 
+  k_{\perp} \sin 4\Theta 
+ Q( 2\cos2\Theta - \cos4\Theta -1)~
\big] ~\vec{n}_{i+1} 
\notag \\
&\qquad
+ ~\big[ ~
- \delta k_{\para} \sin 4\Theta 
+ k_{\perp} \cos 4\Theta 
+ Q( - 2\sin2\Theta + \sin 4\Theta )~
\big] ~\vec{n}_{i+1,\perp} 
\notag \\
&
\simeq
~( \delta k_\para + 4 k_\perp \Theta) ~\vec{n}_{i+1}
~+~ k_\perp  ~ \vec{n}_{i+1, \perp} 
~+ ~
O(\delta k_\para \Theta, ~ Q\Theta^2) \,,
\end{align}
where we did not explicitly write quantities which are much smaller
than $\lf$.  Similarly, let us simplify the expression of the argument
in the form factor.  A decomposition of the momentum,
\begin{align}
& \delta \vp - \delta \vk - Q \vec{n}_i + Q \vec{n}_{i-1} 
\notag \\
&= ~
\big[ ~
\delta p_\para 
- \delta k_{\para} \cos 2\Theta 
-  k_{\perp} \sin 2\Theta 
+ Q( \cos2\Theta -1)~
\big] ~\vec{n}_{i} 
\notag \\
&\qquad
+ \big[ ~
 p_\perp
- k_\perp \cos 2\Theta 
+ \delta k_\para \sin 2\Theta 
- Q \sin2\Theta ~
\big] ~\vec{n}_{i+1,\perp} 
\notag \\
&
\simeq ~
(\delta p_\para - \delta k_\para ) ~ \vec{n}_i 
~ + ~ 
( p_\perp - k_\perp - 2Q\Theta) ~ \vec{n}_{i\perp}
~+ ~
O(\delta k_\para \Theta, ~ Q\Theta^2) \,,
\end{align}
leads to
\begin{equation}
\theta 
\big( \,
\lf^2 - (\delta p_\para - \delta k_\para)^2 
- ( p_\perp -  k_\perp -2Q\Theta)^2 
\,\big) \,.
\end{equation}
The $\theta$ function is non-zero only around the patch boundary, that
is, $p_\perp \sim Q\Theta$ and $k_\perp \sim - Q \Theta$.

Finally let us note that the projection operators in one patch are
different from those in other patches.  Indeed,
\begin{align}
\hspace{-2em}
\frac{1+\Gamma_{i+1,5} }{2} 
\cdot\frac{1- \Gamma_{i-1,5} }{2} 
&=
\frac{1+\gamma_0 \gamma_{i+1,\para} }{2}\cdot 
\frac{1- \gamma_0 
 (\gamma_{i+1,\para} \cos2\Theta - \gamma_{i+1\perp} \sin 2\Theta ) }{2} 
\notag \\
&=
\big(\,
(1- \cos2\Theta) + \sin 2\Theta \gamma_0 \gamma_{i+1,\perp} \,\big)
~\frac{1+ \Gamma_{i+1,5} }{4}\,,
\end{align}
and
\begin{equation}
\hspace{-1.5em}
\frac{1\pm\Gamma_{i+1,5} }{2} 
\cdot\frac{1\pm \Gamma_{i-1,5} }{2} 
=
 \frac{1\pm \Gamma_{i+1,5} }{4}~
\big(\,(1+ \cos2\Theta)
\mp
\sin 2\Theta 
~\gamma_0 \gamma_{i+1, \perp} \,\big)\,.
\end{equation}
In what follows, these slight modifications provide only negligible
contributions of $O(\Theta)$, so we need not care them seriously in
the computations.

\subsection{Perturbative Consideration}
The purpose of this subsection is to get the typical size of
patch-patch interactions, and more importantly, to investigate whether
interactions are attractive or repulsive.  The latter can be done even
without detailed estimates of the gap near the patch boundaries.

The perturbative computations proceed in almost exactly the same way
as before.  After taking the residue, we have an expression analogous
to Eq.~(\ref{eqrestriction2}),
\begin{equation}
\Nc
\int_{\delta k}
M'^2_{\rm B}
( \delta k_\para, k_\perp + 2Q \Theta )
~
\frac{ F_{\rm B}( \delta k_\para, 4 k_\perp \Theta) }
 {~\omega_B( \delta k_\para ) + \omega_B( \delta k_\para + 4k_\perp
   \Theta) ~}
~\ge ~0 \,,
\label{eq:patchint}
\end{equation}
where $0\le F_{\rm B} \le 2$ by definition of
Eq.~(\ref{eqrestriction2}), and $k_\perp \sim - Q\Theta$.  The product
of $M'_{\rm B}$ comes from the condensates (i.e.\ fermion loops) in
the $i$-th patch, and the remaining piece comes from virtual
particle-hole excitations.  We emphasize that for the momentum
conservation to be satisfied, both condensates must come from the same
patch.  The subscript B is attached to remind that the sizes of the
gap and mass function near the boundaries may be considerably
different from those far from the boundaries.

The sign is positive, so this is an energy cost.  Although this is
nothing beyond a generic fact inherent to the second-order
perturbation, it indicates that boundary effects tend to temper the
magnitude of the condensates near the boundaries.

It is very important to notice that the energy cost may be comparable
to the energy gain from forming a condensate within the same phase
space $\sim \lf^2$.  In contrast to the case of subdominant terms, we
do not have $1/Q$ suppression because all fermions can move around the
Fermi surface during the virtual processes.

Let us investigate the order of the magnitude.  If we assumed
$M'_{\rm B} \gg Q\Theta^2$, we can make an approximation that
\begin{equation}
\text{(Integrand in Eq.~(\ref{eq:patchint}))}
\sim
 \frac{M_{\rm B}'^2}{ \sqrt{ \delta k_\para^2 + M_{\rm B}'^2 } } 
\bigg(1 
- \frac{\delta k_\para \, k_\perp \Theta}
{~\sqrt{ \delta k_\para^2 +  M_{\rm B}'^2  }~}
+ \cdots \bigg) \,,
\end{equation}
thus the integral of the above integrand in the IR region gives an
approximate expression as 
\begin{equation}
\lf M_{\rm B}'^2
\bigg[ \ln \bigg( \frac{\lf}{M_{\rm B}'} \bigg)
+ 
\frac{2\, Q \Theta^2}{ \sqrt{ \lf^2 +  M_{\rm B}'^2~} } 
\cdots 
\bigg]
~\sim~  M_{\rm B}'^2 ~ 
\bigg(
\frac{1}{G} + O(Q\Theta^2) \bigg) 
~ \sim ~ \lf^3\,,
\end{equation}
where we have used $G \sim \lf^{-1}$ and
the parametric behavior of the mass gap,
$M_{\rm B}' \sim \lf\, \rme^{-C/G\lf}$.  This expression indeed
confirms that the energy cost is of the same order as the energy gain
from condensation effects within the same phase space.

Unfortunately, any obvious expansion parameter did not appear for this
perturbative expansion, so we have no good reason to cast away
higher-order diagrams.  For a more reasonable estimate, the
non-perturbative computations are necessary to simultaneously treat
patch-patch interactions and the mean-field problem near the patch
boundaries.

\subsection{Some Non-perturbative Considerations:
A (1+1)-Dimensional Example of Two Chiral Spirals}
\label{1+1Dexample}
To get some insights for the 
inter-patch interactions
between several
chiral spirals, let us consider the simplest (1+1)-dimensional
example.  We assume the mean field which has two chiral spirals with
wavevectors $Q_0$ and $Q_1$.  We will see how these two chiral spirals
affect each other.

The mean-field eigenvalue equation is
\begin{equation}
\hspace{-2em}
E_{{\rm MF}} ~ \Psi(x_\para)
=
\begin{pmatrix}
 \rmi \partial_\para  
~&~ M_0\, \rme^{2\rmi Q_0 x_\para} 
+ M_1\, \rme^{2\rmi Q_1 x_\para}   \\ 
M_0\, \rme^{-2\rmi Q_0 x_\para} 
+ M_1\, \rme^{-2\rmi Q_1 x_\para}  
~&~ -\rmi \partial_\para
\end{pmatrix}
\Psi(x_\para) \,, \nonumber 
\\
\end{equation}
where 
$\Psi (x_\para) 
= \big(\psi_{+}(x_\para), \psi_{-}(x_\para) \big)^T$
is defined as before.

We rewrite this expression in the $\psi'$-representation in order to
eliminate the oscillating factors.  When we have two chiral spirals,
those oscillating factors cannot be eliminated simultaneously.  If
$M_0 > M_1$, it is better to eliminate 
$\rme^{\pm 2\rmi Q_0 x_\para}$, as
will be clear in the following.  Using
$\psi_\pm = \psi'_\pm\,\rme^{\pm\rmi Q_0 x_\para}$ 
and multiplying the Hamiltonian
squared, we have the following Schr\"{o}dinger equation,
\begin{equation}
(E_{{\rm MF}} - Q_0)^2 ~ \Psi' (x_\para )
=
\begin{pmatrix}
{\cal H}'_{{\rm diag.}}
~&~   2\delta Q M_1\, \rme^{2\rmi \delta Q x_\para}
\\
2\delta Q M_1\, \rme^{-2\rmi \delta Q x_\para}
~&~ 
{\cal H}'_{{\rm diag.}}
\end{pmatrix}
\Psi' (x_\para) \,,
\end{equation}
where $\delta Q= Q_1- Q_0$, and
\begin{equation}
{\cal H}'_{{\rm diag.}}
=
- \partial_\para^2 + (M_0 - M_1)^2 + 4M_0 M_1
\cos^2 \delta Q x_\para  \,.
\end{equation}
The off-diagonal Hamiltonian has the amplitude proportional to
$\delta Q M_1$.  Therefore, if $|\delta Q|$ or $M_1$ is small enough,
one can ignore the off-diagonal part.  Here one can understand that
our choice to eliminate $Q_0$ rather than $Q_1$ is suited for this
approximation since $M_0 > M_1$.

\begin{figure}[tb]
\begin{center}
\hspace{-0.2cm}
\scalebox{0.5}[0.5] {
  \includegraphics[scale=.36]{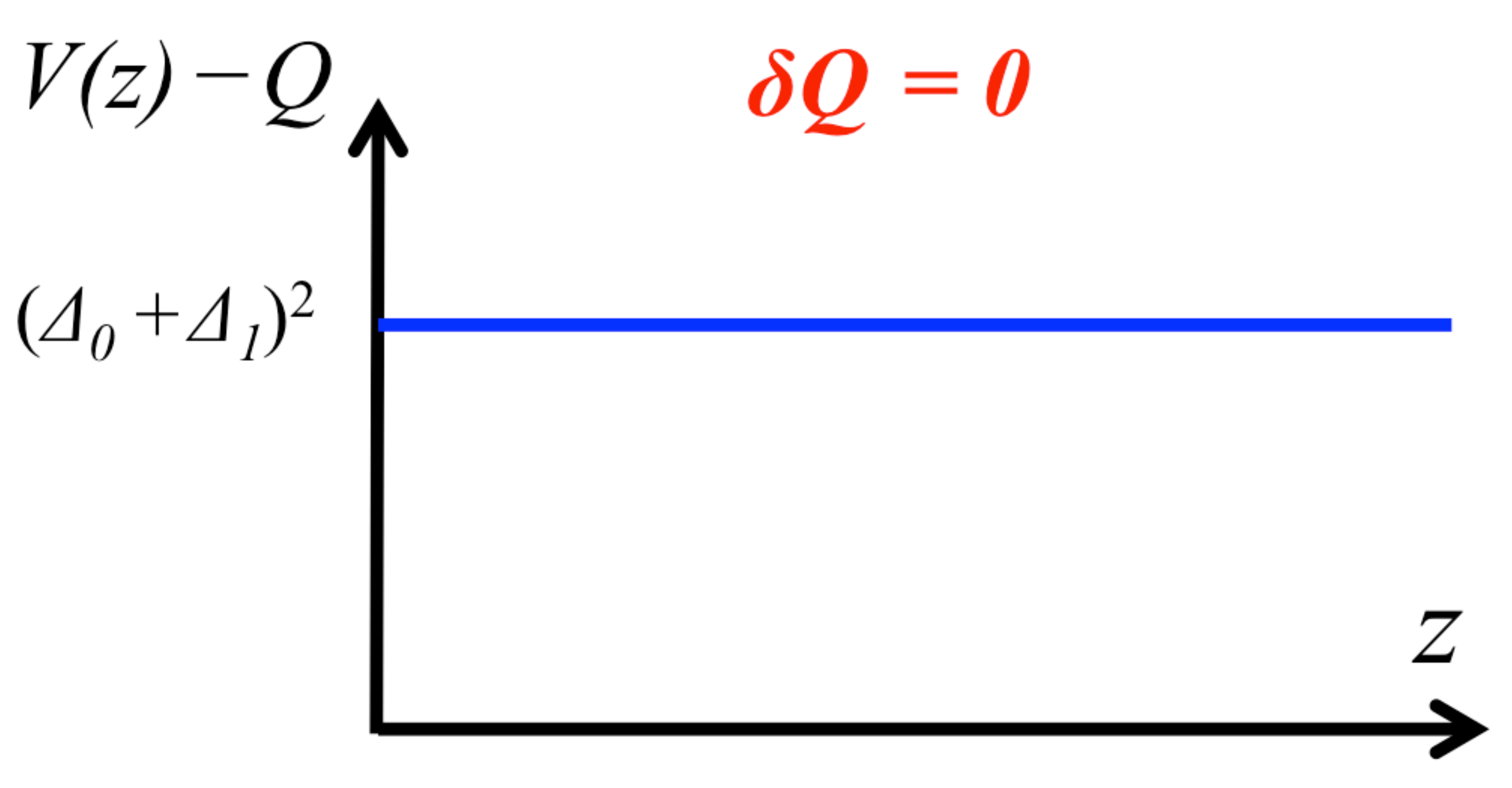} }
\hspace{0.3cm}
\scalebox{0.5}[0.5] {
  \includegraphics[scale=.36]{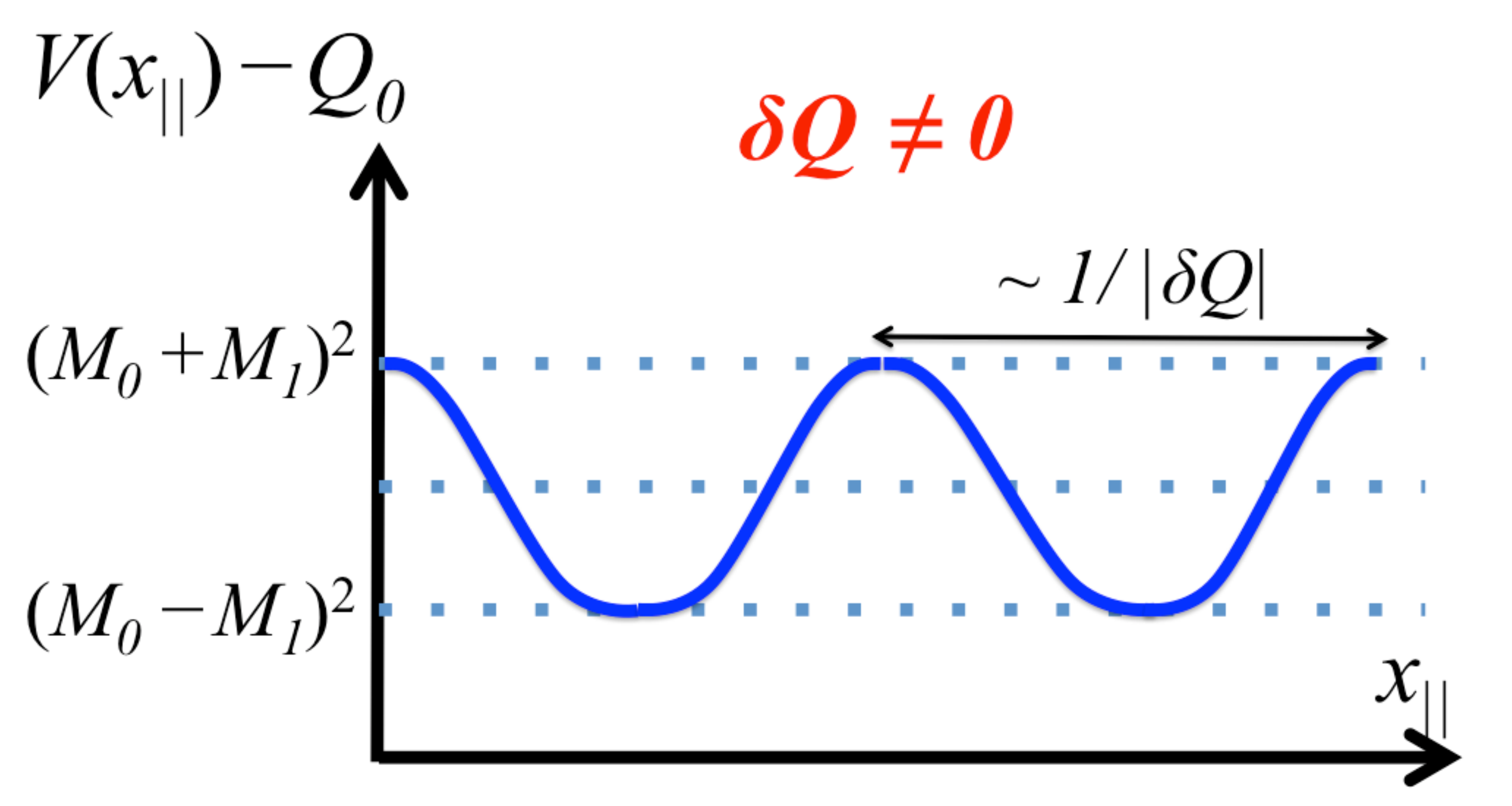} }
\end{center}
\vspace{0.2cm}
\caption{The potential $V(x_\para)-Q_0$.\ \
(Left) $\delta Q=0$ case.  The potential is constant.\ \
(Right) $\delta Q \neq 0$ case.  The potential is oscillating, so
  particles can stay around the valley of the potential.  For very
  small $\delta Q$, its kinetic energy cost is small.
}
\vspace{0.3cm}
\label{fig:pot}
\end{figure}

The diagonal part has a positive oscillating potential whose period is
$1/|\delta Q|$;  see Fig.~\ref{fig:pot}.  A singularity lies at
$\delta Q=0$ which leads to a constant potential, and the energy
spectra are discontinuous from those at $\delta Q\neq 0$.  If
$|\delta Q|$ is small enough but not zero, then we can find the
eigenfunction whose kinetic energy is very small.
That is,
\begin{equation}
E \sim Q_0 \pm |M_0 - M_1| \,,
\end{equation}
%
for $0 < |\delta Q| \ll M_0$.  When $M_0$ is comparable to $M_1$, they
nearly cancel each other, making the effective gap small.  This
analysis indicates that two chiral spirals with different but similar
wave vectors tend to reduce the energy gain in single particle
contributions.

If two chiral spirals have substantially different wavevectors, say,
$|\delta Q| \sim M$, then the valley of the potential becomes narrow,
and the kinetic energy is $\sim M$.  In such a case 
inter-patch interactions
do not strongly reduce the energy gain.  This remark will be
increasingly important when we consider the lower density region where
higher harmonics of chiral spirals start to contribute because of the
subdominant terms.  We will discuss this in the next section.

\section{Discussion}
\label{secdiscussion}
So far we have used the four-Fermi interaction which is strong enough
to induce the chiral symmetry breaking near the Fermi surface.  At the
same time, we have relied on the high-density approximation to make
our discussions simple enough to explore analytic insights.  In
reality with $\Nc=3$, however, gluons will be screened at high
density, reducing the magnitude of our coupling constant.  Perhaps our
approximations may need improvements to be realistic.

Hence we try to extrapolate our insights at high density into the
lower-density domains, by arguing how correction terms grow up.
Essentially interweaving chiral spirals are disturbed by the
transverse dynamics and 
inter-patch interactions, 
and these effects become
increasingly important at lower density.  In this respect, we quote
several elaborated numerical studies in the low-density side, in order
to complement our current studies.  
At the same time, we will make use of
our perturbative corrections to interpret some results in the existing
literatures.

Related to our high density approximation,
it is very important to know
how the $1/\Nc$ corrections grow up at
higher density.  
We will summarize effects which we have ignored by using
the large-$\Nc$ approximation.

Another important question is
how our interweaving chiral spirals look like
in coordinate space.
Up to $\Np=3$, the chiral density
has a periodic translational order and
a orientational order that are classified
by usual crystallography.
Beyond $\Np=3$, however,
the chiral density wave is no longer periodic,
but just shows certain patterns
with orientational symmetry.
We briefly discuss these aspects,
leaving several interesting questions
open for future studies.

The remaining topic is the instanton-induced interaction
\cite{Kobayashi:1970ji}, which is frequently used to introduce strong
diquark correlations \cite{Rapp:1997zu}.  We will see that this
interaction would provide quite different effects below and above the
strange-quark threshold.

\subsection{Comparisons with Other Works in the Low Density Regime}
\label{comparison}

In Ref.~\cite{Rapp:2000zd}, the authors have numerically studied the
chiral crystals in (3+1) dimensions in the density region,
$\mu_{\rm q} = 0.4-0.6\;\text{GeV}$, using the NJL model and the model
with the instanton-induced interactions.  They studied the
scalar-isoscalar channel with the plane-wave oscillations
$\sim \sigma \,\rme^{\rmi\vec{Q} \cdot \vec{r} }$.  In the NJL model,
the strength of the four-Fermi interaction becomes weaker at higher
density because of the model cutoff, so they found only small mass
gaps, $\sim O(10)\;\text{MeV}$.  On the other hand, the
instanton-induced interaction is stronger near the Fermi surface, so
it is possible to have a large mass gap of $O(\lqcd)$.  Importantly,
they showed that the creation of the differently oriented
chiral-density waves (crystals) does not provide much energy benefit,
and the chiral density wave evolved only in one direction is
energetically favored.   This result can be interpreted as a
consequence of 
inter-patch interactions, 
and is consistent with our
current analysis.

In the case of the chiral density wave in one particular direction,
there can be a better solution than the simple plane wave.  Recently,
such a solution in the (3+1)-dimensional NJL model was found by the
authors of Ref.~\cite{Nickel:2009ke} who have used trial functionals
motivated from (1+1)-dimensional studies \cite{Schon:2000qy}.  The
following results are deeply relevant to ours:
(i) The solution appears to be of a solitonic type at low density, and
approaches the plane-wave type at high density.
(ii) A quark number modulation also occurs at low density.  It
smoothly approaches the uniform distribution as the density
increases.
(iii) The chiral spirals with $\sigma$ and $\vec{\pi}$, which is
naively expected, is not energetically favored as compared to single
modulation of $\sigma$, as far as we keep $\mu_u=\mu_d$.

These statements are all consistent with our analyses, and in fact
could be inferred from our framework as follows.  Let us explain this
in order.

(i) The deviation from the plane-wave solution is caused by the
subdominant terms in our formulation,
\begin{equation}
\bar{\psi}_+' \Slash{\partial_\perp} \psi_-' \rme^{-2\rmi Qx_\para}
\,, 
\quad
(\bar{\psi}'_+ \psi'_- )^2 \rme^{ -4\rmi Q x_\para} \,,\quad 
\cdots
\end{equation}
which provide higher harmonics necessary to construct solitonic
solutions at low density\footnote{
In (1+1) dimensions at $T=0$, one can validate this discussion by
investigating the Gross-Neveu (GN) model \cite{Gross:1974jv} with or
without the continuous chiral symmetry.  The former is free from the
subdominant terms, and the chiral spirals can appear at arbitrary low
density and the quark density is always uniform.  (At nonzero $T$, we
have the twisted kink crystal in which the amplitude field also
modulates \cite{Basar:2008im}.)  On the other hand, in the version
with discrete chiral symmetry, solitonic objects first appear at
density beyond some critical chemical potential which is slightly
lower than the constituent quark mass.  As density increases,
subdominant terms stop to disturb the chiral rotation, then quark
distributions smoothly approach those with the chiral spirals and
uniform quark density.  See also Sec.~6 in Ref.~\cite{Kojo:2011fh}.
}.
As we discussed, these terms become unimportant as the density
increases, recovering the plane-wave solutions.

(ii) In the computations of the expectation value of the quark number, the
non-perturbative mean-field propagator gives uniform distribution,
while the perturbations from subdominant terms can generate the
spatial modulation
(for more explanations,
see Sec.\ref{coordinate}).  
The distribution approaches the uniform one as
the density increases\footnote{
Explicit calculations will be reported elsewhere.
}.

(iii) It is quite straightforward to extend our one-flavor studies to
the multi-flavor ones in terms of $\Phi = (u,d,\cdots)^T$, and we can
easily infer that the chiral spirals should emerge as a rotation in
the $U(1)$ quark number sector,
\begin{equation}
\la \bar{\Phi}_+ \Phi_- \ra = \Delta\, \rme^{-2\,\rmi Qx_\para} \,, 
\qquad
\la \bar{\Phi}_- \Phi_+ \ra = \Delta\, \rme^{2\,\rmi Qx_\para} \,, 
\end{equation}
which are equivalent with the following combination,
\begin{equation}
\la \bar{\Phi} \Phi \ra = 2\Delta \cos 2Qx_\para \,, 
\qquad
\la \bar{\Phi} i \gamma_0 \gamma_\para \Phi \ra 
= 2\Delta \sin 2Qx_\para \,. 
\end{equation}
The expectation value of the latter was not calculated in
Ref.~\cite{Nickel:2009ke}, though.  Here we have not found any
particular mechanism to generate flavor rotations, at least in the
high-density limit.  Of course, once we had explicit flavor breaking
coming from conditions such as the charge neutrality and
$\beta$-equilibrium\footnote{
These conditions are driving mechanism to destabilize a homogeneous
color-superconducting phases \cite{Huang:2004bg}
 into the crystalline states
\cite{Alford:2000ze,Casalbuoni:2002pa,Kiriyama:2006ui,Gorbar:2005tx}.
}, 
they are likely to generate other chiral rotations as well.  It would
be interesting to consider astrophysical consequences of such
inhomogeneous distributions.  (See discussions on the implication to the
glitch problem in Ref.~\cite{Alford:2000ze}, for example.)

Assembling these works and insights in this paper, let us infer what
kind of calculations are desirable at low density.  While the chiral
crystals were not favored in the plane-wave Ansatz in
Ref.~\cite{Rapp:2000zd}, we know that higher harmonics become
increasingly important at lower density, as shown in
Ref.~\cite{Nickel:2009ke} and our perturbative calculations.  They are
relevant to describe a localized quark number density as well.  An
interesting question is whether crystals including higher harmonics
are energetically more favored as compared to one-dimensional
solitonic configuration in Ref.~\cite{Nickel:2009ke}.  As discussed at
the end of the previous section, 
inter-patch interactions
are strong for
the chiral spirals with close wavevectors.  That argument, however,
also implies that two chiral spirals with very different wavevectors
do not strongly destroy one another.  Thus, for configurations with
higher harmonics, deconstruction due to 
inter-patch interactions
might be tempered, so that the solitonic crystal structure might be
energetically favored.

\subsection{On the $1/\Nc$ Corrections}

\begin{figure}[tb]
\vspace{0.0cm}
\begin{center}
\scalebox{0.3}[0.3] {
\hspace{-0.6cm}
  \includegraphics[scale=.60]{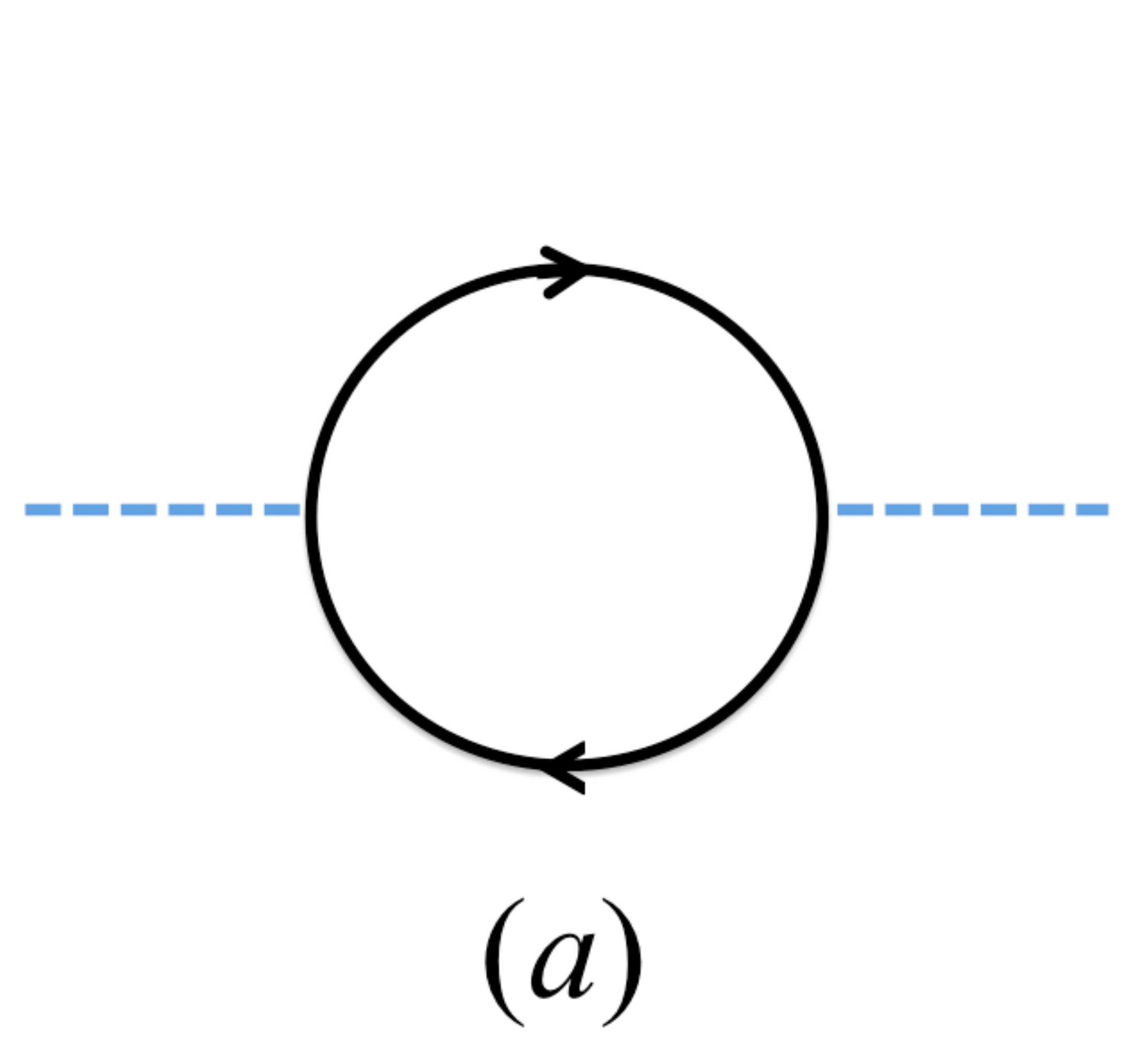} }
\scalebox{0.3}[0.3] {
\hspace{1.5cm}
  \includegraphics[scale=.60]{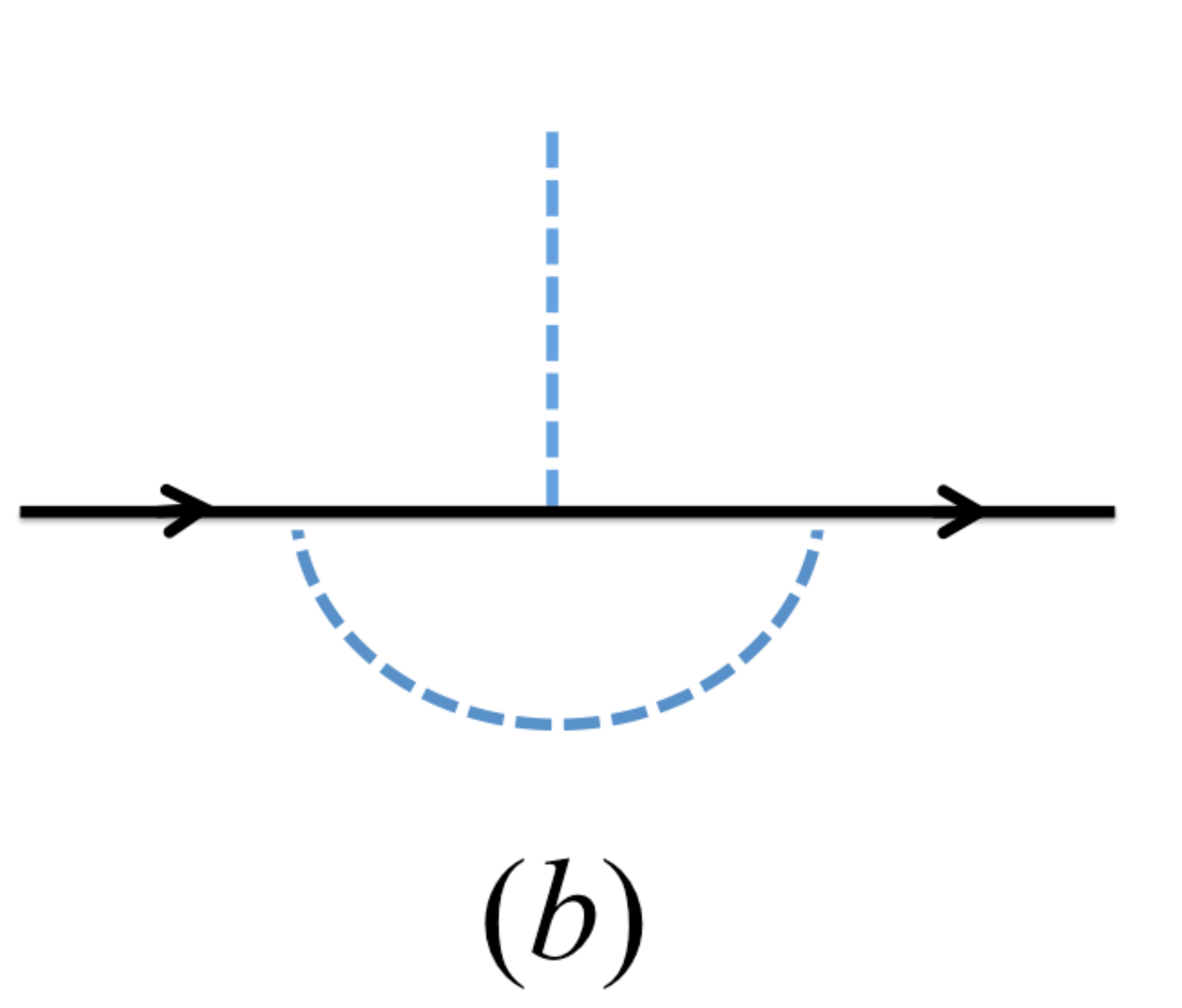} }
\scalebox{0.3}[0.3] {
\hspace{1.5cm}
  \includegraphics[scale=.60]{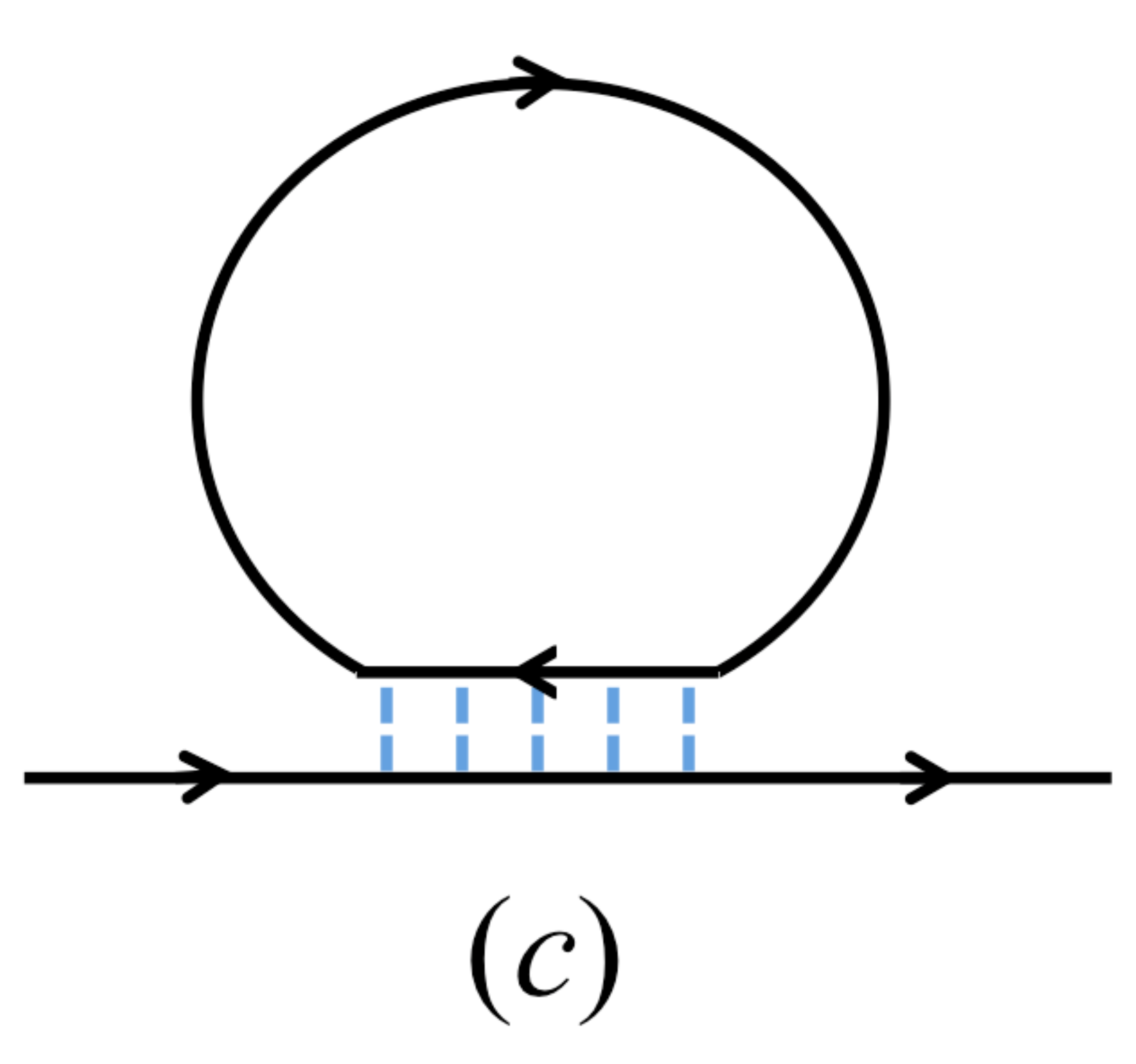} }
\end{center}
\vspace{0.0cm}
\caption{
Some diagrammatic examples of the $1/\Nc$ corrections for:
(a)
the gluon propagator,
(b) the quark-gluon vertex,
(c) our resummation scheme
shown in the left panel in Fig.~\ref{fig:SDeq}.
Modifications by three- and four-gluon vertices should exist as well.
}
\label{fig:1nc}
\end{figure}

As the density increases, the $1/\Nc$ corrections grow up because the
increasing phase space around the Fermi surface enhances low-energy
quark fluctuations\footnote{
Needless to say, when we try to include the $1/\Nc$ corrections, we
have to restart all of the computations from the vacuum problem, in
order to renormalize the theory including fermion loops.  Here we are
arguing the medium-induced modification after the correct
renormalization is made.
}.
The enhancement is parametrically $\sim (\mu_{\rm q}/\lqcd)^{d-1}$
where $d$ represents the number of spatial dimensions.  Such effects
are illustrated in the typical diagrams in Fig.~\ref{fig:1nc}.  The
diagrams (a) and (b) reduce the effective size of our coupling
constant as the density increases.  The diagram (c) modifies our
resummation scheme or the mean-field treatment shown in
Fig.~\ref{fig:SDeq}.  In the terminology of the four-Fermi
interaction, we have treated the Hartree term while ignoring the Fock
term in the large-$\Nc$ limit, which will be modified.

Especially let us note that in the diagram (c), all momenta of the
loop, the incoming quark, and the outgoing quark need not to be close,
in contrast to the leading-$\Nc$ contributions.  Therefore once the
$1/\Nc$ contributions to the condensate become comparable to the
leading order, they will violate the locality of the quark-condensate
interactions in momentum space.  It means that 
inter-patch interactions
among chiral spirals occur not only near the patch boundaries but
everywhere near the Fermi surface.  The situation becomes much more
complicated than that we have treated in this paper.

The quantitative estimation of the $1/\Nc$ corrections is a highly
non-linear problem.  It should be quite sensitive to the effective
masses of quarks, while the mass is determined by non-perturbative
interactions whose strength is reduced by the screening effects by
quarks.  In this respect, it does matter whether we took into account
possibilities of the inhomogeneous condensates or not.  Under the
assumption of homogeneous chiral symmetry breaking, quarks lose the
mass gaps originated from quark-antiquark condensation, and then
fluctuations would rapidly grow up at finite density.  However, if we
included a possibility of chiral symmetry breaking near the Fermi
surface, the quarks acquire the mass gap through the constituent quark
mass, suppressing the fluctuations near the Fermi surface.
Calculations including the latter have not been done yet.

The above issue is also related to the validity of the stationary
phase approximation.  Here we have to distinguish two kinds of the
fluctuations.  One is the amplitude fluctuation, largely related to
the size of the quark mass gap.  Another is the phase fluctuation,
which is massless.

For the latter, one might suspect that, according to
Mermin-Wagner-Coleman's arguments on 
lower dimensions \cite{Mermin:1966fe}, 
the IR
fluctuations of the phases would be strong enough to destroy chiral
spirals with quasi (1+1)-dimensional structure, and therefore quarks
would lose their mass gaps.

A few remarks are in order.  First, the absence of the spontaneous
symmetry breaking in real (1+1) dimensions is extremely sensitive to
the {\it deep} IR region.  Our phase fluctuations cannot access such a
region since the Fermi surface always has finite curvature effects
which provide the IR cutoff\footnote{
One might think that this cutoff argument would be inconsistent with
our (1+1)-dimensional reduction of the gap equations.  This is not so.
The point is that the mass gap is much less sensitive to the
{\it deep} infrared region --- even if we omit such small phase space,
we can find the solution, like the NJL model in vacuum.
}
$\sim p_\perp^2/\pF$.  So even without using the large-$\Nc$ limit, we
conclude that the chiral spirals have the long-range order.
This issue was already addressed in Ref.~\cite{Kojo:2010fe}.

Second, even if the system has only quasi long-range order, the mass
gap is not washed out.  The absence of the condensate does not mean
the presence of a gapless quark, as far as the amplitude field takes a
finite expectation value.  For a (1+1)-dimensional example, see
\cite{Witten:1978qu}.  So a primary question concerned with the
stationary phase approximation is whether the fields strongly
fluctuate or not.

Besides the fluctuation effects, we should also mention on the
possibility of the diquark condensate at $\Nc=3$, which could not be
addressed in the large-$\Nc$ limit.  The Meissner effect would change
our non-perturbative forces.  This issue is beyond our scope, and can
be addressed only by taking the energy competition between the
interweaving chiral spirals and the color-superconducting phases.

\subsection{A Coordinate Space Structure of the Interweaving
Chiral Spirals}
\label{coordinate}

So far we have not discussed coordinate 
space structures of the interweaving chiral spirals.
This is because like BCS theory,
the energy minimization at high density
is turned out to be 
sensitive only to the momentum space structures,
at least in our model.
This situation is very different from
determination of crystal structures in
atomic physics \cite{solid}
where coordinate space
descriptions are useful for the energy minimization.

Nevertheless, it is certainly interesting
to illustrate how various densities in 
the interweaving chiral spirals
look like in coordinate space.
In practice, the coordinate space considerations
have potential relevance for the considerations of
the density domain close to baryonic matter,
in which coordinate space descriptions of quarks
are more appropriate than momentum space ones.

For a quark number density,
the distribution is just uniform
at the leading
order of the high density expansion.
This is because
the quark number density made 
from different patch contributions,
\begin{equation}
\bar{\psi} \gamma_0 \psi 
= \sum_{i=1}^{\Np} 
\left( \bar{\psi}_{i+} \gamma_0 \psi_{i+}
+ \bar{\psi}_{i-} \gamma_0 \psi_{i-} \right)~,
\end{equation}
has no mixture of $(+,-)$ fields.
The spatially modulating contributions
start to appear only after inclusion
of perturbative corrections from subdominant terms
such as $\bar{\psi}_+\,  \rmi\, \Slash{\partial}_\perp  \psi_-$,
thus are suppressed at high density.

On the other hand,
a distribution of the chiral density is
nontrivial.
After summing up contributions from different patches,
we have
\begin{equation}
\left\la \bar{\psi} \psi (x) \right\ra
= \sum_{i=1}^{\Np}
 \left\la \bar{\psi}_{i+} \psi_{i-} (x) 
+ \bar{\psi}_{i-} \psi_{i+} (x) \right\ra
\sim \Delta
\sum_{i=1}^{\Np}
{\rm Re} \left( \rme^{2 \rmi Q\vec{n}_i \cdot (\vx - \vx_i) } 
\right) ~,
\end{equation}
where we did not write the form factor dependence
for notational simplicity,
and we took into account $\vx_i$ 
to make the amplitude field $\Delta$ real.

\begin{figure}[tb]
\vspace{0.0cm}
\begin{center}
\scalebox{1.0}[1.0] {
\hspace{0.2cm}
  \includegraphics[scale=.30]{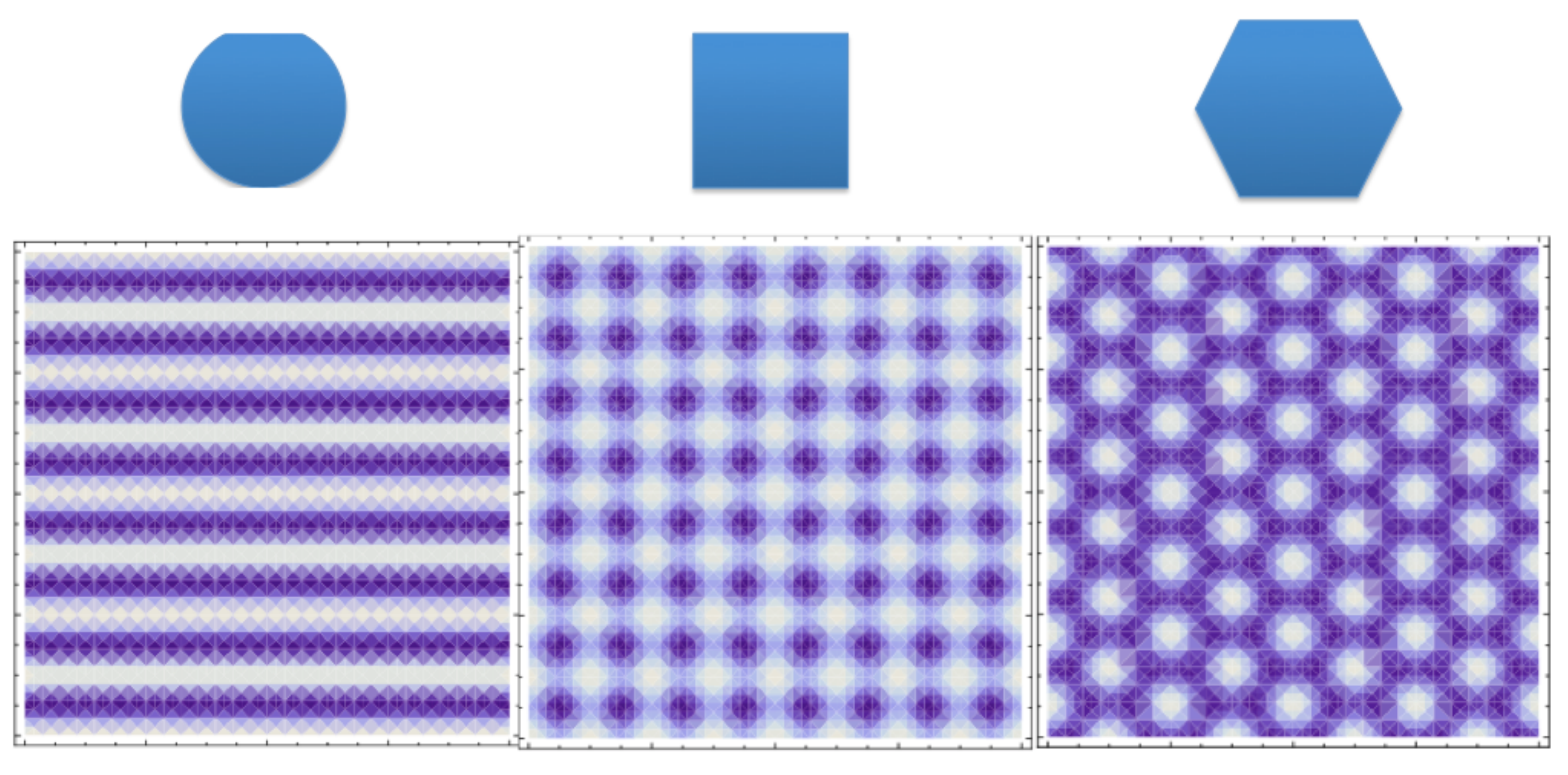} }
\end{center}
\caption{The crystal structures of the chiral density
for $\Np=1,2,3$. The corresponding
shapes of the Fermi sea are also shown.
(For simplicity, we chose $\vx_i=\vec{0}$.)
}
\label{fig:Crystals1}
\vspace{0.2cm}
\end{figure}

Up to a patch number, $\Np=3$,
the chiral density has a periodic structure 
and an orientational symmetry 
known in the conventional crystallography.
Shown in Fig.~\ref{fig:Crystals1} are
the crystal structures of chiral density, 
``chiral density crystals'', for $\Np=1,2,3$.

\begin{figure}[tb]
\vspace{0.0cm}
\begin{center}
\scalebox{1.0}[1.0] {
\hspace{0.2cm}
  \includegraphics[scale=.30]{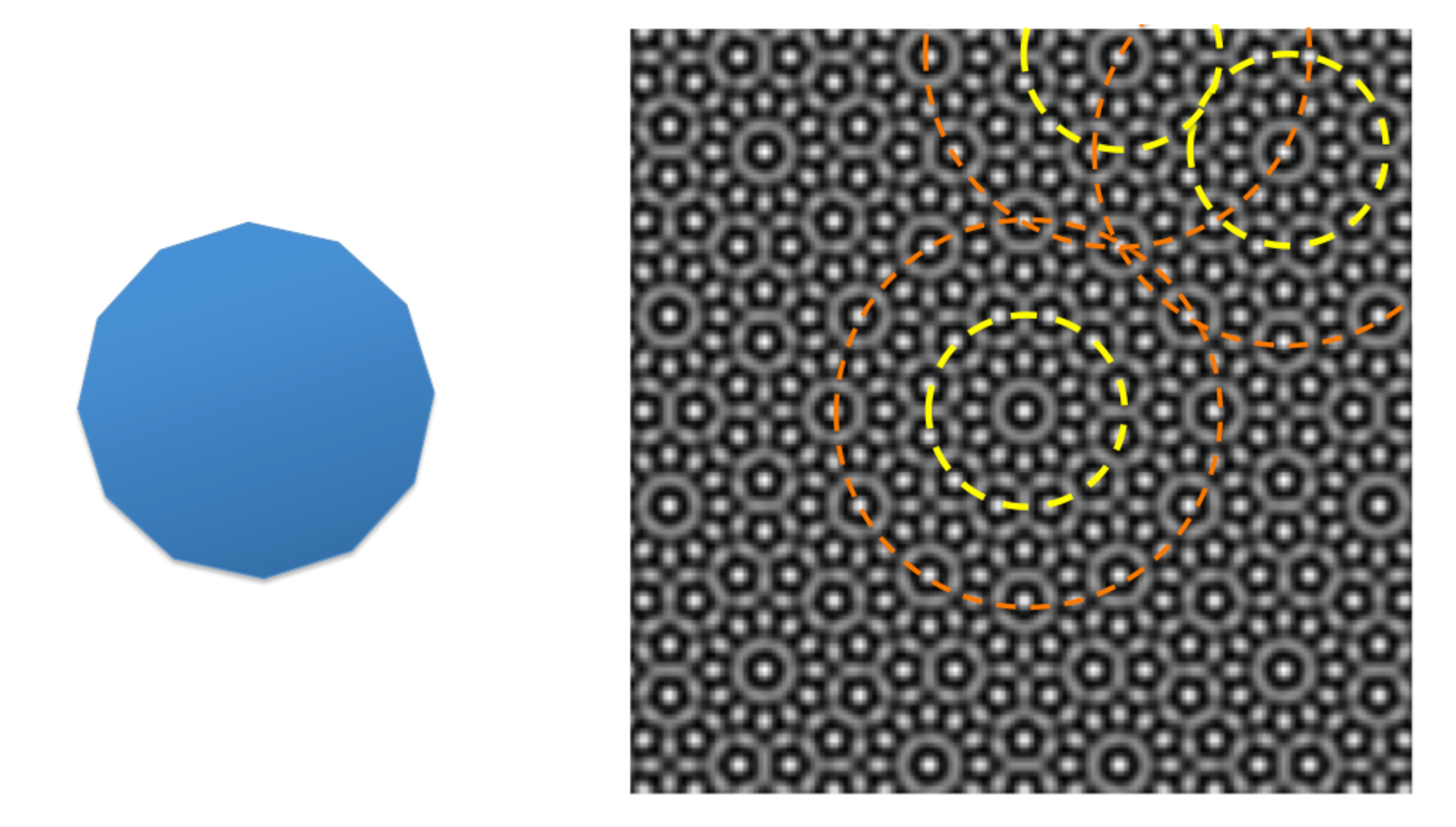} }
\end{center}
\caption{The structures of the chiral density
for $\Np=6$.
(Left) The shape of the Fermi sea.
(Right) The coordinate space distributions
of the chiral density where the center
of figure is chosen as $\vx = \vec{0}$.
(For simplicity, we chose $\vx_i=\vec{0}$.)
Some circles connecting nodes are drawn
as guidelines to look at patterns.
}
\label{fig:Crystals2}
\vspace{0.2cm}
\end{figure}

The situation is, however, different beyond $\Np=3$.
The chiral density has only 
an $Z_{2\Np}$ orientational symmetry,
but does not possess a periodic structure
(The simple proof is given in Appendix~\ref{proof}.)
The spatial distributions show
certain repeating patterns, though.
(See Fig.\ref{fig:Crystals2}).

\begin{figure}[tb]
\vspace{0.0cm}
\begin{center}
\scalebox{1.0}[1.0] {
\hspace{0.2cm}
  \includegraphics[scale=.34]{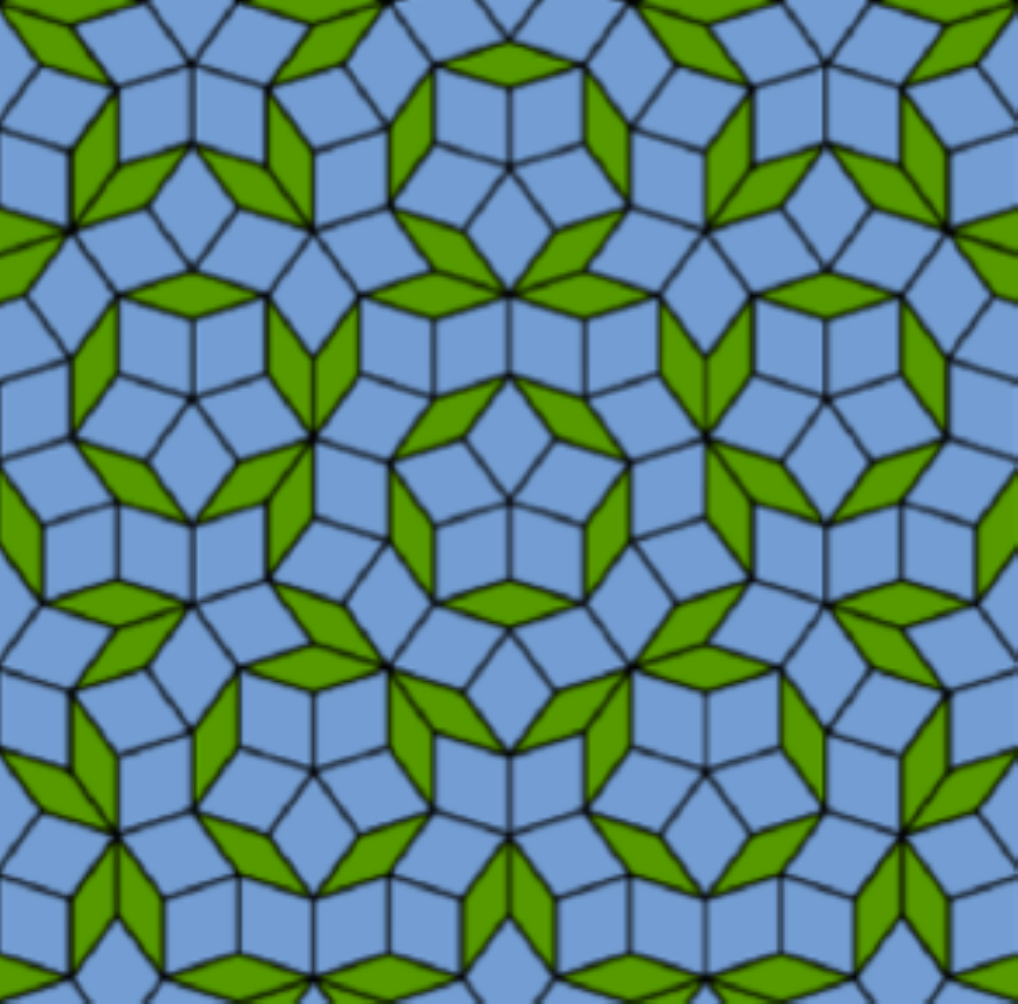} }
\end{center}
\caption{
The Penrose tiling as an example of two dimensional
quasi-crystals.
We have only two types of tiles to cover the entire
space. There is no strict periodicity.
(This figure is taken from 
Ref.~\cite{Penrose}.)
}
\label{fig:Penrose}
\vspace{0.2cm}
\end{figure}

A natural question is
whether interweaving chiral spirals with $\Np \ge 4$
can be classified by structures known 
in the solid state physics.
So far three types of solids have been known:
(i) amorphous (or glassy) structures 
without any of the long-range correlations,
(ii) crystal structures
with periodic translational order
and long-range (near-neighbor bond)
orientational order with special crystallographic
discrete subgroups of the rotational group,
and 
(iii) quasi-crystal structures
that can have arbitrary types of orientational symmetry
and the corresponding 
quasi-periodicity \cite{Shechtman84}.
A nontrivial fact is that in spite of lack of periodicity,
quasi-crystals are made of a finite number of
unit cell species.
In Fig.~\ref{fig:Penrose}
we show a specific example of two dimensional quasi-crystals
which is known as Penrose tiling.

Interweaving chiral spirals state
with $\Np \ge 4$ does not have a periodicity,
so they are not crystals.
Also, the interweaving chiral spirals
have definite rotational properties
and the size of wavevectors,
so perhaps do not belong to amorphous.

It is not clear whether 
our interweaving chiral spirals belong
to the remaining candidate, a quasi-crystal.
In usual solids,
their basic properties can be discussed 
by knowing ion species and their locations.
The separation among ions
are much larger than the size of ions,
so it is good approximation to treat
ion density as a sum of $\delta$-functions.
The electron distribution is determined accordingly.
Then we can attach a unit cell to each ion,
and such cell structures appear repeatedly. 
Therefore studies of unit cells
and their connections to other cells
are enough to know properties of crystals.
The relevant point here is that
a number of cell and bond species is only finite.

In case of interweaving chiral spirals,
it is tempting to assign nodes of
interweaving chiral spirals
as alternatives of positions of ions.
The problem, however, is that
unlike electron density in atomic crystals,
the chiral density around nodes
is not well-determined from
locations of nodes,
perhaps because
the minimum energy configuration at high density
is dominantly determined by
momentum space structures.
We expect that
a number of varieties in cell shapes might be finite
but contents in them might have a lot of varieties.

We leave further discussions about 
the classification of interweaving chiral spirals
for future studies.
We expect that such studies will 
become important for the region where
the system changes from quark to baryonic matter,
or appropriate descriptions for energy minimization
changes from the momentum space to coordinate space one.
 
\subsection{On the Six-Fermi Interaction}

Let us briefly mention on the effects of the six-Fermi interaction
that breaks $U(1)_{\rm A}$ symmetry for three flavors.  When we
discuss dynamics of $u,d$ quarks, we typically take the expectation
value of $\bar{s} s$, and normalize the coupling constant as
\begin{equation}
\sim G_i ~(\bar{u} u) (\bar{d} d) (\bar{s} s) 
~\longrightarrow~
\big( G_i \la\bar{s} s\ra \big) ~ (\bar{u} u) (\bar{d} d)
= G_i^s ~(\bar{u} u) (\bar{d} d) \,,
\end{equation}
where we did not write any explicit structure of the Dirac $\gamma$
matrices.  Thus at zero density, this interaction would renormalize
the coupling constant of our model.

At high density, however, the four-Fermi and the six-Fermi
interactions show interesting differences.
Let us first consider the following situation;
$\mu \simeq \mu_u \simeq \mu_d \simeq \mu_s \gg \lqcd$,
where all light flavors form the quark Fermi sea.  Using the
$\psi'_\pm$-representation, the interaction looks like
\begin{equation}
\sim G_i \big(\,
(\bar{u}'_+ u'_-) (\bar{d}'_+ d'_-) (\bar{s}'_+ s'_-)
\rme^{-6\rmi \mu_{\rm q} x_\para} 
+ 
(\bar{u}'_+ u'_-) (\bar{d}'_+ d'_-) (\bar{s}'_- s'_+)
\rme^{-2\rmi \mu_{\rm q} x_\para} + \cdots \big)\,. 
\end{equation}
Here let us note that all of vertices must have oscillation factors,
i.e., they are subdominant terms.  These terms tend to disturb the
chiral spiral formations, but eventually become negligible at very
high density.

On the other hand, at intermediate density such that only $u$ and $d$
quarks have the Fermi sea, the vertex $\la \bar{s}'s' \ra$ is uniform
and does not include the oscillating factor.  Then we can find the
dominant four-Fermi interactions which help the formation of the
chiral spirals in the $u,d$ channels.  Actually this is the situation
discussed in the aforementioned work~\cite{Rapp:2000zd}.  It is
interesting to see whether any drastic changes happen near the
strange-quark threshold.

\section{Summary}
\label{secsummary}

In this paper we have argued how to construct the interweaving chiral
spirals at high density.  The patch size is determined by the balance
between the energy gain from condensation and the energy cost from the
deformed Fermi surface.  The Fermi surface effects drove the
spontaneous breaking of chiral, translational, and rotational
invariance.

A key ingredient of our discussions was that the interaction among
quarks and the condensate, which is dominant at large $\Nc$, was local
in momentum space.  Because of this property, energy contributions
could be characterized by the limited phase space for the scattering
between quarks and the condensate.  
In particular, complicated interplay
between differently oriented chiral spirals happen only
near the patch boundaries, so we could roughly estimate the energy
cost by just seeing the phase space for patch-patch interactions.

We have argued the (2+1)-dimensional model for which the geometric
shape of the Fermi sea is relatively simple.  In reality with higher
dimensions, however, such simplicity is no longer the case since there
are many ways to interweave chiral spirals.  On the other hand, the
principle to choose the best shape should be relatively simple
according to the arguments in this paper.  We need get the largest
energy gain from condensation by maximizing the area of the Fermi
surface until the kinetic energy cost becomes crucial, and at the same
time, we have to minimize the length of the patch intersection lines
to reduce inter-patch interactions.
To find such a geometric shape would
be a mathematically
well-defined problem.

The interweaving chiral spirals will survive in (3+1) dimensions,
since the key point is the IR enhancement of the interaction, which is
the fundamental property based on QCD.\ \ This situation is quite
different from charge or spin density waves in condensed matter
physics, which are  energetically favored in one-dimensional systems,
but not in higher dimensions \cite{Dagotto}.

The form of the chiral Lagrangian near the Fermi surface was derived
in Ref.~\cite{Kojo:2010fe}, regarding the system near the Fermi
surface as the (1+1)-dimensional chains with the transverse hopping.
In the dispersion relation for the collective modes, transverse
kinetic terms are suppressed by powers of $\lf/Q$.  As the density
increases, therefore, the spectrum will approach the (1+1)-dimensional
one, so that the IR fluctuations will be stronger.  We also anticipate
that the temperature effects strongly enhance the phase 
fluctuations \cite{Baym:1982ca}.
Results on this issue will be reported in the future.

\section*{Acknowledgments}
We thank
G.~Basar,
G.~Dunne,
E.J.~Ferrer,
L.Y.~Glozman,
V.~Incera, 
J.~Liao,
S.~Nakamura,
R.~Rapp,
E.~Shuryak,
G.~Torrieri,
and I.~Zahed for useful comments
and/or raising several important questions
related to multiple (Quarkyonic) chiral spirals.
Special thanks go to A.M.~Tsvelik
for the collaboration \cite{Kojo:2010fe}
with several of the present authors.
We acknowledge the referee 
for constructive questions which have helped us to
improve the original manuscript.
T.K.\ acknowledges to S.~Carignano and M.~Buballa
for explaining their NJL model studies on the chiral crystals 
before the publication.
He also thanks the Asia Pacific Center for Theoretical Physics (APCTP)
and Hashimoto Laboratory in RIKEN Nishina Center for their hospitality
during his visit in March and April 2011.
The research 
of Y.H.\ is supported by RIKEN and
the Grand-in-Aid for the Global COE Program
``The Next Generation of Physics, Spun from
Universality and Emergence'' from the Ministry
of Education, Culture, Science and Technology (MEXT) of Japan;
that of T.K.\ by the Postdoctoral Research Program of RIKEN;
that of L.D.M.\ and R.D.P.\ by the U.S.\ Department of Energy
under contract No.\ DE-AC02-98CH10886.
R.D.P.\ also thanks the Alexander von Humboldt Foundation for their
support.

\appendix
\section{The single particle dispersion for $\vq \neq 2Q\vec{n}$}
\label{vqneq2Q}
\begin{figure}[tb]
\begin{center}
\hspace{-0.0cm}
\scalebox{0.5}[0.5] {
  \includegraphics[scale=.55]{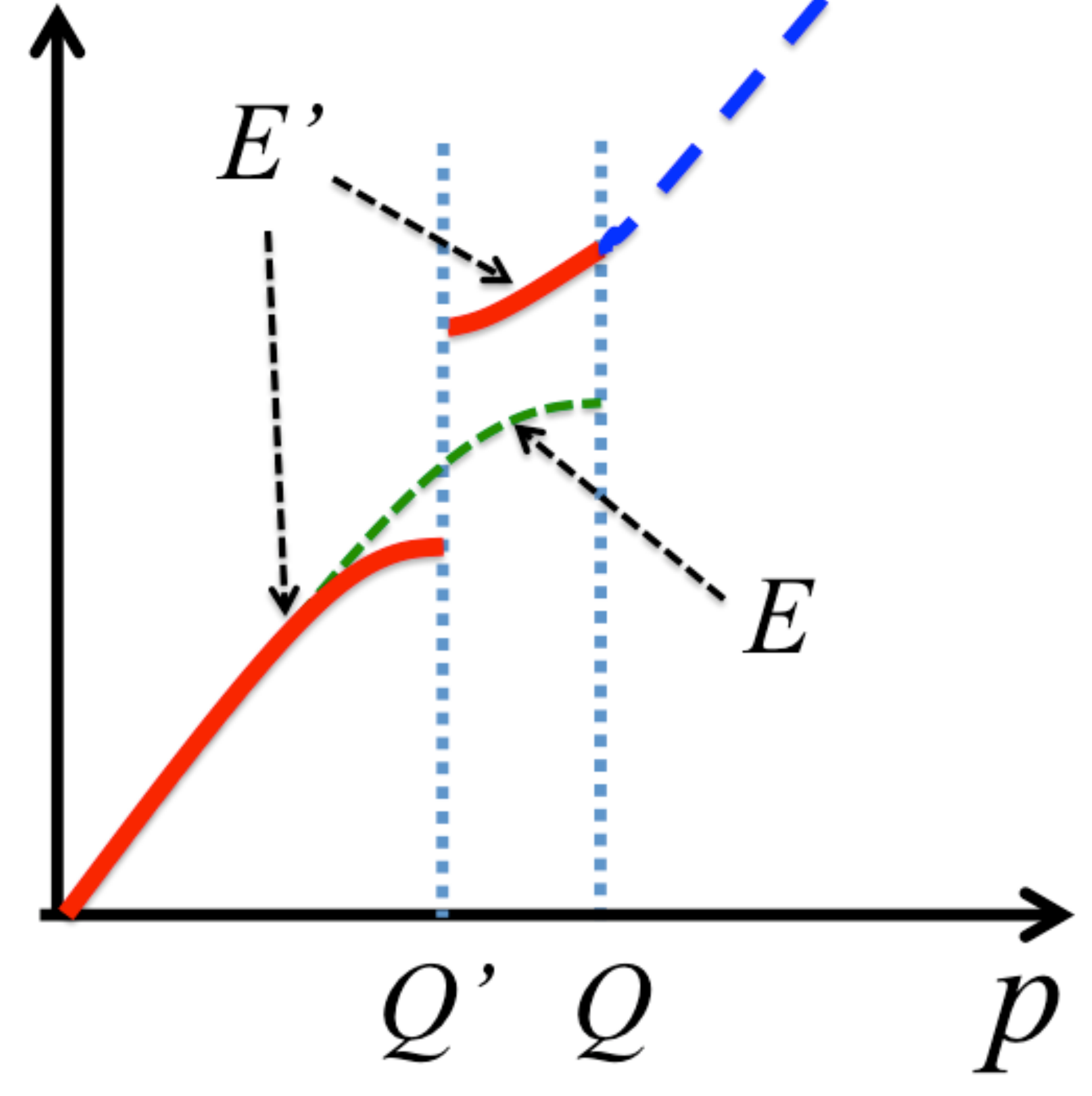} }
\hspace{0.3cm}
\scalebox{0.5}[0.5] {
  \includegraphics[scale=.55]{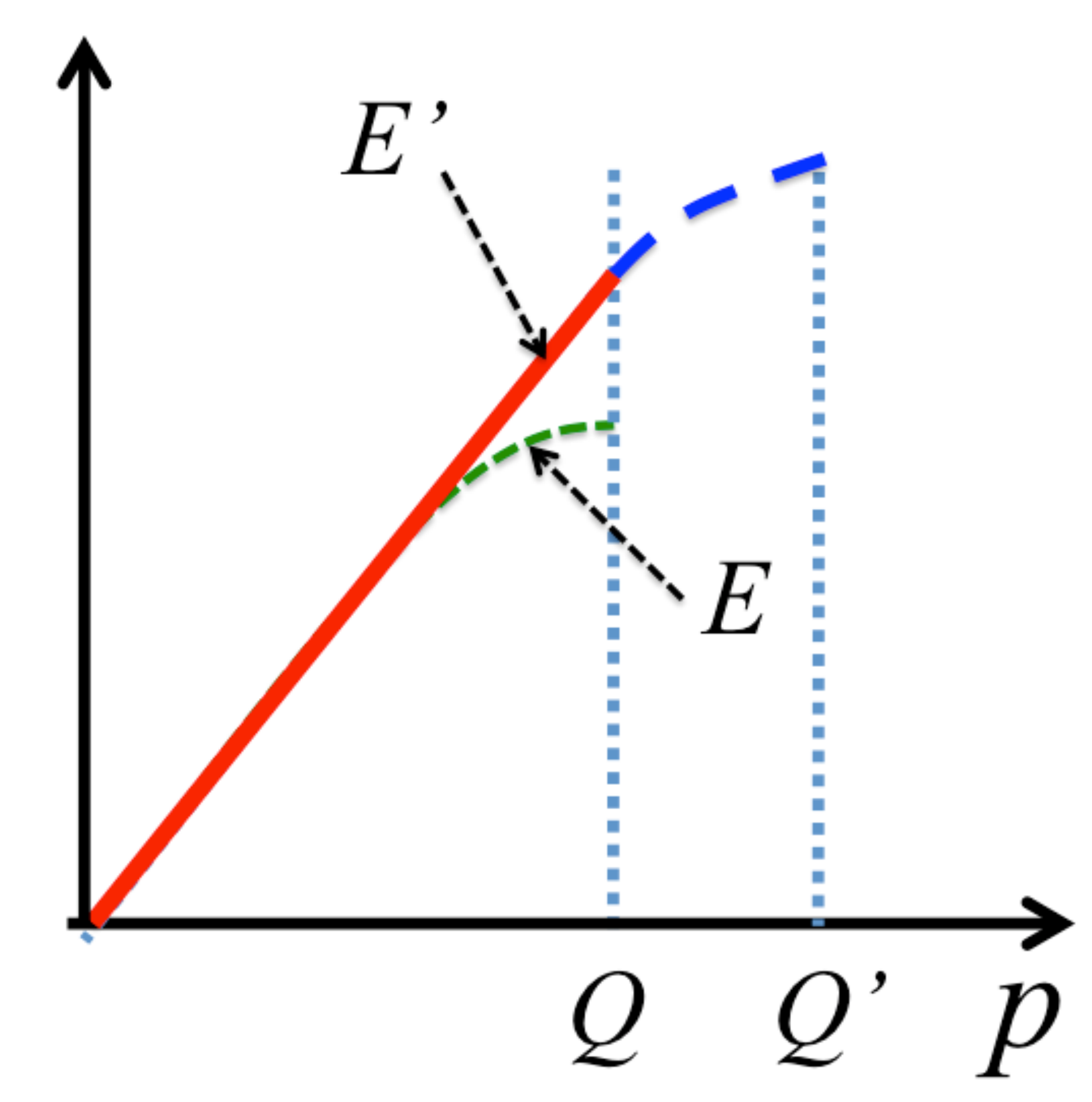} }
\end{center}
\vspace{0.0cm}
\caption{The single particle dispersion $E'$ 
for $Q'<Q$ (left) and $Q'>Q$ (right) cases.
Because of the particle number constraint,
particles must occupy up to $p=Q$.
The $Q'=Q$ case with the energy $E$
is also plotted as a guideline.
}
\vspace{0.3cm}
\label{fig:qneq2Q}
\end{figure}

In this work we have assumed $\vq = 2Q\vec{n}$
in the (1+1) dimensional construction.
Here we show what will happen if we chose a different 
wavevector.
To do this, we go back to the original $\psi$ bases,
and write the eigenvalue equation,
\begin{equation}
\Psi^\dag
\left(
\begin{array}{ccc}
 \rmi \partial_\para  ~&~ M \rme^{2 \rmi Q' x_\para}  \\ 
 M \rme^{-2 \rmi Q' x_\para}  ~&~ - \rmi\partial_\para 
\end{array}
\right) 
\Psi(x_\para) 
= E'_{{\rm MF}}  ~ 
\big( \Psi^\dag \Psi( x_\para ) \big)\,, 
\end{equation}
where we omit the form factor effects on $M$ for the sake of simplicity.
To diagonalize it, we choose 
$\psi_\pm = \psi_\pm'' \rme^{\pm iQ'  x_\para}$, 
and eliminate the oscillating factors.
The energy for the right moving branch is
\begin{equation}
E'_{ {\rm MF} } (p_\para)
= Q' \pm \sqrt{ ( p_\para - Q') ^2 + M^2  } ~ ,
\end{equation}
and its behavior for $Q'<Q$ and $Q'>Q$ is shown 
in Fig.~\ref{fig:qneq2Q}
with the guideline of the $Q'=Q$ case. 
With the particle number constraint,
we have to put particles up to the momentum $Q$.
For $Q'<Q$, the particles occupy the 
upper branch, costing energy.
For $Q'>Q$, the particles occupy only the lower branch,
but the energetic benefit is small
because of the distance from the gap region.
Therefore the case of $Q'=Q$ is the best way to 
minimize the total single particle contributions.
If we add the corrections to the (1+1) dimensional approximations,
it can deviate from $Q'=Q$, though.

%

\section{A component expression of the mean field propagator}
\label{apppro}
In the computation of the perturbative corrections,
it is useful to use the component expressions
which are decomposed into different pole structures.
Explicitly, ($\omega = \omega(\delta \vec{p} )$)
\begin{eqnarray}
&&
S_{++} (\delta p) 
= \rmi \times \frac{1+ \Gamma_5}{2} \times \frac{\gamma^0}{2}~ 
\bigg[ 
\frac{1 + \delta p_\para/\omega } {~\delta p_0 - \omega + i\epsilon~}
+
\frac{1 - \delta p_\para/\omega } {~\delta p_0 + \omega - i\epsilon~}
\bigg]~ , \nonumber \\
&&
S_{--} (\delta p) 
= \rmi \times \frac{1- \Gamma_5}{2} \times \frac{\gamma^0}{2}~
\bigg[ 
\frac{1 - \delta p_\para/\omega } {~\delta p_0 - \omega + i\epsilon~}
+
\frac{1 + \delta p_\para/\omega } {~\delta p_0 + \omega - i\epsilon~}
\bigg]~ ,
\end{eqnarray}
and 
\begin{equation}
S_{+-} (\delta p) = \frac{~1+\Gamma_5~}{2} S_M \,,
~~~~ 
S_{-+} (\delta p) = \frac{~1-\Gamma_5~}{2} S_M \,,
\end{equation}
where
\begin{equation}
S_M(\delta p) = \rmi \times \frac{M}{~2\omega~} ~
\bigg[ 
\frac{1 } {~\delta p_0 - \omega + i\epsilon~}
- \frac{1 } {~\delta p_0 + \omega - i\epsilon~}
\bigg]\,.
\end{equation}
%

\section{There is no periodic structure for $\Np \ge 4$}
\label{proof}
We shall explain why interweaving chiral spirals
with $\Np\ge 4$ can not have
a periodic structure.
For simplicity,
we will consider only wavevectors in differently
oriented chiral spirals have common $Q$.

Suppose that the condensate is 
invariant under the translation,
$\vx \rightarrow \vx - \va$,
\begin{equation}
\left\la \bar{\psi} \psi (\vx) \right\ra
= \left\la \bar{\psi} \psi (\vx+\va) \right\ra
=
\Delta
\sum_{i=0}^{\Np-1}
{\rm Re}
 \left( \rme^{2 \rmi Q \vec{n}_i \cdot \va } ~ \rme^{2 \rmi Q\vec{n}_i \cdot (\vx - \vx_i) } 
\right) ~.
\end{equation}
To satisfy this equation for arbitrary $\vx$, 
the condition
\begin{equation}
2 Q ~\vec{n}_i \cdot \va = 2 \pi m_i ~, ~~~~(m_i: {\rm integer})
\label{condition}
\end{equation}
must be satisfied for all $i$.

We choose $(x,y)$ axis in such a way that
$\vec{n}_k$ is expressed as
\begin{equation}
  \vec{n}_k = \left( \cos \left( \frac{\pi k}{\Np}  \right),~
\sin \left( \frac{\pi k}{\Np}  \right) \right),~
~~( k =0 ,1, \cdots, \Np-1)
\end{equation}
And we take $\va$ as
\begin{equation}
\va = a~ (\cos \varphi, ~\sin \varphi) ~.
\end{equation}
Then the condition (\ref{condition}) is
\begin{equation}
2Qa ~\cos \left( \varphi - \frac{\pi k}{\Np}  \right)
= 2 \pi m_k ~. ~~~~(m_k: {\rm integer})
\label{condition1}
\end{equation}
This condition should be satisfied for all $k$.
We wish to know whether we can choose $a$ and $\varphi$
to satisfy this condition.

The ratio
\begin{equation}
\frac{ \cos \left( \varphi - \frac{\pi k}{\Np}  \right) }
{\cos \varphi}
= \cos \left( \frac{\pi k}{\Np} \right)
+ \tan \varphi ~\sin \left( \frac{\pi k}{\Np} \right)
= \frac{m_k}{m_0} ~,
\end{equation}
must be rational number.
If we also consider $\Np-k$ case,
we obtain
\begin{equation}
\frac{ \cos \left( \varphi - \frac{\pi (\Np-k) }{\Np}  \right) }
{\cos \varphi}
=
-\cos \left( \frac{\pi k}{\Np} \right)
+ \tan \varphi ~\sin \left( \frac{\pi k}{\Np} \right)
= \frac{m_{\Np-k} }{m_0} ~.
\end{equation}
After subtraction, we have
\begin{equation}
2 \cos \left( \frac{\pi k}{\Np} \right)
= \frac{m_k - m_{N-k} }{m_0} ~.
\end{equation}
Therefore the LHS must be rational number
for all $k$.
This is necessary condition.
For $\Np \ge 4$,
$\cos(\pi k/\Np)$ has an irrational number,
and the periodic condition is not satisfied.
%


\end{document}